%% file: main.tex
\documentclass[aps,prb,reprint,superscriptaddress]{revtex4-2}

\usepackage{graphicx}         
\usepackage[percent]{overpic} 
\usepackage{placeins}         
\usepackage{soul}             
\usepackage{siunitx}
\usepackage{mhchem}

\usepackage{tikz}
\usepackage[compat=1.1.0]{tikz-feynman}

\usepackage{amsmath}
\usepackage{amssymb}
\usepackage{amsfonts}
\usepackage{mathtools}
\usepackage{xfrac}
\usepackage[overload]{empheq}
\usepackage{cases}
\usepackage{newtxtext,newtxmath}
\DeclareMathAlphabet\mathbfcal{OMS}{cmsy}{b}{n}
\DeclareSymbolFontAlphabet{\mathcal}{symbols}
\DeclareMathAlphabet{\mathcal}{OMS}{cmsy}{m}{n}
\SetMathAlphabet{\mathcal}{bold}{OMS}{cmsy}{b}{n}

\usepackage[colorlinks=true,citecolor=blue,urlcolor=blue]{hyperref}         
\usepackage{cleveref}

\newcommand{\iu}[0]{\mathrm{i}}
\newcommand{\ec}[0]{\mathrm{e}}
\newcommand{\moatMomentum}[0]{q_{\mathrm{m}}}
\newcommand{\moatEnergyBarrier}[0]{\epsilon_{\mathrm{m}}}
\newcommand{\softCoreRange}[0]{\sigma_{\mathrm{sc}}}
\newcommand{\softCoreStrength}[0]{U_{\mathrm{sc}}}
\newcommand{\intRange}[0]{\sigma}
\newcommand{\intStrength}[0]{U}   

\renewcommand{\vec}[1]{\boldsymbol{#1}}
\newcommand{\dimlessMarker}[1]{\tilde{#1}}

\newcommand{\dimlessMoatMomentum}[0]{\dimlessMarker{q}_{\mathrm{m}}}
\newcommand{\dimlessMoatEnergyBarrier}[0]{\dimlessMarker{\epsilon}_{\mathrm{m}}}
\newcommand{\paramDispKTwo}{a}
\newcommand{\paramDispKFour}{b}
\newcommand{\paramMoatMomVSIntRange}{\lambda}
\newcommand{\paramSSMoat}{\lambda}
\newcommand{\paramSSInt}{\Lambda}
\newcommand{\paramDensityVSIntRange}{\rho}
\newcommand{\paramIntStrengthVSMoatBarrier}{\gamma}
\newcommand{\interactionKernel}{v}
\newcommand{\TMatrixKernel}{t}
\newcommand{\momentumOperator}{\mathbf{p}}

\makeatletter
\g@addto@macro\bfseries{\boldmath}
\makeatother

\begin{document}

\title[Correlated phases of moat-band excitons in two dimensions]{Correlated phases of moat-band excitons in two dimensions}

\author{L. Maisel Licerán}
\email{l.maiselliceran@uu.nl}
\affiliation{Institute for Theoretical Physics and Center for Extreme Matter and Emergent Phenomena, Utrecht University, Princetonplein 5, 3584 CC Utrecht, The Netherlands}

\author{S. H. Boeve}
\affiliation{Institute for Theoretical Physics and Center for Extreme Matter and Emergent Phenomena, Utrecht University, Princetonplein 5, 3584 CC Utrecht, The Netherlands}
 
\author{H. T. C. Stoof}
\affiliation{Institute for Theoretical Physics and Center for Extreme Matter and Emergent Phenomena, Utrecht University, Princetonplein 5, 3584 CC Utrecht, The Netherlands}

\date{\today}

\begin{abstract}
	We study dilute two-dimensional systems of interacting excitons with a moat dispersion, whose ground-state manifold consists of a set of discrete or continuously degenerate energy minima.
	At low densities and in the presence of contact interactions, it is known that the bosons can undergo statistical transmutation and give rise to a chiral spin liquid.
	Here, we show that long-range interactions such as those expected in excitonic systems introduce a rich competition between the chiral spin liquid and different kinds of Bose--Einstein condensates.
	The moat dispersion can favor Bose--Einstein condensation into states occupying multiple momenta, leading to inhomogeneous and supersolid phases with a highly anisotropic superfluid response.
	We demonstrate that a proper $T$-matrix renormalization of the exciton--exciton interaction is essential for describing these phases and show that they can arise even from purely repulsive interactions.
	This formalism is employed to obtain the phase diagram of competing homogeneous and stripe condensates with the chiral spin liquid in an electron--hole bilayer model.
	In addition, we show how the $T$ matrix enters the familiar Gross--Pitaevskii framework and map out extended phase diagrams within a pseudopotential approximation.
	We place our findings in the context of real excitonic systems by discussing the roles of finite lifetimes, disorder, band-structure warping, and nonzero temperatures.
	We conclude that moat bands can drive Bose--Einstein condensation and supersolidity already at weak coupling, in contrast to the case of a standard parabolic dispersion.
\end{abstract}

\maketitle

\section{Introduction}
\label{sec:intro}
\input{sec_introduction}

\section{Theoretical modeling}
\label{sec:modeling}
\input{sec_modeling}

\section{Many-body phases}
\label{sec:MBPhases}
\input{sec_manyBodyPhases}

\section{Superfluidity and BKT physics}
\label{sec:SFResp}
\input{sec_superfluidity}

\section{Gross--Pitaevskii approach to periodic exciton condensates}
\label{sec:GrossPitaevskii}
\input{sec_GrossPitaevskii}

\section{One-dimensional lattices}
\label{sec:1D}
\input{sec_1D}

\section{Two-dimensional lattices}
\label{sec:2D}
\input{sec_2D}

\section{Conclusion and outlook}
\label{sec:conclusion}
\input{sec_conclusion}

\section{Acknowledgments}

This work is supported by the Delta-ITP consortium, which is part of the Netherlands Organisation for Scientific Research (NWO).
We also acknowledge the research program ``Materials for the Quantum Age'' (QuMat) for financial support.
This program (registration number 024.005.006) is part of the Gravitation program financed by the Dutch Ministry of Education, Culture and Science (OCW).
We thank Tigran Sedrakyan and Rui Wang for valuable comments and useful discussions.

\appendix

\input{app_Tmatrix}
\label{app:tmatrix}

\section{Interaction energy of the chiral spin liquid}
\label{app:CSL}
\input{app_CSL}

\bibliography{references}

\end{document}

%% file: sec_introduction.tex

Excitons are bound electron--hole pairs in semiconductors that can display bosonic statistics due to their composite fermionic nature.
Well over half a century ago, this observation led to the proposal of an excitonic Bose--Einstein condensate (BEC) \cite{blatt1962bose,keldysh1968collective} and the related excitonic insulator phase \cite{keldysh1965possible,jerome1967excitonic,halperin1968possible}.
Experimental signatures of excitonic condensation were finally observed over three decades later in coupled quantum wells \cite{butov1994condensation,butov2001stimulated,butov2002macroscopically,butov2002towards} and, shortly thereafter, in quantum-Hall bilayers \cite{eisenstein2004bose,eisenstein2014exciton}.
Two-dimensional (2D) exciton systems have since emerged as promising platforms to realize novel correlated many-body phases even beyond the Bose--Einstein paradigm.
The rise of highly tunable material platforms such as stacked (and possibly twisted) transition-metal dichalcogenides has facilitated the experimental realization of correlated phases even in the absence of strong magnetic fields \cite{brem2020tunable,wilson2021excitons,regan2022emerging,moon2025exciton,qi2025competition}.
Theoretically proposed phases include interlayer exciton superfluids \cite{lozovik1976new,fogler2014high,li2017excitonic,berman2016high,berman2017bose,berman2017superfluidity}, supersolid and Mott-moir{\'e} excitons \cite{julku2022nonlocal,huang2023mott,dai2024strong}, topological flat-band excitons \cite{froese2025topological}, and Wigner-crystal excitons \cite{you2025moire}.
Experimentally, excitonic Mott insulators \cite{xiong2023correlated,gao2024excitonic,lian2024valley}, interlayer superfluid order \cite{cutshall2025imaging}, and strongly correlated exciton--electron (Fermi-polaron) states \cite{miao2021strong,yan2025anomalously} have all been reported.
More recently still, the condensation of excitons composed of fractional charges has been realized in quantum-Hall bilayers \cite{zhang2025excitons,nguyen2026bilayer}, extending the excitonic paradigm to the realm of fractionalized quasiparticles.

These systems all have in common that the kinetic energy of electrons and holes, and consequently that of the excitons themselves, follows the characteristic quadratic dispersion.
However, another possibility that has attracted significant interest is that of bosons on a moat band, where the low-energy manifold comes in the form of a ring-shaped valley in momentum space containing a highly degenerate continuum of ground states.
The realization of moat bands has been explored in bipartite lattices with next-nearest-neighbor hopping \cite{varney2011kaleidoscope,sedrakyan2014absence,sedrakyan2015spontaneous}, degenerate quantum gases with Rashba spin--orbit coupling \cite{campbell2011realistic,gopalakrishnan2011universal,berg2012electronic,ozawa2012stability,galitski2013spin,zhai2015degenerate}, and ultracold atoms in a Floquet-driven optical lattice \cite{bracamontes2022realization}, and two of us have recently shown that the camelback feature of the electron and hole bands in band-inverted topological insulators can be inherited by the excitons to realize an excitonic moat band \cite{maisel2023single}.

The large degeneracy of the moat ground state is expected to drive strong correlations between the excitons.
An exciting avenue first studied in Refs.\ \cite{sedrakyan2013composite,sedrakyan2014absence,sedrakyan2015spontaneous,sedrakyan2015statistical,wei2023chiral} is that bosons on a moat band can undergo statistical transmutation.
In this picture, they are described as fermions with a unit flux-quantum attachment which occupy the lowest Landau level corresponding to the emergent effective magnetic field.
The result is a chiral spin liquid (CSL) state that realizes long-range quantum entanglement and topological order \cite{wang2022emergent,wang2024susceptibility}.
Strong signatures of moat-band-induced excitonic topological order were recently observed in imbalanced \ce{InAs}/\ce{GaSb} semiconductor bilayers at low densities \cite{wang2023excitonic}.

\begin{figure*}[!t]
	\centering
	\includegraphics[width=\textwidth]{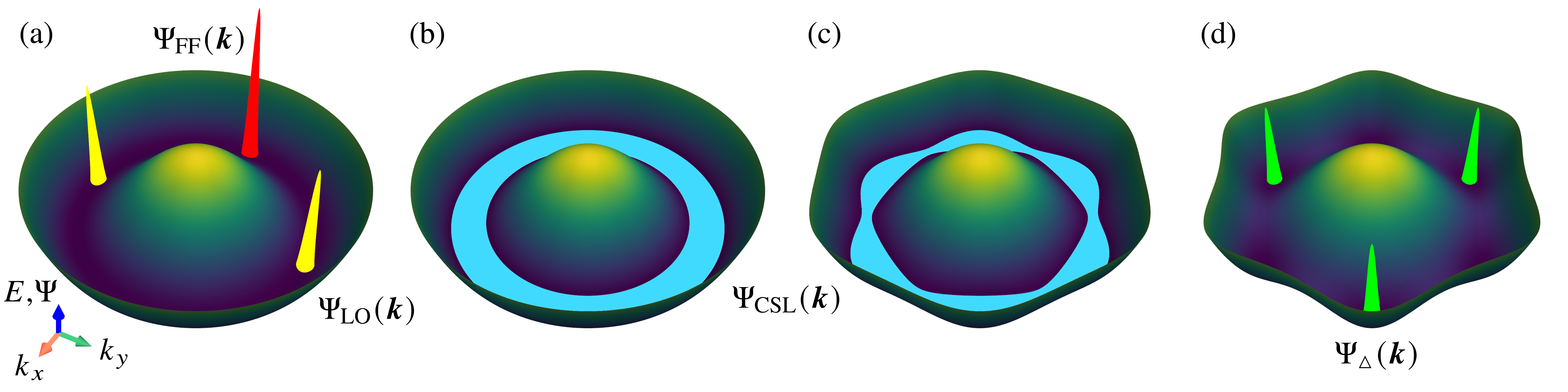}
	\caption{Momentum-space wave functions of different competing many-body phases on a moat band, whose low-energy manifold consists of a ring of degenerate states at some nonzero momentum.
	In (a) we show two kinds of Bose--Einstein condensates, namely the Fulde--Ferrel phase (FF, in red) occupying a single momentum on the moat, and the Larkin--Ovchinnikov phase (LO, in yellow) occupying two opposite momenta.
	The momentum distribution of the latter leads to an inhomogeneous stripe density pattern.
	Depending on the density and the details of the moat and the long-range exciton--exciton interaction, these condensates compete with a chiral spin liquid (CSL, in cyan) shown in (b), a noncondensed topological state where the bosons tend to occupy all available momenta on the moat.
	In real systems, warping effects from the underlying crystalline lattice can deform the moat and lift its degeneracy, resulting in a finite number of discrete minima along the ring.
	This results in a deformation of the CSL, which acquires a degree of anisotropy as shown in (c).
	Meanwhile, Bose--Einstein phases are relatively agnostic to this effect as their momentum distribution is strongly localized at a finite number of points, but the warping will tend to select only those phases whose discrete rotational symmetry is commensurate with that of the lattice.
	For hexagonal warping in a $C_{3}$-symmetric crystal one possibility is the triangle phase shown in (d), combining superfluidity with the inhomogeneous density profile of a crystalline solid.}
	\label{fig:introSummary}
\end{figure*}

Here we propose dilute 2D excitonic systems with a moat dispersion and long-range interactions as a platform for realizing competing exotic phases of correlated bosonic matter.
Besides the excitonic CSL described above, another intriguing possibility is the formation of an excitonic BEC phase of excitons, with the degenerate moat offering the possibility of realizing a coherent state of multiple simultaneously occupied momenta.
This leads to breaking of the translational symmetry and can result in a supersolid, a phase combining the solid-like elastic response to shear stress with the superfluid flow of the BEC \cite{PhysRevA.2.256,1969JETP...29.1107A,PhysRevLett.25.1543,nature08913}.
While the potential supersolidity of excitons has been investigated before in systems with conventional parabolic bands \cite{joglekar2006wigner,conti2023chester,conti2025gross,deoliveira2025striped}, we show that a moat band can lead to inhomogeneous condensates also in the limit of weak interactions and yields an unusual superfluid response.

Our work presents a unified framework for many-body phases of excitons on moat bands.
In particular, we highlight how an appropriate $T$-matrix renormalization of the bare exciton--exciton interaction is crucial to obtain inhomogeneous BEC phases and exemplify this in the case of a simple bilayer potential between dipolar excitons.
Additionally, we use a Gross--Pitaevskii framework to study excitonic supersolids across a larger parameter region.
We also review the theory behind the statistical transmutation of bosons on moat bands and compare the energies of the resulting CSL phases with those of the BECs in the aforementioned electron--hole bilayer model.
We demonstrate that long-range interactions between excitons introduce a direct competition between the BEC phases and the CSL that is largely absent in the case of contact interactions, and obtain associated phase diagrams that can guide future experimental realizations of the proposed systems.
We discuss effects associated with band-structure warping, disorder, and nonzero temperatures that are relevant for the physics of real systems.
Our findings, which are summarized in Fig.\ \ref{fig:introSummary}, establish moat-band excitons as an exciting avenue for the exploration of correlated phases and further motivate the use of highly tunable 2D material platforms for the realization of novel interacting bosonic matter.

This article is organized as follows.
In Sec.\ \ref{sec:modeling} we introduce the system and a convenient set of three dimensionless parameters that fully characterizes the problem.
In Sec.\ \ref{sec:MBPhases} we develop the theory behind the BEC and CSL phases, paying particular attention to the $T$-matrix renormalization.
We compare their energies in the case of a simple bilayer potential and discuss the phenomenon of band-structure warping in real systems.
In Sec.\ \ref{sec:SFResp} we study the superfluid response of a homogeneous or inhomogeneous moat-band BEC.
In Sec.\ \ref{sec:GrossPitaevskii} we introduce a Gross--Pitaevskii framework for the treatment of inhomogeneous and supersolid phases of excitons within a pseudopotential approximation.
In Secs.\ \ref{sec:1D} and \ref{sec:2D} we use this framework to compute the phase diagram of a simplified model for excitonic BECs in one and two dimensions, respectively, and develop an analytical Landau theory that matches the numerical Gross--Pitaevskii results.
Finally, in Sec.\ \ref{sec:conclusion} we conclude our work and give an outlook for further research.

%% file: sec_modeling.tex

We consider a 2D system of $N$ Wannier excitons in the dilute limit, meaning that the exciton number density $n$ and the exciton radius $a_{X}$ satisfy $na_{X}^{2} \ll 1$.
In terms of bosonic field operators $\hat{b}(\vec{r})$ and $\hat{b}^{\dagger}(\vec{r})$ representing the excitons, the Hamiltonian of the system is written as (we set $\hbar = 1$)
\begin{equation}
\label{eq:BosonicH}
   \begin{split}
      \hat{H} &= \int \mathrm{d}^{2} r \, \hat{b}^{\dagger}(\vec{r}) K(\momentumOperator) \hat{b}(\vec{r}) \\
      &+ \frac{1}{2} \int \mathrm{d}^{2} r \, \mathrm{d}^{2} r' \, \hat{b}^{\dagger}(\vec{r}) \hat{b}^{\dagger}(\vec{r}') V(\vec{r} - \vec{r}') \hat{b}(\vec{r}') \hat{b}(\vec{r}) ,
   \end{split}
\end{equation}
where $K$ is the kinetic-energy operator, $\momentumOperator = {-} \iu \vec{\nabla}$ is the momentum operator, and $V$ is the \emph{bare} exciton--exciton interaction, assumed here to depend only on the distance $|\vec{r} - \vec{r}'|$.
As explained below, the bare interaction will have to be appropriately renormalized for the treatment of excitonic BECs.

Note that, while the bosonic approximation for excitons requires them to be relatively dilute, it does not imply that their internal structure must be completely neglected.
Even within the bosonic treatment, the latter can be effectively taken into account to a high degree of accuracy by appropriately modeling the interaction between the excitons.
As shown in Ref.\ \cite{noordman2026variational}, this interaction can naturally include scatterings between all different constituents as well as internal exchanges of identical particles directly associated with their composite nature.
These effects are then naturally incorporated into the $T$-matrix formalism described in this section, and in particular we stress that the dependence of $V$ on $|\vec{r} - \vec{r}'|$ only in Eq.\ \eqref{eq:BosonicH} is a simplification and not a requirement.

In this article we will study the many-body phases of this model via mean-field theory at zero temperature, where the mean-field approach is valid.
We will briefly comment on our expectations for the physics at nonzero temperatures.

\subsection{Moat dispersion}

We consider a kinetic energy of the form
\begin{equation}
\label{eq:KEOperator}
	K(\momentumOperator) = \frac{a^{2}}{4 b} - a \momentumOperator^{2} + b \momentumOperator^{4} ,
\end{equation}
characterized by two parameters $\paramDispKTwo$ and $\paramDispKFour$.
When $\paramDispKTwo > 0$ the single-particle dispersion reads $\varepsilon_{\vec{k}} = b (k^{2} - \moatMomentum^{2})^{2}$, where $k \equiv |\vec{k}|$.
This has the form of a Higgs potential and features a moat in 2D, i.e., a ring-shaped valley with radius $\moatMomentum \equiv \sqrt{\paramDispKTwo / 2 \paramDispKFour}$.
In the 1D case, the ``moat'' is simply a pair of minima at $\pm \moatMomentum$.
Also, the constant shift in Eq.\ \eqref{eq:KEOperator} is such that the kinetic energy vanishes at the moat momentum, so the moat energy barrier is $\moatEnergyBarrier \equiv \varepsilon_{\vec{0}} = \paramDispKTwo^{2} / 4 \paramDispKFour$.
Our task will be to determine how the presence of a moat band can influence the phase diagram of the interacting system.

Note that Eq.\ \eqref{eq:KEOperator} corresponds to the lowest-order terms in a gradient expansion of a moat-shaped kinetic energy and behaves quartically at large momenta.
This poses no issue as a proof of principle, but in realistic settings we expect a quadratic behavior at large $k$.
If desired, one can take a dispersion of the form $(k -\moatMomentum)^{2} / 2 M$ instead, which however corresponds to a nonlocal operator in position space.
Ultimately, the qualitative (and often quantitative) behavior of the system is mostly the same in any scenario, and we will explicitly point out the few differences when they arise.

\subsection{Effective interaction}

Regarding the bare interaction potential $V(r)$, it will be crucial that it is \emph{not} a simple contact interaction.
Indeed, in many systems it is expected on physical grounds that the exciton--exciton interaction is relatively long-ranged.
For instance, the pair potential between two excitons in the $s$-wave ground state in monolayers of conventional semiconductors displays an attractive ${-} 1/r^{6}$ behavior at long distances due to the induced dipole--dipole nature of the interaction, while in electron--hole bilayer systems the excitons form permanent dipoles leading to a repulsive $1/r^{3}$ behavior at large separations.
We thus characterize the exciton--exciton interaction via a potential $V(r)$ with a characteristic strength $\intStrength$ and a characteristic length scale $\intRange$ via
\begin{equation}
	V(r) = \intStrength \hspace{0.1mm} \interactionKernel(r / \intRange) ,
\end{equation}
where the dimensionless function $\interactionKernel$ will be referred to as the interaction kernel.
In momentum space the interaction is written as $V_{\vec{q}} = 2 \pi \intStrength \intRange^{2} \mathfrak{h}_{0}[\interactionKernel](q \intRange)$, where $\mathfrak{h}_{n}[f](x) \equiv \int_{0}^{\infty} \mathrm{d}u \, u f(u) J_{n}(xu)$ is the order-$n$ Hankel transform of $f$.
Due to the rapid decay expected at large distances, we will assume that the interaction is integrable, $\int \mathrm{d}^{2} r \, V(r) < \infty$, such that the zero-momentum Fourier component $V_{0}$ is finite.

In a significant portion of this paper we will be concerned with BEC phases of excitons.
For these kinds of many-body states, where the ground-state wave function $\Psi$ at the mean-field level is given by a product of identical wave functions, it is incorrect to use the bare interaction potential between two particles.
Indeed, the latter usually contains an ultraviolet divergence, which does not accurately reflect the fact that in the real many-body system the particles will tend to avoid each other at short distances.
Thus, it is expected that this divergence of the bare potential in the full theory is canceled by the suppression of the many-body wave function $\Psi(\{\vec{r}_{i}\})$ when two of the coordinates come close to each other.
To account for this, when dealing with a BEC the bare microscopic interaction potential appearing in Eq.\ \eqref{eq:BosonicH} must be replaced by an effective energy-dependent many-body $T$ matrix accounting for interparticle correlations at short distances \cite{bijlsma1997variational,proukakis1998comparison,lee2002energy,gies2005many}.
Furthermore, it has been previously argued \cite{stoof1993kosterlitz,andersen2002phase,alkhawaja2002low,rajagopal2004density} that in 2D the many-body $T$ matrix of a cold Bose gas at low energies can be approximated by the two-body $T$ matrix at energy ${-}2\mu$, where $\mu$ is the chemical potential, as this is precisely the energy cost to excite two atoms from the condensate \footnote{In Ref.\ \cite{lee2002energy}, a somewhat more detailed (but not fully rigorous) argument is used to justify that the 2D many-body $T$ matrix should be replaced by the two-body one at an energy ${-}\mu$, as opposed to ${-}2 \mu$. However, the calculation depends on the precise form of the single-particle dispersion and is therefore not universal. In view of this uncertainty we choose here to consider the ${-}2 \mu$ energy arising from the intuitive physical picture. In any case, we do not expect the results to be affected significantly by this particular aspect of the theory.}.
In Appendix\ \ref{app:TMatrixCorrelationsEquiv} we show that using the $T$ matrix is indeed equivalent to including short-range correlations into the many-body wave function.

\begin{figure}[!t]
\centering
\resizebox{\linewidth}{!}{
\begin{tikzpicture}
  \begin{feynman}
    \vertex (a1) at (0,2);
    \vertex (a2) at (0,0);
    \vertex (b1) at (1,2);
    \vertex (b2) at (1,0);
    \vertex (c1) at (3,2);
    \vertex (c2) at (3,0);
    \vertex (d1) at (4, 2);
    \vertex (d2) at (4, 0);
    \diagram* {
      (a1) -- [plain, line width = 0.75mm] (b1) -- [plain, line width = 0.75mm] (c1) -- [plain, line width = 0.75mm] (d1),
      (a2) -- [plain, line width = 0.75mm] (b2) -- [plain, line width = 0.75mm] (c2) -- [plain, line width = 0.75mm] (d2),
      (b1) -- [plain, line width = 0.75mm] (b2),
      (c1) -- [plain, line width = 0.75mm] (c2),
    };
  \end{feynman}

  \node at (2, 1) {\LARGE $T^{\mathrm{MB}}$};

  \node at (4.5,1) {\Large $=$};

  \begin{feynman}
    \vertex (e1) at (5,2);
    \vertex (e2) at (5,0);
    \vertex (f1) at (6,2);
    \vertex (f2) at (6,0);
    \vertex (g1) at (7,2);
    \vertex (g2) at (7,0);
    \diagram* {
      (e1) -- [plain, line width = 0.75mm] (f1) -- [plain, line width = 0.75mm] (g1),
      (e2) -- [plain, line width = 0.75mm] (f2) -- [plain, line width = 0.75mm] (g2),
      (f1) -- [boson, line width = 0.75mm] (f2),
    };
  \end{feynman}

  \node at (7.5,1) {\Large $+$};
  \node at (11, 1) {\LARGE $T^{\mathrm{MB}}$};

  \begin{feynman}
    \vertex (h1) at (8,2);
    \vertex (h2) at (8,0);
    \vertex (i1) at (9,2);
    \vertex (i2) at (9,0);
    \vertex (j1) at (10, 2);
    \vertex (j2) at (10, 0);
    \vertex (k1) at (12, 2);
    \vertex (k2) at (12, 0);
    \vertex (l1) at (13, 2);
    \vertex (l2) at (13, 0);
    \diagram* {
      (h1) -- [plain, line width = 0.75mm] (i1) -- [plain, line width = 0.75mm] (j1) -- [plain, line width = 0.75mm] (k1)-- [plain, line width = 0.75mm] (l1),
      (h2) -- [plain, line width = 0.75mm] (i2) -- [plain, line width = 0.75mm] (j2) -- [plain, line width = 0.75mm] (k2)-- [plain, line width = 0.75mm] (l2),
      (i1) -- [boson, line width = 0.75mm] (i2),
      (j1) -- [plain, line width = 0.75mm] (j2),
      (k1) -- [plain, line width = 0.75mm] (k2)
    };
  \end{feynman}

\end{tikzpicture}}
\caption{Diagrammatic Lippmann--Schwinger equation for the many-body $T$ matrix.
The solid lines correspond to many-body propagators and the squiggly lines represent the bare interaction.}
\label{fig:TMatrixDiagramMB}
\end{figure}

The many-body $T$ matrix is diagrammatically given by the ladder sum depicted in Fig.\ \ref{fig:TMatrixDiagramMB}.
The two-body $T$ matrix can be obtained by replacing the many-body propagators in this diagrammatic Lippmann--Schwinger equation by noninteracting propagators and neglecting the bosonic occupation factors corresponding to the excited states.
This leads to
\begin{equation}
\label{eq:TMatrixEqWithDimensions}
    \begin{split}
        &T^{\mathrm{2B}}_{\vec{k}\vec{k}'}(\vec{K}, E) = V_{\vec{k}-\vec{k}'} \\
        &+ \frac{1}{\mathcal{A}} \sum_{\vec{p}} V_{\vec{k}-\vec{p}} \frac{1}{E - \varepsilon_{\vec{K}/2 + \vec{p}} - \varepsilon_{\vec{K}/2 - \vec{p}} + \iu 0^{+}} T^{\mathrm{2B}}_{\vec{p} \vec{k}'}(\vec{K}, E) ,
    \end{split}
\end{equation}
where $\mathcal{A}$ is the surface area of the 2D system and the summation can be replaced by an integral in the thermodynamic limit via $\sum_{\vec{p}} \rightarrow \mathcal{A} \int \mathrm{d}^{2} p / (2 \pi)^{2}$.
We see that setting $E = {-} 2 \mu$ results in a fully real $T$ matrix, so the infinitesimal imaginary part in the denominator can safely be neglected for our purposes.
With this definition, the $T$ matrix enters the second-quantized Hamiltonian as
\begin{equation}
\label{eq:HintWithTMatrix}
   \hat{H}_{\text{int}} = \frac{1}{2 \mathcal{A}} \! \sum_{\vec{K} \vec{k} \vec{k}'} T^{2\mathrm{B}}_{\vec{k} \vec{k}'}(\vec{K}, {-}2\mu) \hspace{0.3mm} \hat{b}_{\frac{\vec{K}}{2} + \vec{k}}^{\dagger} \hat{b}_{\frac{\vec{K}}{2} - \vec{k}}^{\dagger} \hat{b}_{\frac{\vec{K}}{2} - \vec{k}'} \hat{b}_{\frac{\vec{K}}{2} + \vec{k}'} .
\end{equation}
This $\hat{H}_{\text{int}}$ is the correct interaction Hamiltonian to consider when dealing with BECs and is meant to replace the interaction term in the Hamiltonian of Eq.\ \eqref{eq:BosonicH} containing the bare interaction.

We also note that the bare local interaction $V_{\vec{k}}$ in Eq.\ \eqref{eq:TMatrixEqWithDimensions} may be straightforwardly replaced by a more general, possibly nonlocal bare interaction $V_{\vec{k} \vec{k}'}(\vec{K})$ depending also on the total momentum $\vec{K}$, which may be more appropriate if one desires to capture the full detailed behavior of excitonic systems \cite{noordman2026variational}.
As already mentioned above, this allows for a systematic inclusion of internal scattering and exchange processes that can accurately reflect the composite nature of the excitons.

\subsection{Dimensionless parameters}
\label{sec:dimlessUnits}

Within our dilute approximation $n a_{X}^{2} \ll 1$ and for $a > 0$, i.e., in the presence of a moat, the problem of interacting excitons involves the three length scales $\moatMomentum^{-1}$, $n^{-1/2}$, and $\intRange$, and the two energy scales $\moatEnergyBarrier$ and $\intStrength$.
By the Buckingham pi theorem, this means that the physics are governed by exactly three dimensionless parameters.
Motivated by this observation, we form the combinations
\begin{subequations}
	\begin{align}
		\label{paramLambda}
		\paramMoatMomVSIntRange &\equiv {-} a\intRange^{2}/2b , \\
		\label{paramRho}
		\paramDensityVSIntRange &\equiv \sqrt{\smash[b]{\pi n \intRange^{2}}} , \\
		\label{paramGamma}
		\paramIntStrengthVSMoatBarrier &\equiv \intStrength \intRange^{4} / b .
	\end{align}
\end{subequations}
Note that these are also well-defined when $a < 0$, thus in the absence of a moat.
When $a > 0$, we can rewrite the first and third in terms of the moat parameters as $\paramMoatMomVSIntRange = {-} (\moatMomentum \intRange)^{2}$ and $\paramIntStrengthVSMoatBarrier = (\intStrength / \moatEnergyBarrier) (\moatMomentum \intRange)^{4} = \paramMoatMomVSIntRange^{2} \intStrength / \moatEnergyBarrier$.
The parameters $\paramMoatMomVSIntRange$ and $\paramDensityVSIntRange$ compare the range of the interaction with the moat radius and the typical interexciton distance, respectively, while $\paramIntStrengthVSMoatBarrier$ describes a competition between the interaction strength and the moat energy barrier and plays the role of an effective interaction strength.

To work with these parameters, we make all positions and momenta dimensionless as $\dimlessMarker{\vec{r}} \equiv \vec{r} / \intRange$ and $\dimlessMarker{\vec{k}} \equiv \vec{k} \intRange$, respectively (thus in particular $\dimlessMoatMomentum \equiv \moatMomentum \intRange$), and rescale all energies via $\dimlessMarker{E} \equiv E / (b / \intRange^{4})$.
Thus, the dimensionless dispersion is $\dimlessMarker{\varepsilon}_{\dimlessMarker{\vec{k}}} = (\dimlessMarker{k}^{2} + \paramMoatMomVSIntRange)^{2}$ and the dimensionless interaction in position space is $\dimlessMarker{V}(\dimlessMarker{r}) = \paramIntStrengthVSMoatBarrier \interactionKernel(\dimlessMarker{r})$.
We accordingly define the interaction kernel in momentum space as $\interactionKernel_{\dimlessMarker{\vec{q}}} \equiv \mathfrak{h}_{0}[v](\dimlessMarker{q})$, and thus the dimensionless interaction in momentum space is $\dimlessMarker{V}_{\dimlessMarker{\vec{q}}} = 2 \pi \paramIntStrengthVSMoatBarrier \interactionKernel_{\dimlessMarker{\vec{q}}}$.
In analogy with $V_{\vec{q}}$, we define a dimensionless $T$-matrix kernel $t$ via $T^{\mathrm{2B}}_{\vec{k} \vec{k}'}(\vec{K}, E) \equiv 2 \pi \intStrength \intRange^{2} t_{\dimlessMarker{\vec{k}} \dimlessMarker{\vec{k}}'}(\dimlessMarker{\vec{K}}, \dimlessMarker{E})$, such that the dimensionless $T$ matrix is $\dimlessMarker{T}^{\mathrm{2B}}_{\dimlessMarker{\vec{k}} \dimlessMarker{\vec{k}}'}(\dimlessMarker{\vec{K}}, \dimlessMarker{E}) = 2 \pi \paramIntStrengthVSMoatBarrier t_{\dimlessMarker{\vec{k}} \dimlessMarker{\vec{k}}'}(\dimlessMarker{\vec{K}}, \dimlessMarker{E})$.
The equation for the kernel $\TMatrixKernel$ reads
\begin{equation}
\label{eq:TMatrixKernelEq}
	\TMatrixKernel_{\dimlessMarker{\vec{k}} \dimlessMarker{\vec{k}}'}(\dimlessMarker{\vec{K}}, \dimlessMarker{E}) = \interactionKernel_{\dimlessMarker{\vec{k}}-\dimlessMarker{\vec{k}}'} + \frac{2 \pi \paramIntStrengthVSMoatBarrier}{\dimlessMarker{\mathcal{A}}} \sum_{\dimlessMarker{\vec{p}}} \frac{\interactionKernel_{\dimlessMarker{\vec{k}}-\dimlessMarker{\vec{p}}} \hspace{0.6mm} \TMatrixKernel_{\dimlessMarker{\vec{p}} \dimlessMarker{\vec{k}}'}(\dimlessMarker{\vec{K}}, \dimlessMarker{E})}{\dimlessMarker{E} - \dimlessMarker{\varepsilon}_{\dimlessMarker{\vec{K}}/2 + \dimlessMarker{\vec{p}}} - \dimlessMarker{\varepsilon}_{\dimlessMarker{\vec{K}}/2 - \dimlessMarker{\vec{p}}}} ,
\end{equation}
where $\dimlessMarker{\mathcal{A}} \equiv \mathcal{A} / \intRange^{2}$ and evidently $\sum_{\dimlessMarker{\vec{p}}} \rightarrow \dimlessMarker{\mathcal{A}} \int \mathrm{d}^2 \dimlessMarker{p} / (2\pi)^{2}$ in the thermodynamic limit.

Finally, we note that if the quadratic moat with dispersion $(k - \moatMomentum)^{2} / 2M$ is used, then it becomes more convenient to define $\paramMoatMomVSIntRange \equiv {-} \moatMomentum \intRange$ and to measure energies in units of $(2 M \intRange^{2})^{{-}1}$, so that $\dimlessMarker{\varepsilon}_{\dimlessMarker{\vec{k}}} = (\dimlessMarker{k} + \paramMoatMomVSIntRange)^{2}$ and $\paramIntStrengthVSMoatBarrier \equiv 2 M \intStrength \intRange^{2} = (\intStrength / \moatEnergyBarrier) (\moatMomentum \intRange)^{2}$, where now $\moatEnergyBarrier \equiv \moatMomentum^{2} / 2M$.

%% file: sec_manyBodyPhases.tex

On an ideal moat band there is a macroscopically large number of available states, leading to a phase competition that we explain in this section.
For concreteness and in order to illustrate the basic physics of the problem, in this section we will consider an electron--hole bilayer system with a simple toy potential modeling the interaction between two excitons, given by
\begin{equation}
\label{eq:BilayerXXInt}
	V_{XX}(r) = \intStrength \bigg(\frac{1}{r / d} - \frac{1}{\sqrt{1 + (r / d)^{2}}} \bigg) .
\end{equation}
Here, the interlayer distance $d$ plays the role of the $\intRange$ parameter, while $\intStrength \equiv e^{2} / 2 \pi \epsilon_{0} \epsilon_{\mathrm{r}} d$ with $\epsilon_{\mathrm{r}}$ the effective relative permittivity of the surrounding environment.
The interaction kernel associated with $V_{XX}$ is $\interactionKernel(\dimlessMarker{r}) = 1/\dimlessMarker{r} - 1/\sqrt{\dimlessMarker{r}^{2} + 1}$ in position space and $\interactionKernel_{\dimlessMarker{\vec{q}}} = (1 - \ec^{{-} \dimlessMarker{q}}) / \dimlessMarker{q}$ in momentum space.

The simple pair interaction of Eq.\ \eqref{eq:BilayerXXInt} has been used before to address interacting exciton problems in bilayer systems \cite{gotting2022moire,dai2024strong,conti2025gross} and serves as a first approximation to the full interaction motivated by its expected dipole--dipole behavior at large distances.
Generally speaking, however, the phase diagram of interacting excitons will depend on the specific form of the bare interaction potential $V$.
Depending on the system, microscopic details such as the angular dependence of the dipole--dipole interaction may be important to take into account \cite{kezerashvili2022charge}.
Here we assume that excitons are tightly bound in the bilayer system, such that the dipole moments of each exciton are approximately perpendicular to the layers and thus this angular dependence can be neglected.
However, we stress that our $T$-matrix formalism can straightforwardly accommodate the incorporation of related and more complicated effects tailored to the studies of specific systems, if desired.

\subsection{Bose--Einstein condensates}
\label{sec:BECs}

A well-known phenomenon in a system of bosons at low temperatures is the formation of a BEC.
Here we consider the possibility that the excitons condense into a single coherent state populating a finite number of momenta, whose magnitude for weak interactions will be the moat momentum.

Let us first consider a many-body state occupying a single momentum $\vec{g}$ described by the state
\begin{equation}
   \vert \Psi_{\mathrm{FF}} \rangle = \frac{(\hat{b}^{\dagger}_{\vec{g}})^{N}}{\sqrt{N!}} \vert 0 \rangle ,
\end{equation}
where $\vert 0 \rangle$ is the vacuum state containing no excitons.
This corresponds to the so-called Fulde--Ferrel (FF) phase \cite{fulde1964superconductivity}, which possesses a homogeneous density profile and has vanishing kinetic energy when the magnitude of $\vec{g}$ coincides with the moat momentum $\moatMomentum$.
With this ansatz, the energy per particle $e \equiv E / N$ of the FF state in the thermodynamic limit is found in dimensionless variables as
\begin{equation}
\label{eq:intEnergyFFDimlessTMatrix}
    \dimlessMarker{e}_{\mathrm{FF}} = \dimlessMarker{\varepsilon}_{\dimlessMarker{g}} + \paramDensityVSIntRange^{2} \paramIntStrengthVSMoatBarrier \hspace{0.2mm} \TMatrixKernel_{\vec{0} \vec{0}}(2 \dimlessMarker{\vec{g}}, {-} 2 \dimlessMarker{\mu}) ,
\end{equation}
where in particular the interaction energy has been obtained via the Hamiltonian of Eq.\ \eqref{eq:HintWithTMatrix}.
Due to the rotational symmetry, $t_{\vec{0} \vec{0}}(2\dimlessMarker{\vec{g}}, {-}2\dimlessMarker{\mu})$ only depends on the magnitude of the ordering momentum, so we take $\dimlessMarker{\vec{g}} = \dimlessMarker{g} \hat{\mathbf{x}}$ without loss of generality and write $\TMatrixKernel_{\vec{0} \vec{0}}(2 \dimlessMarker{g}, {-} 2 \dimlessMarker{\mu})$ henceforth.
Using the definition of the chemical potential $\mu = (\partial E / \partial N)_{\mathcal{A}}$ and also minimizing the total energy with respect to the ordering momentum $\dimlessMarker{g}$, we find the following coupled equations for the FF state:
\begin{subequations}
   \begin{align}
      \label{eq:FFFullTMatrixEqMu}
      \dimlessMarker{\mu} &= \dimlessMarker{\varepsilon}_{\dimlessMarker{g}} + 2 \paramDensityVSIntRange^{2} \paramIntStrengthVSMoatBarrier \hspace{0.2mm} \TMatrixKernel_{\vec{0} \vec{0}}(2 \dimlessMarker{g}, {-} 2 \dimlessMarker{\mu}) , \\
      \label{eq:FFFullTMatrixEqg}
      0 &= \frac{\partial \dimlessMarker{\varepsilon}_{\dimlessMarker{g}}}{\partial \dimlessMarker{g}} + \paramDensityVSIntRange^{2} \paramIntStrengthVSMoatBarrier \frac{\partial}{\partial \dimlessMarker{g}} \hspace{0.2mm} \TMatrixKernel_{\vec{0} \vec{0}}(2 \dimlessMarker{g}, {-} 2 \dimlessMarker{\mu}) .
   \end{align}
\end{subequations}
These must be solved self-consistently for $\dimlessMarker{\mu}$ and $\dimlessMarker{g}$.
In practice, we have found numerically that $g \approx \moatMomentum$ to very good approximation for all combinations of the three dimensionless parameters that determine the behavior of the system, so we use this approximation to speed up the numerical calculations in the FF phase.
In this approximation the FF state has zero kinetic energy, and thus the interaction part of Eq.\ \eqref{eq:intEnergyFFDimlessTMatrix} also coincides with the total energy.

We can also consider many-body phases where more than one momentum state is populated.
Here we focus on the simplest one, the so-called Larkin--Ovchinnikov (LO) or stripe phase, consisting of a coherent superposition of states with momenta $\vec{g}$ and ${-}\vec{g}$ described by \cite{larkin1965inhomogeneous}
\begin{equation}
\label{eq:ansatzLOStateTwoMomenta}
   \vert \Psi_{\mathrm{LO}} \rangle = \frac{1}{\sqrt{N!}} \bigg(\frac{\ec^{\iu \theta} \hat{b}_{\vec{g}}^{\dagger} + \ec^{{-} \iu \theta} \hat{b}^{\dagger}_{{-}\vec{g}}}{\sqrt{2}}\bigg)^{\! N} \vert 0 \rangle ,
\end{equation}
where $\theta$ is an arbitrary phase \footnote{Note that here we follow the nomenclature of Refs.\ \cite{matsuda2007fulde,quan2010interplay} and make the distinction between the FF and LO phases, which are often collectively denoted as the ``FFLO'' phase in the literature. This distinction is important because only the LO state corresponds to a spatially inhomogeneous state, and furthermore their superfluid properties are completely different.}.
More generally, one should also include the components $0, \pm 2 \vec{g}, \pm 3 \vec{g}, \dots$, but we focus here on the simplest case for simplicity.
In this case the energy per particle reads
\begin{equation}
\label{eq:intEnergyLOFullTMatrix}
    \dimlessMarker{e}_{\mathrm{LO}} = \dimlessMarker{\varepsilon}_{\dimlessMarker{g}} + \paramDensityVSIntRange^{2} \paramIntStrengthVSMoatBarrier \hspace{0.2mm} t_{\mathrm{LO}}(\dimlessMarker{\vec{g}}, {-} 2 \dimlessMarker{\mu}) ,
\end{equation}
where for convenience we have defined
\begin{align}
\label{eq:TMatrixEffLO}
      &\TMatrixKernel_{\mathrm{LO}}(\dimlessMarker{\vec{g}}, {-} 2 \dimlessMarker{\mu}) \\
      &\equiv \frac{1}{2} \big[ t_{\vec{0} \vec{0}}(2 \dimlessMarker{\vec{g}}, {-} 2 \dimlessMarker{\mu})+ \TMatrixKernel_{\dimlessMarker{\vec{g}} \dimlessMarker{\vec{g}}}(\vec{0}, {-} 2 \dimlessMarker{\mu})
      + \TMatrixKernel_{\dimlessMarker{\vec{g}}, {-}\dimlessMarker{\vec{g}}}(\vec{0}, {-} 2 \dimlessMarker{\mu}) \big] . \nonumber
\end{align}
As before, we are allowed to take $\dimlessMarker{\vec{g}}$ along an arbitrary direction, say the $x$ axis, so we write $\TMatrixKernel_{\mathrm{LO}}(\dimlessMarker{g}, {-} 2 \dimlessMarker{\mu})$ in what follows.
That the energy of Eq.\ \eqref{eq:intEnergyLOFullTMatrix} is independent of $\theta$ can be understood from the fact that the wave function is proportional to $\cos (\vec{g} \vec{\cdot} \vec{r} + \theta)$, and thus a change in $\theta$ is equivalent to a translation of the coordinate-space density pattern that defines one of the two zero-energy Goldstone modes.
Similar to the FF case, the chemical potential and the magnitude of the ordering momentum in the LO phase must be found self-consistently by solving the coupled equations
\begin{subequations}
   \begin{align}
      \label{eq:selfConsEqMuLO}
      \dimlessMarker{\mu} &= \dimlessMarker{\varepsilon}_{\dimlessMarker{g}} + 2 \paramDensityVSIntRange^{2} \paramIntStrengthVSMoatBarrier \hspace{0.2mm} \TMatrixKernel_{\mathrm{LO}}(\dimlessMarker{g}, {-} 2 \dimlessMarker{\mu}) , \\
      \label{eq:selfConsEqgLO}
      0 &= \frac{\partial \dimlessMarker{\varepsilon}_{\dimlessMarker{g}}}{\partial \dimlessMarker{g}}  + \paramDensityVSIntRange^{2} \paramIntStrengthVSMoatBarrier \frac{\partial}{\partial \dimlessMarker{g}} \hspace{0.2mm} \TMatrixKernel_{\mathrm{LO}}(\dimlessMarker{g}, {-} 2 \dimlessMarker{\mu}) .
   \end{align}
\end{subequations}

For the numerical results of this section we focus on the case when the parameter $\paramSSInt \equiv 2 \paramDensityVSIntRange^{2} \paramIntStrengthVSMoatBarrier$ is kept fixed with $\paramSSInt \ll 1$, as in this case we can take $\dimlessMarker{g} \approx \dimlessMoatMomentum$ for both phases.
This simplifies the problem as it allows us to solve a single self-consistent equation for the chemical potential in each phase, namely Eqs.\ \eqref{eq:FFFullTMatrixEqMu} and \eqref{eq:selfConsEqMuLO}, as Eqs.\ \eqref{eq:FFFullTMatrixEqg} and \eqref{eq:selfConsEqgLO} are approximately satisfied since $\partial \dimlessMarker{\varepsilon}_{\dimlessMarker{g}} / \partial \dimlessMarker{g} = 0$ at the moat momentum.
This parameter $\paramSSInt$ plays a similar role to the well-known diluteness parameter of the physics of cold atoms \cite{lee1957many}.
However, its physical interpretation in our case is not as straightforward as it contains both the dimensionless density $\paramDensityVSIntRange$, measuring the true diluteness of the system, as well as the dimensionless coupling constant $\paramIntStrengthVSMoatBarrier$.
It thus represents a competition between the system diluteness and the relative interaction strength and can in principle assume a broad range of values compared to its cold-atom counterpart.

\begin{figure}[!t]
    \centering
    \includegraphics[width=\linewidth]{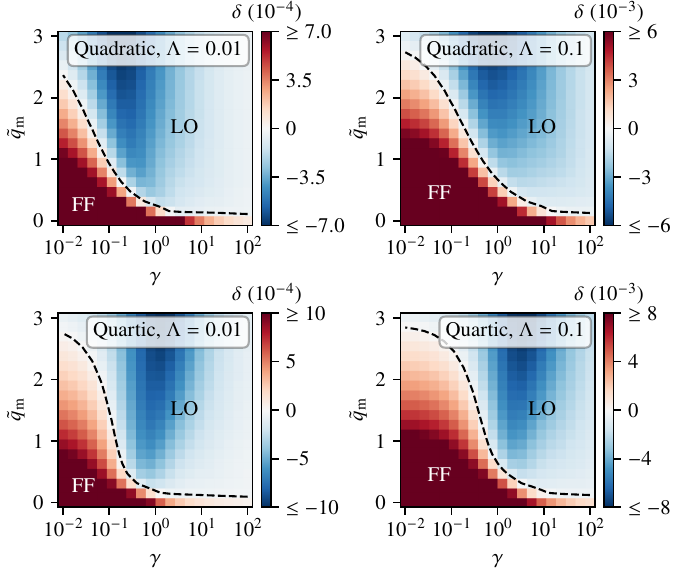}
    \caption{Energy difference $\delta \equiv \dimlessMarker{e}_{\mathrm{LO}} - \dimlessMarker{e}_{\mathrm{FF}}$ for the bilayer exciton model on the $\dimlessMoatMomentum$--$\paramIntStrengthVSMoatBarrier$ plane for a quadratic moat (top) and a quartic moat (bottom) and two different values of the small fixed parameter $\paramSSInt \equiv 2 \paramDensityVSIntRange^{2} \paramIntStrengthVSMoatBarrier$.
    In this regime, the ordering vector has the magnitude of the moat momentum and the two phases have vanishing kinetic energy, so the energies coincide with the interaction energies.
    The dashed black line marks the phase boundary between the homogeneous and inhomogeneous phases.
    As $\paramDensityVSIntRange = \sqrt{\paramSSInt / 2 \paramIntStrengthVSMoatBarrier}$, one observes that a smaller $\paramDensityVSIntRange$ favors the formation of an LO phase at fixed $\dimlessMoatMomentum$ and $\paramIntStrengthVSMoatBarrier$, as the transition line shifts to smaller $\paramIntStrengthVSMoatBarrier$ and smaller $\dimlessMoatMomentum$.
    The calculation takes into account the full $T$ matrix associated with the bilayer exciton--exciton potential of Eq.\ \eqref{eq:BilayerXXInt} and shows that the existence of degenerate minima can give rise to translationally broken phases even in the case of weak interactions and dilute exciton densities.
    }
    \label{fig:FFvsLOSmallLambda}
\end{figure}

In Fig.\ \ref{fig:FFvsLOSmallLambda} we show plots of the energy difference between the FF and LO phases on the $\dimlessMoatMomentum$--$\paramIntStrengthVSMoatBarrier$ plane for both a quadratic moat and a quartic moat (we recall the relations $\paramMoatMomVSIntRange = {-} \dimlessMoatMomentum$ for the former and $\paramMoatMomVSIntRange = {-} \dimlessMoatMomentum^{2}$ for the latter).
The value of $\paramDensityVSIntRange$ is appropriately chosen at each point so as to keep $\paramSSInt \ll 1$ fixed.
The results show that the inhomogeneous stripe phase can have lower energy than the FF phase in a large region of the parameter space, and it can be seen that a large moat favors the formation of an inhomogeneous BEC even in the case of weak interactions, i.e., $\intStrength / \moatEnergyBarrier = \paramIntStrengthVSMoatBarrier / \paramMoatMomVSIntRange^{2} \ll 1$.
The renormalization of the bare interaction via the $T$ matrix, whose numerical treatment is explained in Appendix\ \ref{app:TMatrixNum}, is crucial for these results.
Indeed, the bare momentum-space interaction $V_{\vec{k}}$ is positive everywhere, and thus by itself cannot lead to a stable LO phase or any other inhomogeneous phase \cite{sepulveda2008nonclassical,sepulveda2010superfluid}.
This is also clear from Eqs.\ \eqref{eq:intEnergyFFDimlessTMatrix} and \eqref{eq:TMatrixEffLO} if we replace the full kernel $t_{\vec{k} \vec{k}'}(\vec{K}, {-} 2 \dimlessMarker{\mu})$ by a local, energy-independent function $t_{\vec{k} - \vec{k}'}$ in this case.

For completeness we have also checked the possibility of a hybrid LO state with a nonvanishing zero-momentum component, i.e.,
\begin{equation}
\label{eq:PsiHybrid}
   \vert \Psi_{\text{HLO}} \rangle = \frac{1}{\sqrt{N!}} \big[c_{\vec{0}} \hat{b}^{\dagger}_{\vec{0}} + c_{\vec{g}} (\hat{b}^{\dagger}_{\vec{g}} + \hat{b}^{\dagger}_{{-} \vec{g}} )\big]^{N} \vert 0 \rangle ,
\end{equation}
where $|c_{\vec{0}}|^{2} + 2 |c_{\vec{g}}|^{2} = 1$.
In principle, this state could develop continuously from the LO phase by developing a small nonzero $c_{\mathbf{0}}$ and would display the superfluid properties associated with a homogeneous BEC besides also exhibiting a periodic density pattern (cf. Sec.\ \ref{sec:SFResp}).
By expanding the free energy for an infinitesimal $c_{\mathbf{0}}$ as explained in Appendix \ref{app:GPEFullTMatrix}, however, we conclude that this always results in a positive energy difference with respect to the pure LO state and is thus disfavored.

Even though in this section we have focused specifically on the comparison between the FF and LO phases, higher-order inhomogeneous BECs, i.e., states formed by a superposition of more than two momenta, can be considered as well.
Their general treatment is outlined in Appendix \ref{app:GPEFullTMatrix}.
In particular, in our numerics we have also considered phases for which the wave function for $\paramSSInt \ll 1$ consists of a superposition of three momenta on the moat, whose position-space density profile is then a triangular lattice.
Its self-consistent equations are identical to Eqs.\ \eqref{eq:selfConsEqMuLO} and \eqref{eq:selfConsEqgLO} with the effective kernel replaced by
\begin{equation}
   \begin{split}
      t_{\triangle}(\dimlessMarker{g}, {-} 2 \dimlessMarker{\mu}) &\equiv \frac{1}{3} \big[ t_{\vec{0} \vec{0}}(2 \dimlessMarker{g} \hat{\mathbf{x}}, {-} 2 \dimlessMarker{\mu}) + 2 t_{\frac{\sqrt{3}}{2} \dimlessMarker{g} \hat{\mathbf{y}}, \frac{\sqrt{3}}{2} \dimlessMarker{g} \hat{\mathbf{y}}}(\dimlessMarker{g} \hat{\mathbf{x}}, {-} 2 \dimlessMarker{\mu}) \\
      &+ \, 2 t_{\frac{\sqrt{3}}{2} \dimlessMarker{g} \hat{\mathbf{y}}, {-}\frac{\sqrt{3}}{2} \dimlessMarker{g} \hat{\mathbf{y}}}(\dimlessMarker{g} \hat{\mathbf{x}}, {-} 2 \dimlessMarker{\mu}) \big] .
   \end{split}
\end{equation}
We find that, for $\paramIntStrengthVSMoatBarrier \ll 1$, this phase is stabilized at around $\dimlessMoatMomentum \sim 5.5$--$5.9$ for both the quadratic and quartic moats (note that this is not shown in the parameter region of Fig.\ \ref{fig:FFvsLOSmallLambda}).
Despite the fact that it breaks the translational symmetry, such a triangular phase can exhibit superfluid flow and is therefore called a supersolid.
The superfluid response of these different inhomogeneous BECs is analyzed in detail in Sec.\ \ref{sec:SFResp}.

So far we have only considered wave functions where all particles sit in the same state.
Another intriguing possibility is that of a fragmented condensate, which has been predicted in some systems with a moat band where the renormalized interaction acquires a strong angular dependence \cite{yang2006quantum,gopalakrishnan2011universal}.
As an example, we have computed the energy for a fragmented condensate with wave function
\begin{equation}
   \vert \Psi_{\text{FBEC}} \rangle = \frac{(b^{\dagger}_{\vec{g}})^{N/2}}{\sqrt{(N/2)!}} \frac{(b^{\dagger}_{{-}\vec{g}})^{N/2}}{\sqrt{(N/2)!}} \vert 0 \rangle ,
\end{equation}
but we have found the same interaction energy in terms of the $T$ matrix as for the LO phase.
In a finite system, true spontaneous symmetry breaking into a coherent condensate phase is not possible and the system will tend to fragment.
This is closely related to the physics of phase diffusion \cite{stoof2009ultracold}.
Meanwhile, the states $\vert \Psi_{\text{LO}} \rangle$ and $\vert \Psi_{\text{FBEC}} \rangle$ are in principle degenerate in the thermodynamic limit, but we expect that small spatial inhomogeneities will tend to phase lock the fragmented condensate and give rise to the stripe phase \cite{leggett2006quantum}.
For this reason we do not consider these kinds of fragmented states in the remainder of the article.

\subsection{Composite fermion states}
\label{sec:compFermions}

While the picture of many-body phases of bosons at first naturally leads one to think of Bose--Einstein condensation, another intriguing possibility arises in the case of a moat band.
Namely, as studied in Refs.\ \cite{sedrakyan2013composite,sedrakyan2014absence,sedrakyan2015spontaneous,sedrakyan2015statistical,maiti2019fermionization,wei2023chiral}, the bosons may undergo statistical transmutation into a many-body ground state of effective composite fermions.
The fact that a moat band can cause this effect can intuitively be understood by observing that it exhibits a macroscopic degeneracy, which is reminiscent of the Landau levels (LLs) in the quantum Hall effect (QHE).
As shown explicitly below, we can view the bosons as fermions with a unit flux-quantum attachment generating an effective magnetic field.
Then, at low enough densities, the emergent fermions can populate appropriate LLs and perfectly quench their mean-field kinetic energy due to the familiar quantum-Hall physics.
With both the BEC phases and the composite fermion phases having vanishing kinetic energy, the correlation energy becomes the important point of comparison between both families of states.
The aforementioned articles consider hard-core bosons, i.e., bosons interacting with a repulsive $\delta$-function potential.
In this scenario, it is clear that such a composite state of fermions will have the lowest energy at the mean-field level, because the interaction energy must vanish in the case of a contact interaction due to the Pauli exclusion principle.
However, as we have argued, excitons generally do not exhibit such a short-ranged interaction, and thus we expect an interesting competition between the correlation energies which we explore in this section and the next one.

\begin{figure}[!t]
    \centering
    \includegraphics[width=\linewidth]{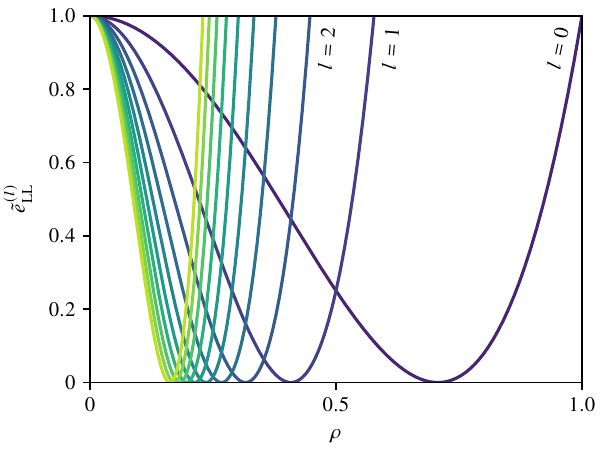}
    \caption{Landau-level energies $\dimlessMarker{e}^{(l)}_{\mathrm{LL}} = [2(2l + 1) \paramDensityVSIntRange^{2} + \paramMoatMomVSIntRange]^{2}$ as a function of the density parameter $\paramDensityVSIntRange$ for $\paramMoatMomVSIntRange = {-}1$ and various values of $l$.
    Below $\paramDensityVSIntRange = \sqrt{-\paramMoatMomVSIntRange/2} \approx 0.707$, the system can always quench the mean-field kinetic energy by distributing particles between two adjacent levels.
    This is not possible when $\paramDensityVSIntRange$ is larger than this value, and the kinetic energy of the system will rapidly increase with the density.
    }
    \label{fig:bilayerModelPhaseDiag}
\end{figure}

Let us first show explicitly how to derive the fermionization of bosons.
Starting from the bosonic Hamiltonian of Eq.\ \eqref{eq:BosonicH}, we perform a flux-attachment transformation for the excitons given by \cite{zhang1989effective,lopez1991fractional,zhang1992chern}
\begin{equation}
	\hat{b}(\vec{r}) = \hat{f}(\vec{r}) \exp \bigg\{\iu \int \mathrm{d}^{2} r' \operatorname{arg}(\vec{r} - \vec{r}') \hat{n}(\vec{r}') \bigg\} ,
\end{equation}
where $\operatorname{arg} (\vec{r} - \vec{r}')$ is the angle sustained by the vector connecting the points $\vec{r}$ and $\vec{r}'$ as measured from some reference direction, say the $x$ axis.
With this definition, the field operator $\hat{f}(\vec{r})$ replacing the excitons obeys fermionic anticommutation relations.
This transformation introduces a statistical Chern--Simons gauge field in the kinetic term via $\momentumOperator \rightarrow \momentumOperator - \hat{\mathbf{a}}$, with
\begin{equation}
	\hat{\mathbf{a}}(\vec{r}) = \int \mathrm{d}^{2} r' \, \frac{\hat{\mathbf{z}} \times (\vec{r} - \vec{r}')}{|\vec{r} - \vec{r}'|^{2}} \hspace{0.3mm} \hat{n}(\vec{r}') .
\end{equation}
Thus, $[\vec{\nabla} \times \hat{\mathbf{a}}(\vec{r})]_{z} = 2 \pi \hat{n}(\vec{r})$ is an effective magnetic field experienced by the emergent fermions.
At the mean-field level we replace $\hat{n}(\vec{r})$ with an average homogeneous density $n$, thus essentially ``smearing'' the magnetic flux equally over all particles in the system.
In this way we obtain the energy of the emergent LLs (in dimensionless units) as
\begin{equation}
	\dimlessMarker{e}^{(l)}_{\mathrm{LL}} = \big[2(2l + 1) \paramDensityVSIntRange^{2} + \paramMoatMomVSIntRange \big]^{2}, \qquad l = 0, 1, 2, \dots
\end{equation}
for the quartic moat.
These LLs are depicted in Fig.\ \ref{fig:bilayerModelPhaseDiag} as a function of $\paramDensityVSIntRange$ for a fixed value of $\moatMomentum$.
It is important to note that, contrary to the usual QHE in a system with parabolic bands, the lowest Landau level (LLL) is not always that with $l = 0$, but instead depends on density and moat size.
We see that the kinetic energy is perfectly quenched at the densities
\begin{equation}
\label{eq:nlFormulaCSL}
	n_{l} = \frac{\moatMomentum^{2}}{2 \pi (2l + 1)} .
\end{equation}

In the fermionized picture, the ground state of the system at density $n_{l}$ will be that of a fully filled LL with index $l$, which will be the LLL at this density.
The bosonic many-body wave function of this state is then constructed as \cite{wei2023chiral}
\begin{equation}
\label{eq:WFPhiBl}
	\Psi^{(l)}_{\mathrm{B}}(\{z_{i}, \bar{z}_{i}\}) = \prod_{i<j}^{N} \frac{z_{i} - z_{j}}{|z_{i} - z_{j}|} \Psi^{(l)}_{\mathrm{F}}(\{z_{i}, \bar{z}_{i}\}) ,
\end{equation}
where $z_{i} = x_{i} + \iu y_{i}$ is the complex 2D coordinate of the $i$th fermion and $\Psi^{(l)}_{\mathrm{F}}$ is a Slater determinant of the single-particle wave functions corresponding to the fully filled LLL.
In symmetric gauge $\mathbf{a}(\vec{r}) = \pi n ({-}y, x, 0)$, the single-particle wave functions adopt the usual forms involving Laguerre polynomials.
The wave function of Eq.\ \eqref{eq:WFPhiBl} describes a chiral spin liquid (CSL) and corresponds to a phase where all momenta on the moat are simultaneously occupied and no condensation occurs \cite{sedrakyan2014absence}.

When the (spinless) bosons interact with a Dirac-delta potential $V_{0} \delta(\vec{r})$, the flux-attachment procedure leads to a system of noninteracting fermions due to the Pauli exclusion principle.
However, when the interaction is somewhat long-ranged, the correlation energy will not vanish.
Instead, for fermionized bosons in the $l$th LL interacting via a potential with momentum-space kernel $v_{\dimlessMarker{q}}$ it is given by
\begin{equation}
\label{eq:CSLEintBilayerXXPot}
    \dimlessMarker{e}^{(l)}_{\text{int},\text{CSL}} = \paramDensityVSIntRange^{2} \paramIntStrengthVSMoatBarrier \hspace{0.3mm} \bigg(1 -  \int_{0}^{\infty} \mathrm{d} u \, v_{2 \paramDensityVSIntRange \sqrt{u}} \hspace{0.3mm} [L_{l}(u)]^{2} \hspace{0.3mm} \ec^{{-} u} \bigg) .
\end{equation}
Note that we have assumed that the zero-momentum kernel $v_{0}$ is $1$, which can always be done by appropriately rescaling the interaction strength $\intStrength$ and in particular holds for the bilayer potential of Eq.\ \eqref{eq:BilayerXXInt} with kernel $v_{\dimlessMarker{q}} = (1 - \ec^{{-}\dimlessMarker{q}}) / \dimlessMarker{q}$.
The derivation of Eq.\ \eqref{eq:CSLEintBilayerXXPot} can be found in Appendix\ \ref{app:CSL}.
We note that this relation is also valid for the contact interaction, as in this case the momentum-space kernel is equal to $1$ for all momenta and Eq.\ \eqref{eq:CSLEintBilayerXXPot} vanishes exactly due to the identity $\int_{0}^{\infty} \mathrm{d} u \, [L_{l}(u)]^{2} \hspace{0.3mm} \ec^{{-}u} = 1$.

Moreover, when the average density $n$ is not exactly equal to one of the $n_{l}$ but rather $n_{l+1} < n < n_{l}$ for some $l \ge 0$, the system will undergo phase separation into two CSL phases on the LLs $l$ and $l + 1$ with densities $n_{1}$ and $n_{2}$, respectively.
We then have $n = \alpha n_{1} + (1 - \alpha) n_{2}$, where $\alpha$ is the volume fraction of the phase with density $n_{1}$.
The resulting effective energy per particle reads
\begin{equation}
   e_{\text{eff}}(n) = \alpha \frac{n_{1}}{n} e(n_{1}) + (1 - \alpha) \frac{n_{2}}{n} e(n_{2}) ,
\end{equation}
where the density ratios arise because $\alpha$ is the volume fraction but $e$ is the interaction energy per particle, not per unit volume.
We note that $e(n)$ here is shorthand for $\min_{l \ge 0} e^{(l)}(n)$, with $e^{(l)}(n) = e_{\mathrm{LL}}^{(l)}(n) + e^{(l)}_{\text{int}, \text{CSL}}(n)$ the energy per particle at density $n$ in the $l$th LL.

The densities $n_{1}$ and $n_{2}$ must be found from the common tangent construction for the total energy, which in terms of the energy per particle yields the coupled equations
\begin{subequations}
   \begin{align}
      e(n_{1}) + n_{1} e'(n_{1}) &= e(n_{2}) + n_{2} e'(n_{1}) , \\
      e(n_{1}) + n_{1} e'(n_{1}) &= \frac{n_{2} e(n_{2}) - n_{1} e(n_{1})}{n_{2} - n_{1}} .
   \end{align}
\end{subequations}
In terms of our dimensionless variables, where $\dimlessMarker{e}$ is a function of $\paramDensityVSIntRange = \sqrt{\pi n \intRange^{2}}$, these read
\begin{subequations}
   \begin{align}
      \dimlessMarker{e}(\paramDensityVSIntRange_{1}) + \frac{\paramDensityVSIntRange_{1}}{2} \dimlessMarker{e}{\hspace{0.1mm} '}(\paramDensityVSIntRange_{1}) &= \dimlessMarker{e}(\paramDensityVSIntRange_{2}) + \frac{\paramDensityVSIntRange_{2}}{2} \dimlessMarker{e}{\hspace{0.1mm} '}(\paramDensityVSIntRange_{2}) , \\
      \dimlessMarker{e}(\paramDensityVSIntRange_{1}) + \frac{\paramDensityVSIntRange_{1}}{2} \dimlessMarker{e}{\hspace{0.1mm} '}(\paramDensityVSIntRange_{1}) &= \frac{\paramDensityVSIntRange_{2}^{2} \dimlessMarker{e}(\paramDensityVSIntRange_{2}) - \paramDensityVSIntRange_{1}^{2} \dimlessMarker{e}(\paramDensityVSIntRange_{1})}{\paramDensityVSIntRange_{2}^{2} - \paramDensityVSIntRange_{1}^{2}} .
   \end{align}
\end{subequations}

Finally, for $n > n_{l=0}$, the energy of the CSL increases indefinitely with $n$ because there are no LLs left to quench the kinetic energy.
In this case there is no possible common tangent construction and the energy of the CSL comes purely from the zeroth LL.

\subsection{Phase competition}
\label{sec:phaseComp}

We are now in the position to compare the energies of the CSL and BEC phases.
It is easily seen from Eqs.\ \eqref{eq:intEnergyFFDimlessTMatrix} and \eqref{eq:intEnergyLOFullTMatrix} that the interaction energy of a BEC phase scales as $\paramDensityVSIntRange^{2} \paramIntStrengthVSMoatBarrier$ and is thus linear in $n$ (insofar as the dilute limit $n a_{X}^{2} \ll 1$ is satisfied) as long as we assume that the $T$ matrix only has a weak dependence on $\paramDensityVSIntRange$.
Meanwhile, the behavior of the CSL with the density must be extracted from Eq.\ \eqref{eq:CSLEintBilayerXXPot} by studying how the integral behaves with $\paramDensityVSIntRange$.
In particular, for low densities we have $l \sim n^{-1} \sim \paramDensityVSIntRange^{-2} \gg 1$ according to Eq.\ \eqref{eq:nlFormulaCSL}, and the Laguerre polynomial assumes a very large order.
As shown in Appendix\ \ref{app:CSL} via a careful asymptotic treatment of the integral, one concludes that $\dimlessMarker{e}_{\text{int},\text{CSL}} \sim \paramDensityVSIntRange^{2} \paramIntStrengthVSMoatBarrier$.
In other words, the CSL scales in exactly the same way as the BEC phases and its interaction energy is linear in $n$.
The only exception to this corresponds to the case of a contact interaction, in which case the correlation energy of Eq.\ \eqref{eq:CSLEintBilayerXXPot} vanishes.
Due to the identical dependence of the BEC phases and the CSL on the dimensionless parameters, we conclude that a moat-band system with long-range interactions will host a rich competition between these phases.

\begin{figure}[!t]
    \centering
    \includegraphics[width=\linewidth]{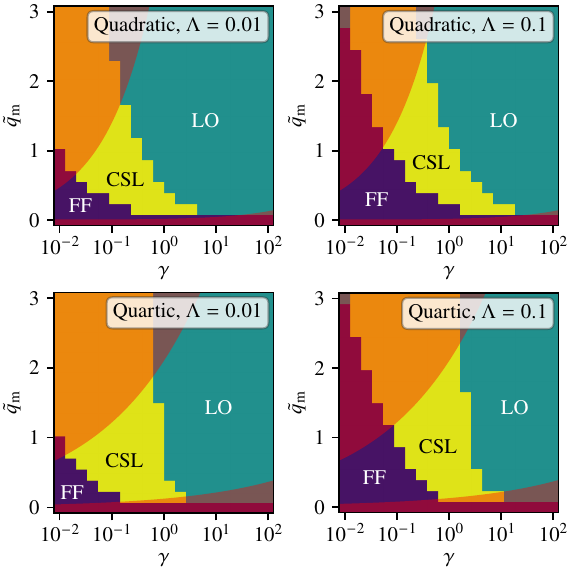}
    \caption{Zero-temperature phase diagram of the bilayer exciton model on the $\dimlessMoatMomentum$--$\paramIntStrengthVSMoatBarrier$ plane for a quadratic moat (top) and a quartic moat (bottom) and two different values of the small fixed parameter $\paramSSInt \equiv 2 \paramDensityVSIntRange^{2} \paramIntStrengthVSMoatBarrier$.
    The CSL phase now occupies a relatively large region between the FF and LO condensate phases shown previously in Fig.\ \ref{fig:FFvsLOSmallLambda}.
    As explained in the main text, the rich competition between the BEC and CSL phases originates from the long-range nature of the potential between the interacting excitons.
    The parts of the plots lying in between the red shaded areas correspond to the ``physical regions'', that is, regions where the dimensionless parameters $\dimlessMoatMomentum$, $\paramIntStrengthVSMoatBarrier$, and $\paramDensityVSIntRange$ translate to a set of reasonably accessible experimental conditions detailed in Sec.\ \ref{sec:expRegion}.
    These include dilute exciton densities, realistic interlayer distances and substrate dielectric constants, and reasonable moat parameters.
    Conversely, the shaded areas represent regions that are likely difficult to access experimentally.
    Our results show that all competing phases should be experimentally reachable under appropriate conditions.
    }
    \label{FIG_CSLvsFFvsLO_phaseDiagBilayer}
\end{figure}

We have performed numerical calculations of the energies of the CSL and BEC phases discussed in the previous sections.
While the comparison between the FF and LO phases was already shown in Fig.\ \ref{fig:FFvsLOSmallLambda}, we now show the full phase diagram including also the CSL in Fig.\ \ref{FIG_CSLvsFFvsLO_phaseDiagBilayer}.
In all cases the FF, LO, and CSL phases are clearly visible and exhibit a nontrivial competition with each other.
Furthermore, as explained in detail in the following section, a large portion of the dimensionless parameter space maps to a combination of realistic physical quantities.
More precisely, the large unshaded areas are such that one can find experimentally accessible combinations of (dilute) densities, interlayer distances, relative dielectric constants, and moat parameters.

While the FF state always has nearly vanishing kinetic energy, that of the CSL can only be quenched as long as $n < \moatMomentum^{2} / 2 \pi$, or equivalently $\paramDensityVSIntRange < \dimlessMoatMomentum / \sqrt{2}$.
Above this density, the total CSL energy acquires a significant contribution coming from the kinetic energy of the zeroth LL.
Thus, using that $\paramDensityVSIntRange = \sqrt{\paramSSInt / 2 \paramIntStrengthVSMoatBarrier}$ we conclude that when $\dimlessMoatMomentum^{2} \paramIntStrengthVSMoatBarrier < \paramSSInt$ the FF condensate will quickly tend to be favored over the CSL.
For small $\dimlessMoatMomentum$ and $\paramIntStrengthVSMoatBarrier$ we thus expect the FF phase to become energetically favored, which is indeed observed in the bottom-left corner of all plots in Fig.\ \ref{FIG_CSLvsFFvsLO_phaseDiagBilayer}.
More precisely, it is seen that the transition from the CSL to the FF phase takes place at smaller $\dimlessMoatMomentum$ with increasing $\paramIntStrengthVSMoatBarrier$, in accordance with the requirement $\dimlessMoatMomentum < \sqrt{\paramSSInt / \paramIntStrengthVSMoatBarrier}$.
Finally, for a fixed value of $\dimlessMoatMomentum$ this phase boundary is expected at increasing $\paramIntStrengthVSMoatBarrier$ with increasing $\paramSSInt$, which is also clearly seen in the plots.

Let us briefly elaborate on the differences between contact and long-range interactions, in particular regarding the effect of density fluctuations.
The interaction energy of the CSL in the former case vanishes exactly due to the Pauli exclusion principle of the emergent fermions, and one would assume that the FF state always has a higher energy.
While this is true at the mean-field level, it turns out that fluctuations can stabilize the FF phase in a region of the parameter space, as explained in Refs.\ \cite{sedrakyan2015statistical,wei2023chiral}.
This can be intuitively understood from the fact that the CSL does not truly have zero kinetic energy even when $n$ coincides with one of the $n_{l}$ of Eq.\ \eqref{eq:nlFormulaCSL}.
Perfect quenching can only take place at the mean-field level, where the statistical magnetic field has been set to a constant.
By contrast, when $\hat{\mathbf{a}}$ is allowed to fluctuate, there can be a transition to the FF state even in the case of a point interaction as the fluctuations lead to a scaling $\mu \sim n^{2} (\ln n)^{2}$ of the CSL chemical potential with the density.
As a result, below a certain density-dependent critical coupling strength, the FF state becomes favored over the CSL also in the case of contact interactions.
Meanwhile, for our specific bilayer interaction we have found $\mu \sim n$, which arises from the interaction energy rather than the kinetic term and thus remains as the relevant scaling in the presence of gauge-field fluctuations.
For this reason we do not expect fluctuations of $\hat{\mathbf{a}}$ to result in significant qualitative changes to the phase diagram of Fig.\ \ref{FIG_CSLvsFFvsLO_phaseDiagBilayer}.
Given that phase competition in the case of long-range interactions already exists at the mean-field level, in this article we do not consider effects beyond the mean-field approach.

\subsection{Experimental accessibility}
\label{sec:expRegion}

Our previous analysis and phase diagrams have all been presented in terms of a set of the three dimensionless variables of Eqs.\ \eqref{paramLambda}--\eqref{paramGamma} that control the physics of the problem.
In this section we discuss how our results so far can be put in the context of real systems by studying which regions of Fig.\ \ref{FIG_CSLvsFFvsLO_phaseDiagBilayer} map to experimentally accessible values of the underlying physical parameters.
We begin by noting that the dimensionless parameter $\paramIntStrengthVSMoatBarrier$ in these plots is related to the density $n$ and the interlayer distance $d$ via
\begin{equation}
   \paramIntStrengthVSMoatBarrier = \frac{\paramSSInt}{2 \pi n d^{2}} .
\end{equation}
We consider a range of exciton densities $n \in (10^{8}, 10^{12}) \, \si{\centi\meter\tothe{-2}}$, which have been realized in various experimental platforms and depending on the system are compatible with the dilute limit where excitons can be treated as perfect bosons \cite{butov2001stimulated,wang2021diffusivity,qi2022molding,bai2023evidence}.
Furthermore, we assume a range of interlayer distances $d \in (0.5, 15) \, \si{\nano\meter}$ corresponding to experimentally realistic scenarios in bilayer systems \cite{laikhtman2009exciton,jiang2021interlayer,chen2025interlayer,liu2025interlayer,zhou2026long}.
These values translate to an experimentally feasible range of $\paramIntStrengthVSMoatBarrier$ lying approximately between $10^{-2} \paramSSInt$ and $10^{6} \paramSSInt$.
For the values of $\paramSSInt$ considered in Figs.\ \ref{fig:FFvsLOSmallLambda} and \ref{FIG_CSLvsFFvsLO_phaseDiagBilayer}, the range of $\paramIntStrengthVSMoatBarrier$ shown in the figures sits comfortably within these limits.
Thus, all of these values are in principle accessible with appropriate combinations of realistic densities and interlayer distances.

On the other hand, we observe that $\intStrength = e^{2} / 2 \pi \epsilon_{0} \epsilon_{\mathrm{r}} d = \paramIntStrengthVSMoatBarrier \moatEnergyBarrier / \dimlessMoatMomentum^{p}$, where $p = 2$ for a quadratic moat and $p = 4$ for a quartic moat.
From here we obtain the moat momentum in dimensionless units in terms of physical parameters as
\begin{equation}
\label{eq:dimlessqmPhysicalParams}
   \dimlessMoatMomentum = \bigg( \frac{2 \pi \paramIntStrengthVSMoatBarrier \epsilon_{\mathrm{r}} \moatEnergyBarrier d}{e^{2} / \epsilon_{0}} \bigg)^{1 / p} ,
\end{equation}
with $e^{2} / \epsilon_{0} \approx \SI{18.095}{\electronvolt\nano\meter}$.
We consider a range of relative dielectric constants $\epsilon_{\mathrm{r}} \in (1, 30)$ for the surrounding environment, as well as moat energies lying in the broad range $\moatEnergyBarrier \in (1, 100) \, \si{\milli\electronvolt}$.
Equation\ \eqref{eq:dimlessqmPhysicalParams} results in a desired range of values of $\dimlessMoatMomentum$ when we consider the minimum and maximum values of the combination $\epsilon_{\mathrm{r}} \moatEnergyBarrier d$.
While this region corresponds to realistic values of $n$, $d$, $\epsilon_{\mathrm{r}}$, and $\moatEnergyBarrier$, we still need to make sure that the resulting \emph{physical} moat momentum $\moatMomentum = \dimlessMoatMomentum / d$ is of reasonable magnitude.
Previous experimental studies of systems with camelback dispersions point to typical moat momenta in the range $\moatMomentum \in (0.01, 1) \, \si{\nano\meter\tothe{-1}}$ \cite{humphreys1978indirect,efros1984exciton,siarkos2000center,takeda2005visualization,minkov2013two,minkov2014hole,ceferino2020crossover,maisel2023single}.
Consequently, there is another desired range of $\dimlessMoatMomentum$ given by the extremal values of $\moatMomentum d$.
Intersecting the two ranges of $\dimlessMoatMomentum$ that we have obtained leads to a $\paramIntStrengthVSMoatBarrier$-dependent interval inside which all physical parameters $n$, $d$, $\epsilon_{\mathrm{r}}$, $\moatEnergyBarrier$, and $\moatMomentum$ attain realistic values.
In Fig.\ \ref{FIG_CSLvsFFvsLO_phaseDiagBilayer} we have shaded out in red the areas that lie outside the realistic range of $\dimlessMoatMomentum$.
This leaves a large ``physical region'' which should be experimentally accessible and spans all the competing phases discussed above.

In addition to the appropriate choice of system parameters, the finite excitonic lifetime is also a significant aspect from an experimental point of view.
Our theory implicitly assumes that excitons can be treated as being in thermodynamic equilibrium.
This requires that their thermalization time is much shorter than their actual lifetime, i.e., that the time it takes them to thermalize to the ground state after their initial excitation is short compared to the time spent in the ground state itself.
This ensures that the many-body phases in question and their potential phase coherence can be properly established and exist for long enough to be considered as being in a quasiequilibrium state.
We note that the use of electron--hole bilayers can drastically enhance the lifetime of excitons to hundreds of nanoseconds and in some cases even a few microseconds \cite{golub1990long,high2008control,high2012spontaneous}, which incidentally justifies our choice for the exciton--exciton potential in this section.
Furthermore, a moat band naturally lengthens the excitonic lifetime, as pure radiative recombination is forbidden due to the small photon momentum compared to the moat momentum.
Finally, even though here we consider excitons as spinless bosons for simplicity, we note that many systems exhibit spin selection rules according to which some excitons are dark.
While these are more difficult to detect, they are also longer lived, potentially yielding an additional option for enhancing the lifetimes \cite{robert2017fine,schmidt2019tracking,chand2023interaction}.
In summary, while excitons are inherently nonequilibrium quasiparticles, we expect that the combination of these mechanisms will be able to generate long recombination times allowing for the observation of the many-body phases presented here.
As signatures of an exciton BEC have been observed previously in systems where the excitonic dispersion is parabolic \cite{butov1994condensation,butov2001stimulated,butov2002macroscopically,butov2002towards}, it is reasonable to assume that moat-band systems with long enough exciton lifetimes can also exist.

Let us also briefly discuss the potential role of random disorder in real samples.
As a topological state, we expect the CSL to be stable against disorder as long as the latter is not strong enough to close the many-body gap.
However, as disorder will tend to introduce a random potential on the moat that will lift the degeneracy, the CSL may become anisotropic as a result.
On the other hand, while BEC phases can be stable against weak disorder, a strong or even moderate amount of it will destabilize them in favor of new emergent phases such as a Bose glass \cite{fisher1989boson,sanchez2010disordered}.
In this article we assume that the samples are clean enough for this to not be the case.

\subsection{The role of band-structure warping}
\label{sec:warping}

In the calculations of the previous sections we have assumed an ideal moat, i.e., one where all states are perfectly degenerate.
However, in real systems where one would hope to observe such excitonic many-body states, another effect is present which can spoil the perfect degeneracy of the moat at large $\moatMomentum$, namely band-structure warping due to the effects of the underlying crystal lattice \cite{fu2009hexagonal}.
Typically, the low-energy effective model for electrons and holes is written as a $\vec{k} \vec{\cdot} \vec{p}$ expansion up to quadratic order in $\vec{k}$.
This often results in dispersions with perfect rotational symmetry.
However, including higher-order terms almost always spoils this symmetry and produces effects reflecting the point-group symmetries of the underlying lattice.
For instance, in Ref.\ \cite{maisel2023single} we considered excitons in a Bernevig--Hughes--Zhang model for a topological insulator with a band inversion.
The latter results in an inverted particle--hole continuum, which in turn results in a moat dispersion for the excitons.
While this is a good approximation to obtain the behavior of the free excitons, we must be careful when making assumptions about the many-body behavior of the system that rely on a significant degree of degeneracy of the emerging moat.

\begin{figure}[!t]
    \centering
    \includegraphics[width=\linewidth]{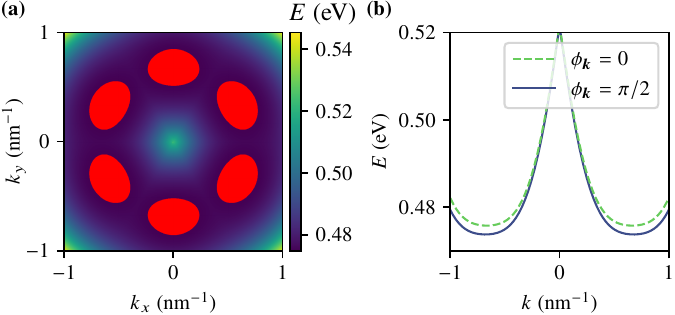}
    \caption{(a) Electron--hole continuum of the bulk-projected Bernevig--Hughes--Zhang model for bismuth selenide of Ref.\ \cite{liu2010model} in the presence of hexagonal warping.
    The projection onto the low-energy bulk bands follows the procedure of Ref.\ \cite{maisel2023single}, and we use the same numerical values for the parameters.
    Due to cubic terms in $k_{x}$ and $k_{y}$, the dispersion reflects the $C_{3}$ symmetry of the underlying lattice, breaking the perfect rotational symmetry of the quadratic model and featuring six minima around the $\Gamma$ point.
    The red blobs mark regions that differ from the minimum by at most $\SI{1}{\milli\electronvolt}$.
    (b) Horizontal and vertical slices of the dispersion in (a) showing the energy in the $k_{x}$ and $k_{y}$ directions, which contain the largest and smallest minima.
    The maximum splitting between the two dispersions is about $\SI{2}{\milli\electronvolt}$ (corresponding to approximately $\SI{23.2}{\kelvin}$) for realistic warping parameters $R_{1} = \SI{0.05}{\electronvolt\nano\meter\tothe{3}}$ and $R_{2} = \SI{-0.1}{\electronvolt\nano\meter\tothe{3}}$ (cf. Ref.\ \cite{liu2010model}).
    }
    \label{fig:warpingBi2Se3}
\end{figure}

In the case of Ref.\ \cite{maisel2023single}, the physical crystal under consideration has a $C_{3}$ rotational symmetry.
Following Ref.\ \cite{liu2010model}, we include in the $\vec{k} \vec{\cdot} \vec{p}$ model terms cubic in the momenta $k_{x}$ and $k_{y}$.
By projecting the resulting Hamiltonian on the first set of bulk bands as described in Ref.\ \cite{maisel2023single} we obtain the effective Hamiltonian to be used for the exciton calculation, now up to order $k^{3}$.
To analyze the effects of such warping terms we obtain the particle--hole continuum in the cubic model by minimizing the electron--hole gap $\epsilon^{c}_{\vec{K}/2 + \vec{k}} - \epsilon^{v}_{\vec{K}/2 - \vec{k}}$ for fixed total exciton momentum $\vec{K}$.
Note that, while this is not the actual exciton energy, we expect a very similar splitting for the latter as the excitons will tend to follow the continuum.
The result is plotted in Fig.\ \ref{fig:warpingBi2Se3} and shows that the warping terms lead to the lifting of the moat degeneracy into six equal minima separated by an angle of $\pi/3$, resulting in a finite effective mass in the direction tangential to the moat.
In general, if the dispersion shows $N_{\text{min}}$ degenerate minima separated by a small energy barrier $\delta \epsilon$, this tangential effective mass can be estimated as
\begin{equation}
\label{eq:effMassTangential}
   m^{\parallel}_{\text{eff}} \sim \bigg(\frac{\pi}{N_{\text{min}}} \bigg)^{2} \frac{\moatMomentum^{2}}{2 \delta \epsilon} .
\end{equation}
For realistic values of the parameters governing the warping perturbation, we find that the energy barrier separating the different minima of Fig.\ \ref{fig:warpingBi2Se3} is of the order of a few meV.
This corresponds to a few tens of Kelvin, meaning that the warping perturbation in this and similar systems should be enough to lock the bosons into the minima at low enough temperatures.

Let us consider the evolution of the system when starting from a CSL on a fully degenerate moat and slowly turning on a warping perturbation.
For very weak warping, the wave function of Eq.\ \eqref{eq:WFPhiBl} remains a reliable variational ansatz and the CSL looks qualitatively the same.
As the degree of warping increases, the bosons undergo a significant reallocation on the manifold of thermally allowed energies, but the system remains in a CSL state in view of the fact that this is a topologically protected state.
However, Eq.\ \eqref{eq:WFPhiBl} is no longer expected to describe the ground state of the system and the warping effect becomes nonperturbative from the point of view of this particular CSL wave function.
Instead, the new CSL ground state may be nematic, with an anisotropic moat occupation such as in Ref.\ \cite{sedrakyan2013composite}.
Finally, when the characteristic warping energy scale becomes comparable to the many-body gap of the CSL, a transition to a BEC takes place as the moat begins to behave more akin to a multivalley quadratic band.
To determine the phase diagram precisely in the presence of warping, a calculation taking into account the full energetics of the optimal CSL and the appropriate BEC is needed.
However, the transition to a BEC at strong enough warping can be established on general grounds in view of the fact that its energy remains largely unaffected by the angular anisotropy as long as its discrete rotational symmetry is compatible with that of the warped moat.
This is because the BEC tends to occupy only a finite number of momenta, so it can be allocated on the minima of the warped dispersion and remain unaffected by the energy barriers appearing between them.
In other words, we expect that band-structure warping can lower the formation threshold of BEC phases in some systems, so that the phase diagram of Fig.\ \ref{FIG_CSLvsFFvsLO_phaseDiagBilayer} may slightly overestimate the CSL region.

While the warping of the moat dispersion will not destroy the condensate phases, it can affect the form of their ground state and their stability.
First, it will only be possible to obtain inhomogeneous BECs whose discrete rotational symmetry is commensurate with that of the underlying lattice.
For instance, for a $C_{3}$ system this means that a triangular phase is in principle possible, but a square lattice becomes unlikely as it would require occupying momenta which do not all correspond to the minima on the warped moat.
Second, warping will tend to weaken the effects of fluctuations in a BEC.
As explained in Refs.\ \cite{brazovskiui1996phase,sedrakyan2015statistical}, on a perfectly degenerate moat fluctuations may destabilize the BEC due to the 1D-like divergence of the density of states, a problem that becomes less severe under the effects of warping.
Finally, we note that if the degree of warping is strong enough to completely gap out the CSL, then the plots of Fig.\ \ref{fig:FFvsLOSmallLambda} become the actual phase diagrams in this scenario.

All in all, while strong warping effects can shape the phase diagram, we expect the CSL and BEC phases to be robust against relatively weak deviations of a perfectly degenerate moat.
On the one hand, this is expected for the CSL due to its topological nature, which will merely lead to a momentum redistribution of the excitons within the CSL state itself.
On the other hand, BEC phases with a finite number of populated momenta can simply acquire a form that completely avoids the higher-energy regions of an inhomogeneous moat.

%% file: sec_superfluidity.tex

A natural question regarding the condensate solutions we have found is whether these configurations can exhibit superfluid flow and how the latter is affected by the moat band compared to the parabolic case.
The superfluid response of a system can be quantified by the superfluid stiffness $\mathbf{D}_{\mathrm{s}}$, a tensor measuring the rigidity of the coherent many-body phase responsible for the existence of a nondissipative flow.
By considering an overall phase difference ${(\Delta \varphi)}_{i}$ applied over the system in direction $i$, the components of $\mathbf{D}_{\mathrm{s}}$ are obtained from the quadratic part of the energy shift via
\begin{equation}
   D_{\mathrm{s}}^{ij} = \frac{\partial^{2} E[\varphi_{\text{min}}]}{\partial {(\Delta \varphi)}_{i} \partial {(\Delta \varphi)}_{j}} \bigg\vert_{\Delta \varphi = 0} .
\end{equation}
Here, $E[\varphi_{\text{min}}]$ is the energy corresponding to the many-body configuration $\Psi_{\varphi}(\vec{r}) \equiv \ec^{\iu \varphi(\vec{r})} \Psi_{0}(\vec{r})$ when the phase is chosen as $\varphi_{\text{min}}(\vec{r})$, i.e., the phase profile that minimizes the energy functional under the boundary conditions imposed by the external phase difference, and $\Psi_{0}(\vec{r})$ is the (phase-independent) macroscopic wave function defining the density profile.

\subsection{Superfluid stiffness on moat bands}

A nontrivial phase profile only affects the kinetic energy, which for a quartic moat now reads
\begin{equation}
\label{eq:DeltaEPhaseTwistKE}
   K[\varphi] = b \int \mathrm{d}^{2} r \, \big| (\momentumOperator^{2} - \moatMomentum^{2}) \Psi_{\varphi}(\vec{r}) \big|^{2} .
\end{equation}
We will focus on the case $\paramSSInt \ll 1$, so that $\Psi(\vec{r})$ can be taken as a linear superposition of plane waves with momenta distributed along the moat.
Explicitly, this means that
\begin{equation}
   \Psi_{0}(\vec{r}) = \sqrt{n} \sum_{i} c_{i} \hspace{0.3mm} \ec^{\iu \vec{q}_{i} \vec{\cdot} \vec{r}} ,
\end{equation}
where $|\vec{q}_{i}| \approx \moatMomentum$ for all $i$ and $\sum_{i} |c_{i}|^{2} = 1$.

We assume the long-wavelength limit for the phase $\varphi$, meaning that the latter varies over distances much larger than the characteristic moat length $L_{\mathrm{m}} \sim 1 / \moatMomentum$ which defines the period of the position-space density pattern, $L_{\mathrm{m}} \ll L_{\mathrm{\Delta \varphi}}$.
In this long-wavelength limit, the action of the operator appearing in Eq.\ \eqref{eq:DeltaEPhaseTwistKE} on a plane wave making up the condensate wave function $\Psi$ can be approximated as
\begin{equation}
\label{eq:approxLWLGradientPlaneWave}
   (\momentumOperator^{2} - \moatMomentum^{2}) \ec^{\iu \vec{q} \vec{\cdot} \vec{r} + \iu \varphi(\vec{r})} \approx 2 (\vec{q} \vec{\cdot} \vec{\nabla} \varphi) \ec^{\iu \vec{q} \vec{\cdot} \vec{r} + \iu \varphi(\vec{r})} ,
\end{equation}
where we have neglected terms quadratic in the phase gradient as these are slowly varying.
From Eq.\ \eqref{eq:approxLWLGradientPlaneWave} we see that the superfluid behavior in the presence of a moat band will contain fundamental differences with respect to the parabolic case.
Indeed, the latter simply gives $n(\vec{r}) (\vec{\nabla} \varphi)^{2}$ for the part quadratic in the phase gradient.
Meanwhile, the energy functional in the moat-band case depends on $\big|\sum_{i} (\vec{q}_{i} \vec{\cdot} \vec{\nabla} \varphi) \ec^{\iu \vec{q}_{i} \vec{\cdot} \vec{r} + \iu \varphi}\big|^{2}$, which couples the different momenta to the components of $\vec{\nabla} \varphi$.
As a result, we expect a strongly anisotropic superfluid stiffness tensor and nonzero off-diagonal components in triangular and higher-order supersolids.

We are now in the position to compute the superfluid stiffness for various BEC phases.
To this end, we must first solve the equation of motion for $\varphi$ that minimizes Eq.\ \eqref{eq:DeltaEPhaseTwistKE} and compute the resulting expression for $\Delta E$.
We begin with the FF phase, thus $\Psi_{\mathrm{FF}}(\vec{r}) = \sqrt{n} \ec^{\iu q x + \iu \varphi(\vec{r})}$ by orienting the $x$ axis of our coordinate system along the symmetry-breaking direction.
We apply a phase difference $(\Delta \varphi)_{\perp}$ along this direction, where the subscript serves as a reminder that it corresponds to the direction perpendicular to the moat.
The equation of motion trivially reads $\partial_{x}^{2} \varphi = 0$, so that $\partial_{x} \varphi = (\Delta \varphi)_{\perp} / L$, where $L$ is the length of the system in either direction.
This results in the energy shift
\begin{equation}
   K_{\mathrm{FF}} = \frac{1}{2} (8 b \moatMomentum^{2} n) (\Delta \varphi)_{\perp}^{2} ,
\end{equation}
from which we read off $D_{\mathrm{s}, \mathrm{FF}}^{\perp} = 8 b \moatMomentum^{2} n$ and all other components are zero.
This result implies that the phase stiffness is nonzero in the longitudinal direction and vanishes in the transverse direction relative to the ordering vector $\vec{q}$.
This is to be expected, as for a perfectly degenerate moat there can only be phase stiffness in the direction perpendicular to the ring, and in fact our obtained $D^{\perp}_{\mathrm{s}, \mathrm{FF}}$ corresponds exactly to the familiar result $n / m^{\perp}_{\mathrm{eff}}$ with $1 / m^{\perp}_{\mathrm{eff}} = 8 b \moatMomentum^{2}$ the inverse effective mass in the radial direction.
By contrast, the flatness of the moat in the tangential direction results in $D_{\mathrm{s}, \mathrm{FF}}^{\parallel} = 0$.
These results can be understood further by computing the Bogoliubov quasiparticle spectrum of the FF phase, which yields a gapless dispersion that behaves linearly in the perpendicular direction and parabolically in the tangential direction.
According to Landau's criterion, this corresponds to a superfluid response in the perpendicular direction only.

Next we turn to the LO phase, which features two plane waves with momenta which we choose to lie at $\pm \moatMomentum \hat{\mathbf{x}}$.
The equation of motion for the phase given a suitable choice of the arbitrary phase $\theta$ appearing in Eq.\ \eqref{eq:ansatzLOStateTwoMomenta} reads $\partial_{x} (\cos^{2} \moatMomentum x \, \partial_{x} \varphi) = 0$, hence $\partial_{x} \varphi = C {(\Delta \varphi)}_{\perp} / \cos^{2} \moatMomentum x$ with $C \equiv \big(\int_{0}^{L} \mathrm{d} x / \cos^{2} \moatMomentum x \big)^{-1}$.
Plugging this into the expression for the energy reveals that $D^{\perp}_{\mathrm{s}, \mathrm{LO}} \propto C$, which vanishes, so there is no superfluidity in the LO phase.
That the superfluid stiffness must vanish for the LO phase can also be understood from Leggett's bound on the superfluid fraction $f_{\mathrm{s}} \equiv n_{\mathrm{s}} / n$, which states that \cite{PhysRevLett.25.1543,rabec2026superfluid}
\begin{equation}
   f_{\mathrm{s}} \leq \frac{1}{n} {\bigg\langle \frac{1}{{\langle n(\vec{r})^{-1} \rangle}_{\hat{\mathbf{e}}_{1}}}\bigg\rangle}_{\hat{\mathbf{e}}_{2}} ,
\end{equation}
where $\hat{\mathbf{e}}_{1}$ and $\hat{\mathbf{e}}_{2}$ are orthogonal directions and ${\langle \cdots \rangle}_{\hat{\mathbf{e}}_{i}}$ stands for the spatial average over each of them \footnote{Note that Leggett's bound in the form given here does not universally apply to the moat-band case due to the mixing of the ordering vectors and the components of the phase gradient in Eq.\ \eqref{eq:approxLWLGradientPlaneWave}. However, for the LO phase in particular, the derived expression coincides with the one by Leggett as $n(\vec{r}) \propto \cos^{2} \moatMomentum x$.}.
The vanishing of $\mathbf{D}_{\mathrm{s}}$ for the stripe phase can be understood intuitively, because the density pattern exhibits nodal lines along which $n(\vec{r})$ vanishes.
On these lines it is possible to perform a finite phase slip at no additional energy cost, so that adjacent stripes are not phase-coherent.

In the triangle phase, the equation of motion for the phase profile cannot be solved analytically.
Instead, we perform a variational calculation to obtain bounds on the eigenvalues of the superfluid stiffness tensor.
To this end we must take into account the orientation $\theta$ of one of the momentum basis vectors with respect to the axes of the physical sample.
Defining $\vec{r}(\theta) = \mathbf{R}(\theta) \vec{r}$, with $\mathbf{R}(\theta)$ the 2D rotation matrix with angle $\theta$, the kinetic energy in the long-wavelength limit reads
\begin{equation}
   K_{\triangle} = 4 b \moatMomentum^{2} n \int \mathrm{d}^{2} r \, (\vec{\nabla} \varphi)^{\mathsf{T}} \mathbf{F}(\vec{r}(\theta), \theta) (\vec{\nabla} \varphi)  ,
\end{equation}
where $\mathbf{F}(\vec{r}(\theta), \theta) = \mathbf{R}(\theta)^{\mathsf{T}} \mathbf{F}(\vec{r}(\theta), 0) \mathbf{R}(\theta)$ with the components $F_{ij}$ of $\mathbf{F}(\vec{r}, 0)$ given by
\begin{subequations}
   \begin{align}
      F_{11}(\vec{r}) &= \frac{1}{2} - \frac{2}{3} \cos \frac{3 \moatMomentum x}{2} \cos \frac{\sqrt{3} \moatMomentum y}{2} + \frac{1}{6} \cos \sqrt{3} \moatMomentum y , \\
      F_{22}(\vec{r}) &= \frac{1}{2} (1 - \cos \sqrt{3} \moatMomentum y), \\
      F_{12}(\vec{r}) &= \frac{\sqrt{3}}{3} \sin \frac{3 \moatMomentum x}{2} \sin \frac{\sqrt{3} \moatMomentum y}{2} .
   \end{align}
\end{subequations}
In the spirit of Leggett \cite{PhysRevLett.25.1543}, we perform a variational calculation for a phase profile $\varphi(\vec{r})$ that depends separately on $x$ or $y$ only.
We can then bound the eigenvalues of $\mathbf{D}_{\mathrm{s}}$ by looking at the largest and smallest values (as a function of $\theta$) of either of the quantities
\begin{subequations}
   \begin{align}
      A^{xx}(\theta) &\equiv 8 b \moatMomentum^{2} n \, {\bigg\langle \frac{1}{{\langle F_{11}(\vec{r}(\theta), \theta) \rangle}_{\hat{\mathbf{y}}}}\bigg\rangle}_{\hat{\mathbf{x}}}^{-1} , \\
      A^{yy}(\theta) &\equiv 8 b \moatMomentum^{2} n \, {\bigg\langle \frac{1}{{\langle F_{22}(\vec{r}(\theta), \theta) \rangle}_{\hat{\mathbf{x}}}}\bigg\rangle}_{\hat{\mathbf{y}}}^{-1} .
   \end{align}
\end{subequations}
There are three directions in which $A^{xx}$ and $A^{yy}$ vanish, which coincide with the angles at which one of the three ordering momenta is perpendicular to the $x$ and $y$ directions, respectively.
Meanwhile, the largest eigenvalue is at most $4 b \moatMomentum^{2} n$, precisely half of the longitudinal stiffness of the FF case. 

Altogether, these results imply that the superfluid response of a condensate on a moat band can be very different from that of a traditional BEC.
The FF phase, while spatially homogeneous like a normal zero-momentum BEC, exhibits a highly anisotropic superfluid behavior whose directional dependence is dictated by the ordering momentum of the many-body wave function.
More precisely, the superfluid current is maximal along the direction specified by $\vec{q}$ and vanishes in the perpendicular direction.
Similarly, in the triangle phase, the superfluid flow vanishes in the three directions perpendicular to one of the ordering momenta and becomes maximal in between.

\subsection{Phase transition at nonzero temperatures}

Until now we have been considering BEC phases at zero temperature, where mean-field theory is valid also in 2D.
However, it is well known that in 2D systems true off-diagonal long-range order is not possible at $T > 0$.
Rather, in a homogeneous 2D superfluid, correlation functions decay algebraically and thus the phase coherence is not perfect, but a \emph{quasi}condensate phase still exists and in particular exhibits superfluidity.
At nonzero temperatures, this superfluid contains bound vortex--antivortex pairs that become less tightly bound with increasing temperature.
Above a certain temperature, all of these pairs become unbound and their proliferation destroys the superfluidity of the quasicondensate \cite{berezinskii1971destruction,berezinskii1972destruction,kosterlitz1973ordering}.
This temperature is known as the Berezinskii--Kosterlitz--Thouless temperature and is related to the superfluid stiffness tensor via $k_{\mathrm{B}} T_{\mathrm{BKT}} = \frac{\pi}{2} \sqrt{\det \mathbf{D}_{\mathrm{s}}}$.

Let us now turn to our case of moat-band condensates.
The strong anisotropy of $\mathbf{D}_{\mathrm{s}}$ will have significant consequences on the nonzero-temperature properties of the BEC phases.
Indeed, for the FF case only $D_{\mathrm{s}, \mathrm{FF}}^{\perp}$ is nonzero for a perfectly degenerate moat, and thus the determinant of the superfluid stiffness tensor vanishes.
This implies that the FF phase cannot exist at nonzero temperatures in the presence of a perfect moat degeneracy.
However, once the degeneracy is lifted via the warping mechanism explained previously or some other perturbation, this issue disappears and we expect a phase transition at some nonzero temperature.
In the presence of warping this temperature will be given by $k_{\mathrm{B}} T_{\mathrm{BKT}} = \pi n_{\mathrm{s}} / 2 ({m_{\text{eff}}^{\perp} m_{\text{eff}}^{\parallel}})^{1/2}$, where $m_{\text{eff}}^{\parallel}$ is the effective mass in the direction tangential to the moat estimated in Eq.\ \eqref{eq:effMassTangential} and $n_{\mathrm{s}}$ is the superfluid density.
This results in
\begin{equation}
\label{eq:kBTBKTMoat}
   k_{\mathrm{B}} T_{\mathrm{BKT}} = \frac{p N_{\min} \sqrt{\moatEnergyBarrier \delta \epsilon}}{2 \moatMomentum^{2}} n_{\mathrm{s}} ,
\end{equation}
where again $p = 2$ for a quadratic moat and $p = 4$ for a quartic moat.

We draw attention to the presence of the \emph{superfluid} density $n_{\mathrm{s}}$ in the expression for $T_{\mathrm{BKT}}$, as opposed to the exciton density $n$.
While in the FF phase the two coincide at $T = 0$, $n_{\mathrm{s}}$ is temperature-dependent and must in principle be calculated via a suitable renormalization procedure taking into account the bound vortex--antivortex pairs \cite{kosterlitz1974critical}.
Hence, Eq.\ \eqref{eq:kBTBKTMoat} must be understood as a transcendental equation $k_{\mathrm{B}} T_{\mathrm{BKT}} \propto n_{\mathrm{s}}(T_{\mathrm{BKT}})$ that must be solved for $T_{\mathrm{BKT}}$.
However, when $T_{\mathrm{BKT}}$ is low enough (typically of the order of a few Kelvin), one can estimate it by simply employing the unrenormalized superfluid density in what is known as the Nelson--Kosterlitz criterion \cite{bighin2018renormalization}.
Note that the unrenormalized superfluid density (i.e., the superfluid density at $T = 0$) coincides with the exciton density $n$ in the FF case, but for higher-order supersolids $n$ and $n_{\mathrm{s}}$ are not equal in the zero-temperature limit as only a fraction of the condensate is superfluid.
For instance, triangular supersolids of rubidium atoms with superfluid fractions $f_{\mathrm{s}} \sim 0.4$--$1$ were realized in a recent experiment \cite{rabec2026superfluid}.

Using the Nelson--Kosterlitz criterion in Eq.\ \eqref{eq:kBTBKTMoat} with the range of physical variables specified in Sec.\ \ref{sec:expRegion} and a warping energy range $\delta \epsilon \in (0.1, 3) \, \si{\milli\electronvolt}$, we obtain BKT temperatures from the order of a few tens of microkelvin up to $10^{5} \, \si{\kelvin}$.
The upper limit is of course incompatible with the Nelson--Kosterlitz approximation and must be disregarded as unphysical.
Nevertheless, it indicates that temperatures of a few Kelvin or even tens of Kelvin are in principle possible with appropriate parameters.

We expect the $T > 0$ physics of inhomogeneous condensates and supersolids to be more complicated than the relatively simple BKT picture of the FF phase.
While the transition from a superfluid to a normal fluid will still take place via the BKT mechanism of vortex--antivortex unbinding, this now requires a more careful analysis as it will generally be intertwined with the destruction of the crystal order.
In this regard we expect a melting transition in the spirit of Kosterlitz, Thouless, Halperin, and Nelson, where 2D crystal melting occurs via the successive unbinding of dislocations and disclinations giving rise to an intermediate hexatic phase \cite{kosterlitz1972long,kosterlitz1973ordering,halperin1978theory,nelson1979dislocation}.
The order parameters of the BKT and melting transitions will generally be coupled together and likely modify the simple BKT analysis.
Whether the crystalline and superfluid order are destroyed simultaneously or in progressive steps, and the temperatures at which this happens, are extremely interesting questions that lie outside the scope of the present work focusing mainly on the zero-temperature phase diagram.

%% file: sec_GrossPitaevskii.tex

In Sec.\ \ref{sec:MBPhases} we have presented the competition between BEC and noncondensate CSL phases.
In the remainder of the article we assume that the conditions are such that the BEC phases are more stable than the CSL, possibly because of the warping effects due to the discrete rotational symmetry of the underlying lattice.
To explore the qualitative behavior of inhomogeneous and supersolid exciton BECs in a larger region of the parameter space, in this and following sections we will replace the full $T$ matrix by a local pseudopotential that can be used as a first approximation to a more accurate full calculation along the lines of the previous sections.
This greatly simplifies the numerical treatment and allows for analytical results that offer more direct physical insight.
Under these approximations, we give necessary conditions for supersolid formation, present numerical phase diagrams, and develop an analytical Landau theory that matches them accurately.

\subsection{Interaction conditions and pseudopotential approximation}
\label{sec:intCondPseudopotApprox}

As we explained in detail in Sec.\ \ref{sec:modeling}, the presence or absence of inhomogeneous and supersolid BEC states crucially depends on the features of the renormalized interaction or $T$ matrix.
We have seen that even a fully repulsive (but long-ranged) exciton--exciton interaction can give rise to such states when the correct $T$ matrix is computed.
However, the full self-consistent problem is numerically expensive and not particularly physically transparent.
To connect with the more familiar treatment of atomic BECs, in this and the following sections we replace the full $T$ matrix by a local function, which allows us to use a standard Gross--Pitaevskii framework.
To this end we must make sure that our choice for $T(\vec{r})$ is compatible with the formation of inhomogeneous BECs, which leads to the following conditions.
First, for any condensate to be stable, the zero-momentum component $T_{0}$ must be positive.
Furthermore, it is easy to show that an inhomogeneous BEC can only arise in the presence of a negative minimum in Fourier space \cite{sepulveda2008nonclassical,sepulveda2010superfluid}.
In summary, we need $T_{0} > 0$ and $T_{\vec{k}} < 0$ for some nonzero $\vec{k}$.

A negative Fourier component in the $T$ matrix can appear when the bare interaction potential supports the formation of bound states, similar to the situation in Ref.\ \cite{degoey1986three}, where oscillations in the $T$ matrix are observed at energies close to the formation of a bound (or quasi-bound) state.
Furthermore, the excitonic $T$-matrix approach of Refs.\ \cite{conti2023chester,conti2025gross} also introduces an oscillatory behavior to the effective interaction by cutting off the bare potential at distances smaller than some $R_{\mathrm{c}}$ below which the pair-correlation function vanishes at low densities.
In view of these observations and for simplicity, in the remainder of this paper we will replace the full two-exciton $T$ matrix by a soft-core pseudopotential of the form
\begin{equation}
\label{eq:RealSpaceSoftCore}
    T_{\text{sc}}(r) = 
    \begin{dcases}
        \, \softCoreStrength & \text{if } r \leq \softCoreRange, \\
        \, 0 & \text{if } r > \softCoreRange .
    \end{dcases}
\end{equation}
Here, $\softCoreStrength$ is the strength of the soft-core repulsion and $\softCoreRange$ the effective interaction range.
The associated $T$-matrix kernels in 1D and 2D are given by
\begin{equation}
    t^{\text{sc}}_{\tilde{\vec{q}}} = 
    \begin{dcases}
        \, \sin \tilde{q} / \tilde{q} & \text{in 1D,} \\
        \, J_{1}(\tilde{q}) / \tilde{q} & \text{in 2D.}
    \end{dcases}
\end{equation}
These are oscillating and feature a series of decaying negative minima away from the origin.
As can be seen from Eq.\ \eqref{eq:GPEdimless} below, making this pseudopotential approximation has the added benefit of reducing the problem to just two dimensionless variables, namely the parameter $\paramMoatMomVSIntRange$ characterizing the moat properties and the effective interaction strength $\paramSSInt \equiv 2 \paramDensityVSIntRange^{2} \paramIntStrengthVSMoatBarrier$.

\subsection{Gross--Pitaevskii equation and periodic BECs}

Having introduced a local $T$ matrix $T(\vec{r})$ we can now derive a Gross--Pitaevskii framework for the exploration of periodic BEC states on a moat band.
As in Sec.\ \ref{sec:BECs}, at the mean-field level we assume that all $N$ excitons have condensed and the many-body wave function of the system is written as a product of identical single-particle wave functions as $\Psi(\{\vec{r}_{i}\}) = \prod_{i=1}^{N} \psi(\vec{r}_{i})$.
For convenience we define the so-called condensate wave function $\phi(\vec{r}) \equiv \sqrt{N} \psi(\vec{r})$, which satisfies the normalization condition
\begin{equation}
\label{eq:normalizationPhiOfX}
    \int \mathrm{d}^{2} r \, |\phi(\vec{r})|^2 = N .
\end{equation}
The mean-field energy of a spinless Bose gas with a moat dispersion interacting via an effective local $T$ matrix $T(\vec{r})$ then reads
\begin{equation}
\label{eq:FreeEnergy}
    \begin{split}
        E &= \int \mathrm{d}^{2} r \, \phi^{*}(\vec{r}) \bigg(\frac{\paramDispKTwo^2}{4 \paramDispKFour} - \paramDispKTwo \momentumOperator^2 + \paramDispKFour \momentumOperator^4 \bigg) \phi(\vec{r}) \\ 
        &+ \frac{1}{2}\int \mathrm{d}^{2} r \, \mathrm{d}^{2} r' \phi^{*}(\vec{r}) \phi^{*}(\vec{r}') T(\vec{r} - \vec{r}') \phi(\vec{r}') \phi(\vec{r}) .
    \end{split}
\end{equation}
Minimizing Eq.\ \eqref{eq:FreeEnergy} while including a Lagrange multiplier $\mu N$ to impose the normalization constraint yields the Gross--Pitaevskii equation (GPE)
\begin{equation}
\label{eq:condensate_equation}
    \bigg(\frac{\paramDispKTwo^2}{4 \paramDispKFour} - \paramDispKTwo \momentumOperator^2 + \paramDispKFour \momentumOperator^4 + T_{\phi}^{\text{eff}}(\vec{r}) - \mu \bigg)\phi(\vec{r}) = 0 ,
\end{equation}
where we have defined
\begin{equation}
    T_{\phi}^{\text{eff}}(\vec{r}) \equiv \int \mathrm{d}^{2} r' \, T(\vec{r}-\vec{r}') \hspace{0.3mm} |{\phi}(\vec{r}')|^2 .
\end{equation}

We are interested in lattice solutions for the condensate wave function, i.e., solutions where the local density $n(\vec{r}) = |\phi(\vec{r})|^2$ displays a periodic pattern.
It is straightforward to show that a periodic $|\phi(\vec{r})|^{2}$ leads to a periodic effective potential $T_{\phi}^{\text{eff}}$, as long as the integral kernel depends on the difference of coordinates only.
Thus, the GPE of Eq.\ \eqref{eq:condensate_equation} takes the form of a Schr\"odinger equation with a periodic potential.
The solution is therefore given by a Bloch wave function which we write as
\begin{equation}
\label{eq:phiExpansionBloch}
    \phi(\vec{r}) = \sqrt{n} \hspace{0.3mm} \ec^{\iu \vec{g} \boldsymbol{\cdot} \vec{r}} u_{\vec{g}}(\vec{r}) ,
\end{equation}
where the factor of $\sqrt{n}$ has been explicitly factored out for convenience.
The Bloch factor $u_{\vec{g}}(\vec{r})$ is a function with the same periodicity as the emerging density profile, and $\vec{g}$ is a crystal momentum lying in the first Brillouin zone of the associated reciprocal space.
The periodicity of $u_{\vec{g}}(\vec{r})$ allows us to expand it in terms of reciprocal lattice vectors $\vec{G}$ via
\begin{equation}
    \label{eq:BlochWavefunction}
    u_{\vec{g}}(\vec{r}) = \sum_{\vec{G}} u_{\vec{g} \vec{G}} \hspace{0.3mm} \ec^{\iu \vec{G} \vec{\cdot} \vec{r}} ,
\end{equation}
With $\phi(\vec{r})$ written in this way the normalization condition of Eq.\ \eqref{eq:normalizationPhiOfX} becomes
\begin{equation}
    \sum_{\vec{G}} |u_{\vec{g} \vec{G}}|^{2} = 1 .
\end{equation}

If there are $d'$ directions in which translational symmetry is broken, the reciprocal space will be spanned by $d'$ basis vectors.
Our task will be to find the crystal momentum $\vec{g}$, the reciprocal lattice basis vectors $\vec{G}$, and the coefficients $u_{\vec{g} \vec{G}}$ that minimize the free energy.
In momentum space, the GPE is an equation for the expansion coefficients $u_{\vec{g} \vec{G}}$ and takes the form
\begin{equation}
    \varepsilon_{\vec{g} + \vec{G}} \hspace{0.2mm} u_{\vec{g} \vec{G}} + n \sum_{\mathclap{\vec{G}' \vec{G}''}} T_{\vec{G}'' - \vec{G}'} u^{*}_{\vec{g} \vec{G}'} u_{\vec{g} \vec{G}''} u_{\vec{g}, \vec{G} + \vec{G}' - \vec{G}''} = \mu u_{\vec{g} \vec{G}} ,
\end{equation}
where $\varepsilon_{\vec{k}} = \paramDispKTwo^{2} / 4 \paramDispKFour - \paramDispKTwo k^{2} + \paramDispKFour k^{4}$ is the moat dispersion.
In the dimensionless units of Sec.\ \ref{sec:dimlessUnits}, the GPE reads
\begin{equation}
    \label{eq:GPEdimless}
    \dimlessMarker{\varepsilon}_{\dimlessMarker{\vec{g}} + \dimlessMarker{\vec{G}}} \hspace{0.2mm} u_{\dimlessMarker{\vec{g}} \dimlessMarker{\vec{G}}} + \paramSSInt \sum_{\mathclap{\dimlessMarker{\vec{G}}' \dimlessMarker{\vec{G}}''}} t_{\dimlessMarker{\vec{G}}' - \dimlessMarker{\vec{G}}''} \hspace{0.2mm} u^{*}_{\dimlessMarker{\vec{g}} \dimlessMarker{\vec{G}}'} u_{\dimlessMarker{\vec{g}} \dimlessMarker{\vec{G}}''} u_{\dimlessMarker{\vec{g}}, \dimlessMarker{\vec{G}} + \dimlessMarker{\vec{G}}' - \dimlessMarker{\vec{G}}''} = \dimlessMarker{\mu} u_{\dimlessMarker{\vec{g}} \dimlessMarker{\vec{G}}} ,
\end{equation}
where we recall that $\paramSSInt \equiv 2 \paramDensityVSIntRange^{2} \paramIntStrengthVSMoatBarrier$.
Note that in 1D, which we will also consider for intuition, the dimensionless parameter $\paramDensityVSIntRange$ is defined as $\paramDensityVSIntRange = \sqrt{n_{1 \mathrm{D}} \intRange}$, with $n_{1 \mathrm{D}}$ the exciton number per unit \emph{length}.

%% file: sec_1D.tex

In this section we present the phase diagram of moat-band excitons with a soft-core pseudopotential in 1D, a qualitative explanation of the underlying physics, and an effective Landau theory validating the numerical calculations.

\subsection{Numerical results}

\begin{figure*}[!t]
    \includegraphics[width=\textwidth]{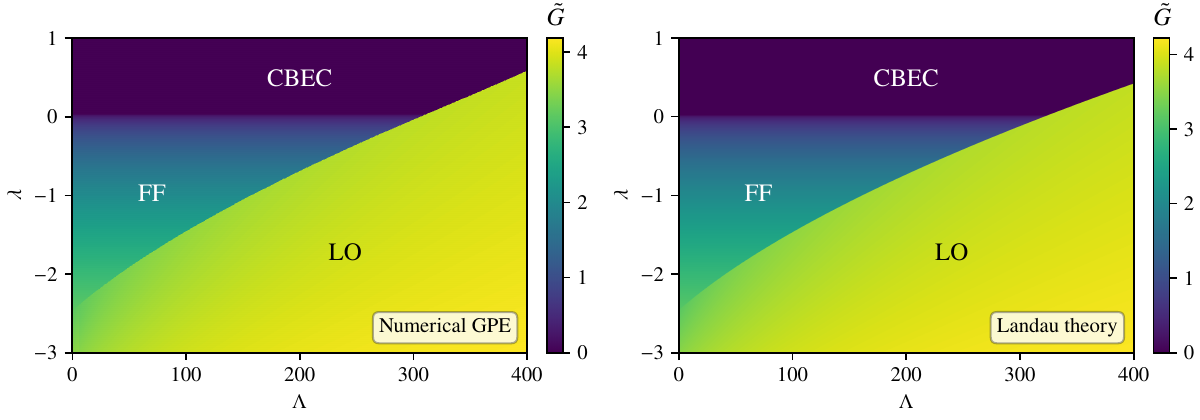}
    \caption{Phase diagram of 1D moat-band excitons with a soft-core $T$ matrix $T_{\mathrm{sc}}(x) = \softCoreStrength \Theta(1 - |x| / \softCoreRange)$ on the $\paramSSMoat$--$\paramSSInt$ plane.
    The left plot shows the result of the full numerical GPE calculation, while the right plot shows the result of the analytical Landau theory.
    The color coding represents the length of the momentum vector $G$ characterizing each phase, made dimensionless via $\dimlessMarker{G} = G \softCoreRange$.
    The constant BEC (CBEC) has a position-independent wave function, $\phi_{\mathrm{CBEC}}(x) = \sqrt{n}$, with $n$ the exciton number density.
    Meanwhile, the FF phase has $\phi_{\mathrm{FF}}(x) = \sqrt{n} \hspace{0.2mm} \ec^{\iu (G / 2) x}$ with $G = 2 \moatMomentum$.
    These two phases give a homogeneous density profile.
    In the LO phase, the order parameter near the transition behaves as $\phi_{\mathrm{LO}}(x) \propto \cos (G x / 2)$, leading to a lattice constant $2 \pi / G$ for the density profile $|\phi(x)|^{2}$.
    The transition between the CBEC and FF phases at $\paramSSMoat = 0$ is continuous, while that between the homogeneous phases and the LO phase is discontinuous.
    For a large enough moat, i.e., negative enough $\paramSSMoat$, an inhomogeneous BEC exists even in the limit of weak interactions and low exciton density.
    The simple Landau theory is in excellent quantitative agreement with the full numerical result in the region of this plot.
    Deviations start to occur at large $\paramSSInt$, where the density peaks localize and the order parameter cannot be written as a simple cosine.
    }
    \label{fig:1DPhaseDiagrams}
\end{figure*}

We have numerically solved Eq.\ \eqref{eq:GPEdimless} as a function of $\paramSSMoat$ and $\paramSSInt$ in 1D, and the resulting phase diagram is plotted in the first panel of Fig.\ \ref{fig:1DPhaseDiagrams}.
In the 1D case, there are three possible phases.
The purple region corresponds to a condensate with a constant order parameter, and is simply dubbed the constant BEC (CBEC) phase for simplicity.
By contrast, the turquoise region corresponds to the homogeneous FF phase, while the yellow region is the LO or stripe phase, featuring an inhomogeneous periodic density profile.

We can qualitatively understand the phase diagram as follows.
For positive $\paramSSMoat$ the dispersion is convex and its minimum lies at zero momentum.
For small values of $\paramSSInt$, i.e., low densities and weak interactions, the bosons will tend to occupy this single momentum state and the resulting phase will be the CBEC.
However, as $\paramSSInt$ increases, the situation can change in the presence of a negative minimum of the interaction potential in Fourier space.
For large enough $\paramSSInt$ this minimum can favor the occupation of two opposite nonzero momentum states $\pm g$, as the resulting increase in kinetic energy can eventually be fully compensated by the attractive coupling induced by a momentum transfer of $2g$.
This marks the onset of the LO phase at positive $\paramSSMoat$.
The transition from the BEC to the LO phase is discontinuous, which can be understood from the fact that the ordering momentum $g$ jumps from zero in the BEC phase to a nonzero value such that $\TMatrixKernel_{2 \dimlessMarker{g}} < 0$.
When $\paramSSMoat$ is small and negative and $\paramSSInt$ is also small, all particles again sit at the state of lowest kinetic energy, which is now spontaneously chosen between the two minima which constitute the moat in the 1D case.
In this situation, it is unfavorable for the excitons to occupy both minima simultaneously because this yields a positive energy shift due to the potential at momentum transfer $2 \moatMomentum$, which is positive for a small moat momentum.
Thus, in this regime we find the FF phase.
The transition from the CBEC to the FF phase as we tune $\paramSSMoat$ from positive to negative is a Lifshitz transition and thus continuous, as in fact the CBEC can simply be understood as an FF phase with zero momentum.
For small and negative $\paramSSMoat$, increasing $\paramSSInt$ leads to the same physics explained above, and we eventually reach a discontinuous phase transition from the FF to the LO state.
As $\paramSSMoat$ is made more negative and the moat increases in size, this transition takes place at increasingly small values of $\paramSSInt$, because the momentum transfer $2 \moatMomentum$ becomes closer to the point where the interaction potential changes sign.
Eventually we reach a point where $\TMatrixKernel_{2 \dimlessMoatMomentum} = 0$, which for the 1D case with $\TMatrixKernel_{\dimlessMarker{k}} = \sin \dimlessMarker{k} / \dimlessMarker{k}$ occurs at $2 \dimlessMoatMomentum = \pi$, or $\paramSSMoat = {-}\pi^{2} / 4 \approx {-}2.47$.
At this point the barrier for the creation of an inhomogeneous BEC is completely removed and any positive $\paramSSInt$, no matter how small, will stabilize the LO phase.
We thus conclude that a moat dispersion can lead to a stripe phase already in the regime of low densities and weak interactions.

We also note that decreasing $\paramSSMoat$ beyond the region shown in the phase diagram of Fig.\ \ref{fig:1DPhaseDiagrams} eventually results in a reentrant FF phase and a subsequent alternating behavior of the FF and LO phases.
This can be fully understood analytically and is explained in the following section.

\begin{figure}[!t]
    \includegraphics[width=\linewidth]{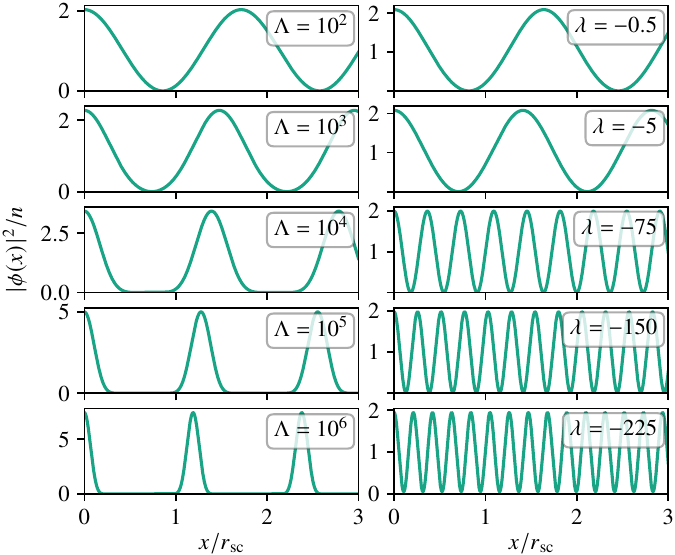}
    \caption{Density profile of the 1D LO phase for different values of the parameters.
    The left column corresponds to a fixed value of $\paramSSMoat = {-2}$ while the right column corresponds to $\paramSSInt = 300$.
    The lattice constant $c$ is related to the 1D ordering vector $G$ via $c = 2\pi / G$.
    For modest values of $\paramSSInt$ the profile is very close to a perfect cosine and the Landau theory developed in Sec.\ \ref{sec:LandauTheory1D}  gives quantitatively accurate results.
    For very large values of $\paramSSInt$ the peaks localize and the Landau theory breaks down as we go deep inside the LO phase.}
    \label{fig:1DDensities}
\end{figure}

In Fig.\ \ref{fig:1DDensities} we have plotted the numerically obtained density profiles for several values of $\paramSSInt$ at fixed $\paramSSMoat$, and vice-versa.
As $\paramSSInt$ increases, the high-density regions become more peaked while the regions between the peaks flatten out.
This is because for large $\paramSSInt$ the interaction dominates and the condensate can lower its energy by forming regions of vanishing density and localizing the latter in increasingly small patches.
Meanwhile, as $\moatMomentum$ grows and thus $\paramSSMoat$ becomes more negative, the lattice constant of the density profile decreases.
This can be easily understood for small $\paramSSInt$, as in this case the ordering momentum will be close to the moat momentum $\moatMomentum$.

\subsection{Analytical Landau theory}
\label{sec:LandauTheory1D}

All phase transitions in 1D can be further understood analytically by writing down an effective Landau theory for the condensate wave function.
Inspired by the numerical results, we write it in terms of only two nonzero coefficients as
\begin{equation}
\label{eq:LO1DAnsatz}
    \phi(x) = \sqrt{n} \hspace{0.2mm} \big(u_{1} \hspace{0.3mm} \ec^{\iu g_{1} x} + u_{2} \hspace{0.3mm} \ec^{\iu g_{2} x} \big),
\end{equation}
where we assume that $g_{1} \neq g_{2}$.
Note that we have absorbed the momentum $g$ of the expansion in Eq.\ \eqref{eq:phiExpansionBloch} inside the reciprocal lattice vectors $G$.
The (dimensionless) energy per particle with this ansatz reads
\begin{equation}
\label{eq:LandauTheory1D}
    \begin{split}
        \dimlessMarker{e} &= |u_{1}|^{2} \dimlessMarker{\varepsilon}_{\dimlessMarker{g}_{1}} + |u_{2}|^{2} \dimlessMarker{\varepsilon}_{\dimlessMarker{g}_{2}} \\
        &+ \frac{\paramSSInt}{2} \TMatrixKernel_{0} \big(|u_{1}|^{2} + |u_{2}|^{2}\big)^{2} + \paramSSInt \TMatrixKernel_{\dimlessMarker{g}_{1} - \dimlessMarker{g}_{2}} |u_{1}|^{2} |u_{2}|^{2} ,
    \end{split}
\end{equation}
which we must minimize with respect to $u_{1}$, $u_{2}$, $\dimlessMarker{g}_{1}$, and $\dimlessMarker{g}_{2}$ under the constraint $|u_{1}|^{2} + |u_{2}|^{2} = 1$.
The latter is introduced via a Lagrange multiplier which is simply the chemical potential, $\dimlessMarker{e} \rightarrow \dimlessMarker{e} - \dimlessMarker{\mu} (|u_{1}|^{2} + |u_{2}|^{2} - 1)$.
It is clear from Eq.\ \eqref{eq:LandauTheory1D} that a relative phase between $u_{1}$ and $u_{2}$ does not influence the free energy, which can be seen from the fact that such a shift simply corresponds to a horizontal displacement of the density profile generated by $\phi(x)$.
Thus, we assume without loss of generality that both $u_{1}$ and $u_{2}$ are real and positive.

Firstly, we find trivial solutions which have $u_{1} = 1$ and $u_{2} = 0$, or $u_{1} = 0$ and $u_{2} = 1$.
For these, the momentum associated with the nonzero component is zero when $\paramSSMoat > 0$ and $\pm \dimlessMoatMomentum$ when $\paramSSMoat < 0$, which correspond to the CBEC and FF phases, respectively.
The chemical potential is given by $\dimlessMarker{\mu} = \paramSSMoat^{2} + \paramSSInt \TMatrixKernel_{0}$ for $\paramSSMoat > 0$ and $\dimlessMarker{\mu} = \paramSSInt \TMatrixKernel_{0}$ for $\paramSSMoat < 0$.
Whether or not these homogeneous phases correspond to the ground state of the system for positive or negative $\paramSSMoat$ will depend on the specific value of $\paramSSInt$, which is present in the two terms on the second line of Eq.\ \eqref{eq:LandauTheory1D}.
The $\TMatrixKernel_{0}$ term has no effect due to the normalization constraint, which effectively results in an overall shift irrespective of $u_{1}$ and $u_{2}$.
However, from the last term of Eq.\ \eqref{eq:LandauTheory1D} we see that $\paramSSInt$ couples to the potential at the momentum transfer $\dimlessMarker{g}_{1} - \dimlessMarker{g}_{2}$.
Thus, it is clear that a negative Fourier component at nonzero momentum can lower the overall energy and thus give rise to periodic solutions.

Second, we find nontrivial solutions where both coefficients are nonzero and satisfy $u_{1} = u_{2} = 1/\sqrt{2}$.
These have $\dimlessMarker{g}_{1} = {-} \dimlessMarker{g}_{2} \equiv \dimlessMarker{G} / 2$ and $\dimlessMarker{\mu} = \dimlessMarker{\varepsilon}_{\dimlessMarker{G}/2} + \paramSSInt (\TMatrixKernel_{0} + \frac{1}{2} \TMatrixKernel_{\dimlessMarker{G}})$ \footnote{The reason for writing $\dimlessMarker{g}_{1}$ and ${-} \dimlessMarker{g}_{2}$ as $\pm \dimlessMarker{G} / 2$ is that in our numerical procedure we obtain the order parameter as $\phi(x) = \ec^{\iu g x} (u_{0} + u_{1} \ec^{\iu G x})$, where $g = {-} G/2$ lies in the first Brillouin zone. In this way, the Bloch factor is identified with the term inside parentheses, which has the same periodicity as the emerging density lattice. Because in the Landau theory we group $g$ and $G$ together, ultimately $g_{1}$ and $g_{2}$ must be identified with $\pm G/2$ in order to compare with the numerical result.}.
Without loss of generality we take $\dimlessMarker{G} > 0$.
Unlike for the trivial solutions with one of the coefficients equal to zero, the value of $\dimlessMarker{g}$ now depends on the details of the interaction kernel $\TMatrixKernel$ and is found by solving the transcendental equation
\begin{equation}
\label{eq:transcendentalEqLandau1D}
    \dimlessMarker{G} (\dimlessMarker{G}^{2} + 4 \paramSSMoat) + \paramSSInt \TMatrixKernel'_{\dimlessMarker{G}} = 0 ,
\end{equation}
where $\TMatrixKernel'$ stands for the derivative of $\TMatrixKernel$ with respect to its argument.
In the case of a soft-core potential in 1D, we have $\TMatrixKernel'_{\dimlessMarker{G}} = (\dimlessMarker{G} \cos \dimlessMarker{G} - \sin \dimlessMarker{G}) / \dimlessMarker{G}^{2}$.
The values of $\paramSSMoat$ and $\paramSSInt$ determine whether the transcendental equation \eqref{eq:transcendentalEqLandau1D} has a nontrivial solution for $\dimlessMarker{G}$.
Plotting the function for $\paramSSMoat > 0$, i.e., in the absence of a moat, reveals that nontrivial solutions for $\dimlessMarker{G}$ exist only when $\paramSSInt$ exceeds some ($\paramSSMoat$-dependent) critical value.
Therefore, periodic solutions in the absence of a moat require relatively high densities or strong interactions.
Meanwhile, for $\paramSSMoat < 0$ there always exists a nontrivial solution for $\dimlessMarker{G}$, even in the limit of $\paramSSInt \rightarrow 0$, where evidently $\dimlessMarker{G} = 2 \dimlessMoatMomentum$.
This indicates that periodic solutions exist regardless of the density or interaction strength.
For small $\paramSSInt$, we can find the momentum of the density lattice to linear order as
\begin{equation}
    \dimlessMarker{G} = 2 \dimlessMoatMomentum + \frac{\paramSSInt}{2 (2 \dimlessMoatMomentum)^{4}} (\sin 2 \dimlessMoatMomentum - 2 \dimlessMoatMomentum \cos 2 \dimlessMoatMomentum) + \mathcal{O}(\paramSSInt^{2}) .
\end{equation}
Thus, in this particular model, turning on the interactions leads to an initial growing or shrinking of the lattice constant of the BEC, depending on the size of the moat.

\begin{figure}
    \centering
    \includegraphics[width=\linewidth]{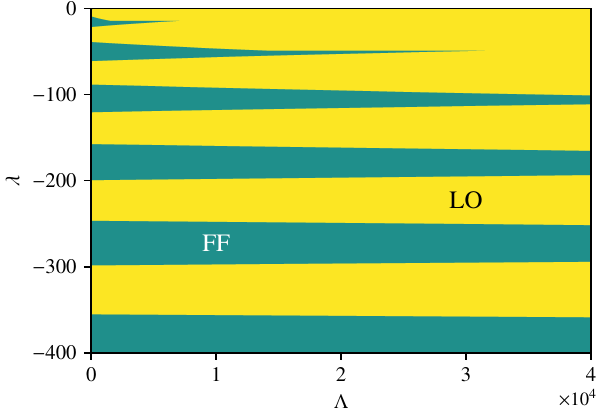}
    \caption{Extended phase diagram of 1D moat-band excitons with a soft-core potential resulting from the analytical Landau theory developed in the main text.
    For $\paramSSMoat < 0$, the oscillations of the interaction potential in momentum space lead to a reentrant behavior between the FF and LO phases.
    For small $\paramSSInt$ these transitions take place at $\paramSSMoat = {-} m^{2} \pi^{2} / 4$ with integer $m > 0$, which are the points where the moat momentum transfer $2 \moatMomentum$ causes the potential kernel to change sign.
    Note that here the color coding only serves to distinguish the phases and does not represent the length of the characteristic momentum.
    }
    \label{fig:1DPhaseDiagramExtended}
\end{figure}

By comparing the free energy of the solutions we have discussed we obtain the phase diagram of the Landau theory, which is plotted in the second panel of Fig.\ \ref{fig:1DPhaseDiagrams}.
We see that the agreement with the full numerical calculation is excellent and only starts deviating slightly for large $\paramSSInt$, the reason being that in this regime the peaks of $\phi(x)$ start to localize and we would need to include more coefficients in the expansion of Eq.\ \eqref{eq:LO1DAnsatz}.

We can use the analytical Landau theory to extend the phase diagram outside of the region for which the phase diagram was obtained numerically.
In Fig.\ \ref{fig:1DPhaseDiagramExtended} we plot the phase diagram for large negative values of $\paramSSMoat$.
While the latter are most certainly unrealistic, the extended phase diagram further helps in understanding the underlying physics.
As $\paramSSMoat$ decreases, the LO phase and the FF phase begin to alternate, which is a consequence of the oscillating nature of the soft-core potential in momentum space.
As we make $\paramSSMoat$ more negative and thus increase the moat radius, the interaction potential vanishes at specific values of the moat momentum transfer $2 \moatMomentum$.
This leads to a change of sign in the interaction energy associated with coupling both sides of the moat and thus results in a back-and-forth behavior between the FF and LO phases.
In 1D and for a soft-core interaction, these transitions take place at $\paramSSMoat = {-} m^{2} \pi^{2} / 4$, where $m$ is a positive integer.
Furthermore, every time we reenter the LO phase the associated reciprocal lattice vector becomes larger and therefore the periodicity of the density pattern is smaller.
If the Fourier transform of the interaction potential were only to change sign once and stay negative above some value of the momentum, then there would be no reentrant behavior and the LO phase would be present for all values below the critical $\paramSSMoat$.

It is worth mentioning that we have also considered a Landau theory with three momentum components, $\phi(x) = \sqrt{n} (u_{0} + u_{+} \ec^{\iu g x} + u_{-} \ec^{{-} \iu g x})$, which would describe a hybrid zero-momentum--LO phase similar to that of Eq.\ \eqref{eq:PsiHybrid}.
The energy in this case reads
\begin{equation}
    \begin{split}
        \dimlessMarker{e} &= |u_{0}|^{2} \dimlessMarker{\varepsilon}_{0} + \big(|u_{+}|^{2} + |u_{-}|^{2}\big) \dimlessMarker{\varepsilon}_{\dimlessMarker{g}} + \paramSSInt \TMatrixKernel_{\dimlessMarker{g}} |u^{*}_{0} u_{+} + u^{*}_{-} u_{0}|^{2}  \\
        & + \frac{\paramSSInt}{2} \TMatrixKernel_{0} \big(|u_{0}|^{2} + |u_{+}|^{2} + |u_{-}|^{2}\big)^{2} + \paramSSInt \TMatrixKernel_{2 \dimlessMarker{g}} |u_{+}|^{2} |u_{-}|^{2} .
    \end{split}
\end{equation}
However, we have found that the energy of such a hybrid phase is always higher than that of one of the pure phases, and thus it does not appear in the considered region of the parameter space, in accordance with the full numerical calculation.
We note that this is an artifact of the quartic dispersion that we are using.
In 1D, as pointed out in Ref.\ \cite{sepulveda2008nonclassical}, a quadratic dispersion with a long-range interaction identical to the one considered here yields a mixed phase in a restricted region of the parameter space.
We have checked that this phase disappears upon the use of a quartic dispersion, so there is no contradiction.

%% file: sec_2D.tex

In this section we provide the phase diagram of two-dimensional moat-band excitons within the pseudopotential approximation.
Having built intuition via the 1D case, it is relatively straightforward to understand the physics in a 2D scenario.
Here the analysis is complicated slightly by the fact that there are several distinct candidate lattice structures.
These must in principle be compatible with the discrete rotational symmetry of the underlying lattice in order to accommodate the warping effects described in Sec.\ \ref{sec:warping}.
In our calculations, where for simplicity we use a perfect moat, we have considered a broken translational symmetry in one direction and broken translational symmetry in two directions for a triangular lattice and a square lattice.
While the former (which leads to a stripe phase) is compatible with any lattice structure, the latter must be chosen depending on the discrete rotational symmetry of the lattice.
Thus, a $C_{3}$- or $C_{6}$-invariant system will be compatible with a triangular phase, while a $C_{4}$-invariant system will be compatible with a square phase.

\subsection{Numerical results}

\begin{figure}[!t]
    \includegraphics[width=\linewidth]{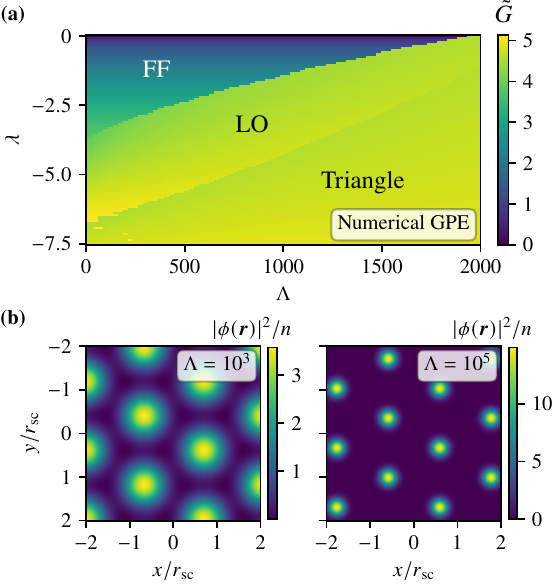}
    \caption{(a) Phase diagram of 2D moat-band excitons with a soft-core interaction $T_{\mathrm{sc}}(r) = \softCoreStrength \Theta(1 - r / \softCoreRange)$.
    The parameters $\paramSSMoat$ and $\paramSSInt$ have the same interpretation as in the 1D case and we do not show the constant BEC present at $\paramSSMoat > 0$.
    The FF and LO phases have the same wave function as in the 1D case, while the triangle phase has three coefficients located at the vertices of an equilateral triangle.
    The artifacts in the bottom left corner are due to numerical instabilities at small $\paramSSInt$.
    (b) Density profiles of the triangle phase for $\paramSSMoat = {-}5$ and two values of the interaction parameter $\paramSSInt$.
    }
    \label{fig:2DPhaseDiagram}
\end{figure}

We numerically obtain the phase diagram involving the three aforementioned lattices by solving the momentum-space GPE with such an ansatz and minimizing the energy with respect to the length of the reciprocal lattice vector, and then comparing the energies obtained for the different structures.
The result is plotted in Fig.\ \ref{fig:2DPhaseDiagram}, where we focus on the $\paramSSMoat < 0$ case as for positive $\paramSSMoat$ the situation is similar to that in the 1D case.
As in the 1D case, for small $\paramSSMoat$ and $\paramSSInt$ we obtain a homogeneous FF phase which is eventually replaced by a stripe or LO phase when either $\paramSSInt$ becomes larger or $\paramSSMoat$ becomes more negative.
Furthermore, for even larger $\paramSSInt$ or larger moats we obtain a triangular phase where the translational symmetry is broken in both directions.
Even deeper in the phase diagram we have obtained a square lattice, although this region is not shown in the plot.
These transitions are all of first order.
We also plot two examples of the density profile in the triangular phase for a fixed value of $\paramSSMoat$.
A similar behavior as for the 1D case can be seen when increasing $\paramSSInt$, by which the peaks tend to localize and the regions between the peaks flatten out.

\subsection{Analytical Landau theories}

In this section we consider Landau theories for the transitions between the 2D periodic BECs at small $\paramSSInt$, which if desired can be extended to all $\paramSSInt$ via the methods introduced in Sec.\ \ref{sec:LandauTheory1D}.
The transition between the FF and LO phases can be understood analytically via the same Landau theory that we used in 1D by simply adding a second component to the momentum and assuming that there is no symmetry breaking along that direction.
The only difference is that the soft-core interaction kernel is now $\TMatrixKernel_{\dimlessMarker{G}} = J_{1}(\dimlessMarker{G}) / \dimlessMarker{G}$, whose derivative entering Eq.\ \eqref{eq:transcendentalEqLandau1D} reads $\TMatrixKernel'_{\dimlessMarker{G}} = {-} J_{2}(\dimlessMarker{G}) / \dimlessMarker{G}$.
This leads to the same qualitative behavior explained in the previous section and only changes the value of $\paramSSMoat$ at which the transition occurs.
This value can be determined by finding the point at which $t_{2 \dimlessMoatMomentum}$ changes sign and yields $\paramSSMoat \approx {-} 3.67$.
This is somewhat higher than for the 1D case, where $\paramSSMoat \approx {-} 2.47$.
Once again the results are in excellent agreement with the numerics and for this reason we do not display them further.

We can also obtain the critical value of $\paramSSMoat$ at which the transition from the LO to the triangle phase occurs at small $\paramSSInt$.
In this case we must compare the interaction energy in each of the two phases as a function of the moat size.
In the LO phase, the interaction energy (without the zero point common to all phases) is $\dimlessMarker{e}^{(2)}_{\text{int}} = \paramSSInt \hspace{0.3mm} \TMatrixKernel_{2 \dimlessMoatMomentum} |u_{+}|^{2} |u_{-}|^{2}$, with $|u_{+}| = |u_{-}| = 1/\sqrt{2}$ the coefficients of $\phi(\vec{r})$ at opposite sides of the moat of radius $\moatMomentum$.
In the triangle phase, it is given by $\dimlessMarker{e}^{(3)}_{\text{int}} = \paramSSInt \hspace{0.3mm} \TMatrixKernel_{\sqrt{3} \dimlessMoatMomentum} (|u_{1}|^{2} |u_{2}|^{2} + |u_{2}|^{2} |u_{3}|^{2} + |u_{3}|^{2} |u_{1}|^{2})$, with $|u_{1}| = |u_{2}| = |u_{3}| = 1/\sqrt{3}$ the coefficients of three points on the moat forming an equilateral triangle with the length of each side equal to $\sqrt{3} \moatMomentum$.
Comparing the two, we conclude that the moat momentum at which the LO-to-triangle transition occurs is found by solving the transcendental equation
\begin{equation}
	\TMatrixKernel_{2 \dimlessMoatMomentum} - \frac{4}{3} \TMatrixKernel_{\sqrt{3} \dimlessMoatMomentum} = 0 .
\end{equation}
For the 2D soft-core interaction this leads to $\paramSSMoat \approx {-} 6.65$, in perfect agreement with the numerical calculation of Fig.\ \ref{fig:2DPhaseDiagram}.

While we have not extended the phase diagram to very negative values of $\paramSSMoat$, we can use the procedure above to infer the presence of other lattice structures deeper in the phase diagram.
For instance, let us consider a square phase with coefficients $u_{1}$, $u_{2}$, $u_{3}$, $u_{4}$ corresponding to the points $(\moatMomentum, 0)$, $(0, \moatMomentum)$, $({-}\moatMomentum, 0)$, and $(0, {-}\moatMomentum)$ in momentum space, respectively, which form a square with diagonal $2 \moatMomentum$.
The interaction energy reads
\begin{equation}
	\begin{split}
		\dimlessMarker{e}^{(4)}_{\text{int}} &= \paramSSInt \hspace{0.3mm} \TMatrixKernel_{2 \dimlessMoatMomentum} \big(|u_{1}|^{2} |u_{3}|^{2} + |u_{2}|^{2} |u_{4}|^{2} \big) \\
		&+ \paramSSInt \hspace{0.3mm} \TMatrixKernel_{\sqrt{2} \dimlessMoatMomentum} \big(|u_{1}|^{2} |u_{2}|^{2} + |u_{2}|^{2} |u_{3}|^{2} + |u_{3}|^{2} |u_{4}|^{2} \\
		&+ |u_{4}|^{2} |u_{1}|^{2} + 2 u_{1}^{*} u_{3}^{*} u_{2} u_{4} + 2 u^{*}_{2} u^{*}_{4} u_{1} u_{3} \big) .
	\end{split}
\end{equation}
This can be found geometrically by considering all scattering processes with nonzero momentum transfer and equal total ingoing and outgoing momenta and counting their multiplicities.
Particular attention must be paid to the last two terms, which depend on the relative phases of the four coefficients.
While we expect that $|u_{1}| = |u_{2}| = |u_{3}| = |u_{4}| = 1/2$, depending on the sign of $\TMatrixKernel_{\sqrt{2} \dimlessMoatMomentum}$ we will have to choose the sign of $u_{1}^{*} u_{3}^{*} u_{2} u_{4}$ to be $+1$ or ${-}1$ in order to lower the total energy.
Taking this into account, we find that at small $\paramSSInt$ there will be a transition from the triangle phase to the square phase at the moat momentum that solves the equation
\begin{equation}
	\TMatrixKernel_{2 \dimlessMoatMomentum} + 2 \big(\TMatrixKernel_{\sqrt{2} \dimlessMoatMomentum} - |\TMatrixKernel_{\sqrt{2} \dimlessMoatMomentum}|\big) - \frac{8}{3} \TMatrixKernel_{\sqrt{3} \dimlessMoatMomentum} = 0 .
\end{equation}
This gives $\paramSSMoat \approx {-}8.75$ for the 2D soft-core potential.

We stress that this triangle-to-square transition is theoretically only valid in the limit of a perfectly degenerate moat, i.e., in the absence of warping, so that it is simultaneously compatible with both a triangle phase and a square phase.
Since some slight degree of warping is likely always present, in a $C_{3}$- or $C_{6}$-symmetric system we expect a hexagonal phase after the triangular one, while in a $C_{4}$-symmetric system it is reasonable to expect the LO to directly give rise to a square phase.
For this reason, we now focus on a $C_{3}$-symmetric system with six minima around the moat as in Fig.\ \ref{fig:warpingBi2Se3}.
We assume that $\paramSSInt$ is small enough so that the many-body wave function is indeed locked into the minima of the dispersion, and we consider the small interaction energy as a perturbation lifting the degeneracy of the different possible phases.
There are three possibilities for a stripe phase, which arise from the three inequivalent ways in which we can distribute two nonzero coefficients among the six minima of the moat.
We can summarize this by writing the stripe interaction energy as
\begin{equation}
	\label{eq:fint2}
	\frac{\dimlessMarker{e}^{(2)}_{\text{int}}}{\paramSSInt} = \frac{1}{4} \operatorname{min} \big(\TMatrixKernel_{\dimlessMoatMomentum}, \TMatrixKernel_{\sqrt{3}\dimlessMoatMomentum}, \TMatrixKernel_{2\dimlessMoatMomentum}\big) .
\end{equation}
Assuming equal weights for the coefficients, there are also three possible inequivalent triangle phases, so that
\begin{equation}
	\label{eq:fint3}
	\frac{\dimlessMarker{e}^{(3)}_{\text{int}}}{\paramSSInt} = \frac{1}{9} \operatorname{min} \big(\TMatrixKernel_{\dimlessMoatMomentum} +  \TMatrixKernel_{\sqrt{3}\dimlessMoatMomentum} + \TMatrixKernel_{2\dimlessMoatMomentum} , 2 \TMatrixKernel_{\dimlessMoatMomentum} + \TMatrixKernel_{\sqrt{3} \dimlessMoatMomentum}, 3 \TMatrixKernel_{\sqrt{3} \dimlessMoatMomentum}\big) ,
\end{equation}
where the last possibility corresponds to the equilateral case.
Finally, the interaction energy of the $C_{3}$-compatible hexagonal phase reads
\begin{equation}
	\label{eq:fint6}
	\frac{\dimlessMarker{e}^{(6)}_{\text{int}}}{\paramSSInt} = \frac{\TMatrixKernel_{2 \dimlessMoatMomentum}}{12} + \frac{5}{24} \big(\TMatrixKernel_{\dimlessMoatMomentum} + \TMatrixKernel_{\sqrt{3} \dimlessMoatMomentum} \big) - \frac{1}{8} \big|\TMatrixKernel_{\dimlessMoatMomentum} + \TMatrixKernel_{\sqrt{3} \dimlessMoatMomentum} \big| ,
\end{equation}
where again we have assumed equal $|u_{i}|$.
Note that we have chosen the phases of the coefficients in such a way that they will always yield the lowest possible interaction energy \footnote{Writing the coefficients as $u_{i} = |u_{i}| z_{i}$, where $|z_{i}| = 1$, an optimal choice for the phases is $z_{i} = 1$ for all $i$ when $\TMatrixKernel_{\dimlessMoatMomentum} + \TMatrixKernel_{\sqrt{3} \dimlessMoatMomentum} < 0$, and $z_{1} z_{4} = 1$, $z_{2} z_{5} = \ec^{2 \iu \pi / 3}$, $z_{3} z_{6} = \ec^{4 \iu \pi / 3}$ when $\TMatrixKernel_{\dimlessMoatMomentum} + \TMatrixKernel_{\sqrt{3} \dimlessMoatMomentum} > 0$. This choice leads to the expression of Eq.\ \eqref{eq:fint6}}.
The assumption of equal $|u_{i}|$ is justified in view of the observation that, in a situation where having unequal coefficients lowers the energy, there will be a stripe phase that is even more stable and which we are already considering.
This relies on the fact that our soft-core interaction in momentum space has a single dominant minimum.
In situations with several competing minima the situation is more complicated, and for a perfectly degenerate moat could lead to anisotropic phases or even quasicrystals owing to the presence of two or more competing length scales in the interaction potential \cite{mkhonta2013exploring,heinonen2019quantum,mendoza2022exploring,grossklags2024self}.
Such an interaction seems rather unnatural for an excitonic system and we do not explore these possibilities any further in this article.

\begin{figure}[!t]
	\includegraphics[width=\linewidth]{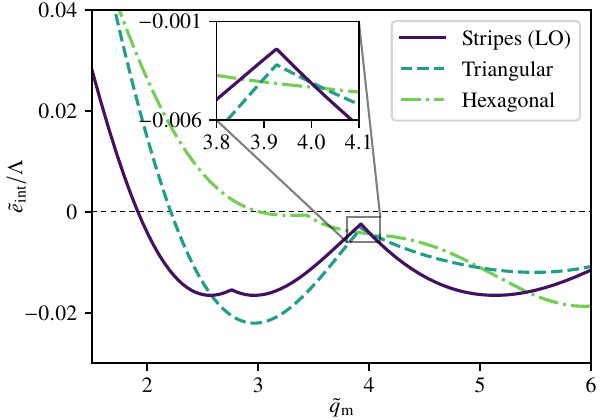}
	\caption{Soft-core interaction energies in 2D for different $C_{3}$-compatible periodic lattices for small $\paramSSInt$ as a function of the moat momentum, corresponding to Eqs.\ \eqref{eq:fint2}--\eqref{eq:fint6} of the main text.
	The different phases correspond to a momentum-space order parameter with $2$, $3$, or $6$ coefficients of equal magnitude distributed on the minima of the moat.
	For $\dimlessMoatMomentum < 1.92$, which corresponds to $\paramSSMoat \approx {-}3.67$, the most stable phase is the homogeneous FF phase with vanishing energy.
	Above this value, the state of lowest energy changes as a function of $\dimlessMoatMomentum$, giving rise to inhomogeneous and supersolid phases with different degrees of discrete rotational symmetry.
	The kinks in the square and hexagonal interaction energy curves correspond to changes in the optimal choice of phases for the coefficients as well as different combinations of the populated minima, as explained in the main text.
	In this range of $\dimlessMoatMomentum$, a stable triangular phase is always found to be equilateral.
	The inset shows a small region where a hexagonal phase is stabilized.}
	\label{fig:intEnergyLattices}
\end{figure}

In Fig.\ \ref{fig:intEnergyLattices} we have plotted the interaction energies of Eqs.\ \eqref{eq:fint2}--\eqref{eq:fint6}.
As long as $\dimlessMarker{e}_{\text{int}} < 0$, the lowest curve will correspond to the most stable phase.
In practice we find that when the triangular phase is stabilized this is always in the form of the equilateral phase, which relates to our comment about the potential possessing a single absolute minimum in Fourier space.
It is worth mentioning that we have also considered a rectangular phase on four minima of the dispersion, but its energy is always higher and we do not show the corresponding curve.
As explained earlier, we indeed find that there is a transition from the triangle phase to the hexagonal phase, and we can solve for the associated critical $\paramSSMoat$ by solving the transcendental equation
\begin{equation}
	\TMatrixKernel_{\sqrt{3} \dimlessMoatMomentum} + \big|\TMatrixKernel_{\dimlessMoatMomentum} + \TMatrixKernel_{\sqrt{3} \dimlessMoatMomentum} \big| - \frac{2}{3} \TMatrixKernel_{2 \dimlessMoatMomentum} - \frac{5}{3} \TMatrixKernel_{\dimlessMoatMomentum} = 0 .
\end{equation}
This leads to $\paramSSMoat \approx {-}15.16$ and corresponds to $\dimlessMoatMomentum \approx 3.89$, which is clearly seen in the inset of Fig.\ \ref{fig:intEnergyLattices}.
Similarly we can also find the LO-to-square transition expected in a $C_{4}$-symmetric system by solving
\begin{equation}
	\TMatrixKernel_{2 \dimlessMoatMomentum} + 2 \big| \TMatrixKernel_{\sqrt{2} \dimlessMoatMomentum} \big| - 2 \TMatrixKernel_{\sqrt{2} \dimlessMoatMomentum} = 0 ,
\end{equation}
which yields $\paramSSMoat \approx {-} 7.92$.

\subsection{Estimates for real systems}

Let us finally give a rough estimate of the parameter $\paramSSMoat$ corresponding to the Bernevig--Hughes--Zhang model for bismuth selenide explored in Ref.\ \cite{maisel2023single}, which would be subject to hexagonal warping due to the $C_{3}$ symmetry of the lattice.
We obtained a moat momentum of the order of $0.5$--$\SI{1}{\nano\meter\tothe{-1}}$, with an energy barrier of a few tens of meV.
While the exciton--exciton interaction is expected to be complicated, we match it to the soft-core potential by using an effective range $\softCoreRange \sim 1$--$\SI{10}{\nano\meter}$, which is of the order of the exciton diameter for the lowest-lying modes.
This gives $\paramSSMoat = {-} (\moatMomentum \softCoreRange)^{2}$ ranging from ${-}0.25$ in the most conservative case to ${-}100$ in the most optimistic one.
The magnitude of $\paramSSInt$ is more difficult to estimate because of the uncertainty regarding the effective exciton--exciton interaction strength.
Nevertheless, our estimate of $\paramSSMoat$ shows that a real exciton system has the chance of sitting in the inhomogeneous (and possibly even supersolid) region of the phase diagram in Fig.\ \ref{fig:2DPhaseDiagram}, even in the case of low densities and weak interactions.

%% file: sec_conclusion.tex

In this article we have studied the phase diagram of a dilute system of excitons with a moat dispersion, where the low-energy degrees of freedom lie on a circle around some high-symmetry point in the Brillouin zone.
We have shown that the naturally long-ranged interactions between excitons give rise to a competition between homogeneous and inhomogeneous Bose--Einstein condensates and a chiral spin liquid of emergent fermions in an effective magnetic field generated by the collection of attached flux quanta.
A self-consistent $T$-matrix renormalization of the bare exciton--exciton interaction has been shown to be crucial for the emergence of periodic and supersolid exciton BECs, whose superfluid response is strongly anisotropic due to the presence of the moat.
We have further explored the phase diagram of supersolid excitonic condensates via a soft-core pseudopotential approximation.
While in the case of a normal parabolic dispersion these periodic states can only exist in the regime of high densities and strong interactions, we have shown that a moat band can help stabilize them even in the regime of low densities and weak interaction strengths.
We have also discussed the effects of band-structure warping, disorder, and finite temperatures on the fate of these correlated states.
Finally, by considering appropriate ranges of values for the physical parameters, we have demonstrated that the BEC and CSL phases compete with each other in an experimentally accessible region of the parameter space.

The work presented in this article serves as a proof of concept and constitutes a minimal model for condensation of moat-band excitons.
Our theory can be expanded to include various effects which may be relevant in a variety of experimental systems.
One example would be to include also the spin of the excitons, which has been ignored here under the assumption that the underlying electronic system is appropriately spin-polarized.
The resulting problem can be mapped to a multi-channel Gross--Pitaevskii equation, which may exhibit richer behavior than the spinless case.
For instance, in the presence of two spin species with a different interspin interaction, a $C_{6}$-symmetric supersolid phase may exhibit frustration akin to that of an antiferromagnetic Heisenberg model on a triangular lattice.
From a quantitative point of view, an immediate (though numerically demanding) way to improve the calculation of the excitonic phase diagram for supersolid phases would be the implementation of a self-consistent $T$-matrix loop in the solution of the full Gross--Pitaevskii equation.
A calculation of this type may also employ a more realistic bare exciton--exciton interaction, for instance based on a combination of the different scattering diagrams derived in Ref.\ \cite{noordman2026variational} which incorporate the excitonic wave functions and which may be generalized to include also band-geometrical, topological, or moir\'e effects.

Our results and speculations on further work indicate that moat-band excitons constitute a promising platform to explore correlated bosonic phases with long-range entanglement or superfluid order even in the case of dilute systems at both strong and weak coupling.
The potential presence of an excitonic moat band can often be inferred from the shape of the particle-hole continuum, which the excitonic energies will tend to follow under the assumption that the binding energies do not depend too strongly on the total exciton momentum.
In Ref.\ \cite{maisel2023single} we asserted that a Bernevig--Hughes--Zhang model with a band inversion for the underlying electrons constitutes a promising candidate system with this feature.
Furthermore, the qualitative conclusions of our work regarding inhomogeneous and supersolid phases of excitons are not strictly limited to a moat: as long as the excitonic energy landscape contains two or more degenerate minima the system will be in the regime susceptible to periodic order.
In view of the general nature of our proposal, its potential as a new pathway to the realization of exotic bosonic phases, and the rapid development of highly tunable electron--hole bilayers and related 2D material platforms, we hope that our work will inspire experimental efforts toward the realization of correlated states of moat-band excitons.

%% file: app_Tmatrix.tex

\section{Equivalence between $T$-matrix theory and short-range correlations}
\label{app:TMatrixCorrelationsEquiv}

In this Appendix we show that renormalizing the effective interaction in such a way that it is softened at the origin is equivalent to incorporating the effect of short-range correlations into the many-body wave function.
Let us start with a more refined ansatz for the latter, namely
\begin{equation}
	\Psi(\{\vec{r}_{i}\}) = \prod_{i} \psi(\vec{r}_{i}) \prod_{i \neq j} f(r_{ij}) ,
\end{equation}
where $r_{ij} \equiv \vert\vec{r}_{i} - \vec{r}_{j}\vert$.
Here, $f(r_{ij})$ is a Jastrow factor which goes to zero when $r_{ij} \rightarrow 0$ and assumes a value of $1$ when $r_{ij}$ exceeds some correlation distance \cite{jastrow1955many}.
Written in this way, the wave function becomes a product of uncorrelated single-particle wave functions when the interparticle distances are large enough, but vanishes when any two particles come too close to each other.
Thus, we are effectively introducing the desired effect of short-range correlations into the variational ansatz.

Let us now calculate the interaction energy associated with this ansatz.
This will have $N(N - 1)/2$ equal terms given by
\begin{equation}
	\int \mathrm{d}^{d} r_{1} \, \mathrm{d}^{d} r_{2} \, |\psi(\vec{r}_{1})|^{2} \hspace{0.2mm} V_{\text{eff}}(\vec{r}_{1}, \vec{r}_{2}) \hspace{0.2mm} |\psi(\vec{r}_{2})|^{2} ,
\end{equation}
where $d$ is the number of spatial dimensions and the effective interaction reads
\begin{equation}
	\begin{split}
		V_{\text{eff}}(\vec{r}_{1}, \vec{r}_{2}) = V(r_{12}) |f(r_{12})|^{2} \int \mathrm{d}^{d} r_{3} \dots \mathrm{d}^{d} r_{N} & \\
		\times \prod_{i > 2} |\psi(\vec{r}_{i})|^{2} |f(r_{1i})|^{2} |f(r_{2i})|^{2} \prod_{i > j > 2} |f(r_{ij})|^{2} & .
	\end{split}
\end{equation}
If the correlation distance is relatively small, then the integrals of the terms in the second line will be close to unity, so their role is to slightly renormalize the interaction strength of $V$.
The most important effect is that $V$ is now accompanied by the square of the Jastrow factor, as seen in the first line.
Now, assuming that $f$ does not vary too fast, its derivative can be neglected in the kinetic term.
In this way, the resulting GPE for the single-particle wave function is the same as that obtained with the simpler ansatz $\Psi(\{\vec{r}_{i}\}) = \prod_{i} \psi(\vec{r}_{i})$, except that the interaction potential $V$ is replaced by $V_{\text{eff}}$, which modifies its behavior at the origin.
Ultimately, the many-body $T$ matrix turns out to be a suitable choice for this effective potential, as it precisely softens the short-range behavior of the bare interaction.

\section{$T$-matrix formulation for a general BEC}
\label{app:GPEFullTMatrix}

In Sec.\ \ref{sec:BECs} we have presented self-consistent equations for the FF, LO, and triangle phases taking into account the full microscopic renormalization of the bare interaction via the $T$ matrix.
In this Appendix we derive similar equations for a general coherent-state ansatz
\begin{equation}
	\vert \Psi \rangle = \frac{1}{\sqrt{N!}} \bigg(\sum_{\vec{g}} c_{\vec{g}} b^{\dagger}_{\vec{g}}\bigg)^{N} \vert 0 \rangle ,
\end{equation}
which contains an arbitrary number of momenta.
In the thermodynamic limit, the energy per particle of this state reads
\begin{equation}
\label{eq:EintFullTGeneralCSAnsatz}
	\begin{split}
			\dimlessMarker{e}[\Psi] = \sum_{\dimlessMarker{\vec{g}}} |c_{\dimlessMarker{\vec{g}}}|^{2} \dimlessMarker{\varepsilon}_{\dimlessMarker{\vec{g}}} + \paramDensityVSIntRange^{2} \paramIntStrengthVSMoatBarrier \sum_{\{\dimlessMarker{\vec{g}}_{i}\}} c^{*}_{\dimlessMarker{\vec{g}}_{1}} c^{*}_{\dimlessMarker{\vec{g}}_{2}} c_{\dimlessMarker{\vec{g}}_{3}} c_{\dimlessMarker{\vec{g}}_{1} + \dimlessMarker{\vec{g}}_{2} - \dimlessMarker{\vec{g}}_{3}} & \\
			\times \, t_{\frac{\dimlessMarker{\vec{g}}_{1} - \dimlessMarker{\vec{g}}_{2}}{2}, \frac{\dimlessMarker{\vec{g}}_{1} + \dimlessMarker{\vec{g}}_{2}}{2} - \dimlessMarker{\vec{g}}_{3}}(\dimlessMarker{\vec{g}}_{1} + \dimlessMarker{\vec{g}}_{2}, {-} 2 \dimlessMarker{\mu}) & .
	\end{split}
\end{equation}
It is easy to check that this expression reproduces the energies of the FF and LO phases of the main text if only the coefficients $c_{\vec{g}}$ and $c_{\pm \vec{g}}$ are allowed to be nonzero, respectively.
From Eq.\ \eqref{eq:EintFullTGeneralCSAnsatz}, together with the constraint $\sum_{\dimlessMarker{\vec{g}}} |c_{\dimlessMarker{\vec{g}}}|^{2} = 1$, one easily derives the self-consistent equation
\begin{equation}
	\begin{split}
		&\dimlessMarker{\mu} c_{\dimlessMarker{\vec{g}}} = \dimlessMarker{\varepsilon}_{\dimlessMarker{\vec{g}}} c_{\dimlessMarker{\vec{g}}} \\
		&+ \, \paramSSInt \sum_{\dimlessMarker{\vec{g}}' \dimlessMarker{\vec{g}}''} t_{\frac{\dimlessMarker{\vec{g}} - \dimlessMarker{\vec{g}}'}{2}, \frac{\dimlessMarker{\vec{g}} + \dimlessMarker{\vec{g}}'}{2} - \dimlessMarker{\vec{g}}''}(\dimlessMarker{\vec{g}} + \dimlessMarker{\vec{g}}', {-} 2 \dimlessMarker{\mu}) c^{*}_{\dimlessMarker{\vec{g}}'} c_{\dimlessMarker{\vec{g}}''} c_{\dimlessMarker{\vec{g}} + \dimlessMarker{\vec{g}}' - \dimlessMarker{\vec{g}}''} ,
	\end{split}
\end{equation}
where we have used the fact that $t_{\vec{k} \vec{k}'}(\vec{K}, E) = t_{{-}\vec{k}, {-}\vec{k}'}(\vec{K}, E)$ when the bare interaction is a function of $|\vec{k} - \vec{k}'|$.
If the coherent state is assumed to correspond to a lattice solution, one can also minimize the energy with respect to the lengths of the reciprocal lattice basis vectors to yield additional self-consistent equations corresponding to the analogs of Eqs.\ \eqref{eq:FFFullTMatrixEqg} and \eqref{eq:selfConsEqgLO} for these higher-order lattices.
Furthermore, in the case of a lattice solution one can connect with the Gross--Pitaevskii treatment and the corresponding Eq.\ \eqref{eq:GPEdimless}, which assumes Bloch's theorem.
This is accomplished by redefining $\vec{g} \rightarrow \vec{g} + \vec{G}$ and $c_{\vec{g} + \vec{G}} \equiv u_{\vec{g} \vec{G}}$, where now $\vec{G}$ is a reciprocal lattice vector and $\vec{g}$ lies in the first Brillouin zone of the emerging reciprocal space, and simultaneously approximating $t_{\vec{k} \vec{k}'}(\vec{K}, E)$ by a local, energy-independent function $t_{\vec{k} - \vec{k}'}$.

In Sec.\ \ref{sec:BECs} we have briefly discussed the possibility of a hybrid zero-momentum--LO phase given by the state $\vert \Psi_{\text{HLO}} \rangle$ of Eq.\ \eqref{eq:PsiHybrid}.
Its free energy follows directly from Eq.\ \eqref{eq:EintFullTGeneralCSAnsatz}.
In this case, the energy depends on the relative phase between the coefficients $c_{\mathbf{0}}$ and $c_{\vec{g}}$.
Without loss of generality we choose $c_{\vec{g}}$ as real, whence it is easy to show that the most optimal choice for the zero-momentum component is $c_{\vec{0}} = \iu |c_{\vec{0}}|$, i.e., $c_{\vec{0}}$ is purely imaginary.
We can then verify if an infinitesimal $c_{\vec{0}}$ is able to develop from the LO phase anywhere on the parameter space of Fig.\ \ref{fig:FFvsLOSmallLambda}.
To lowest order in $c_{\vec{0}}$, the energy difference between the hybrid phase and the LO phase is given by $\dimlessMarker{e}_{\text{HLO}} - \dimlessMarker{e}_{\text{LO}} = \Delta |c_{\vec{0}}|^{2}$, where
\begin{equation}
	\begin{split}
		\Delta &= (\dimlessMoatEnergyBarrier - \dimlessMarker{\mu}) \\
		&+ \paramSSInt \big[ t_{\frac{\dimlessMarker{\vec{g}}}{2}, \frac{\dimlessMarker{\vec{g}}}{2}}(\dimlessMarker{\vec{g}}, {-} 2 \dimlessMarker{\mu}) + t_{\frac{\dimlessMarker{\vec{g}}}{2}, {-}\frac{\dimlessMarker{\vec{g}}}{2}}(\dimlessMarker{\vec{g}}, {-} 2 \dimlessMarker{\mu}) - t_{\dimlessMarker{\vec{g}}, \vec{0}}(\dimlessMarker{\vec{g}}, {-} 2 \dimlessMarker{\mu}) \big] .
	\end{split}
\end{equation}
Numerically we find that $\Delta$ is positive everywhere, thus the LO phase remains pure in the entire region depicted in Figs.\ \ref{fig:FFvsLOSmallLambda} and \ref{FIG_CSLvsFFvsLO_phaseDiagBilayer}.

\section{Numerical treatment of the $T$ matrix}
\label{app:TMatrixNum}

Here we outline our numerical procedure to obtain the full $T$ matrix required for the plots of Fig.\ \ref{fig:FFvsLOSmallLambda}.
To efficiently compute the $T$ matrix numerically, we expand the associated kernel in partial waves as
\begin{equation}
   t_{\dimlessMarker{\vec{k}} \dimlessMarker{\vec{k}}'}(\dimlessMarker{\vec{K}}, \dimlessMarker{E}) = \sum_{mm'} t_{mm'}(\dimlessMarker{k}, \dimlessMarker{k}', \dimlessMarker{K}, \dimlessMarker{E}) \frac{\ec^{\iu m \phi_{\dimlessMarker{\vec{k}}}}}{\sqrt{2 \pi}} \frac{\ec^{{-}\iu m' \phi_{\dimlessMarker{\vec{k}}'}}}{\sqrt{2 \pi}} .
\end{equation}
Here we have set $\phi_{\vec{K}} = 0$ without loss of generality for the rotationally invariant system, but we note that if the effects of warping are explicitly included in the single-particle dispersion, then $\phi_{\vec{K}}$ must be chosen in the direction of one of the minima.
Using this expansion in the dimensionless $T$-matrix equation of Eq.\ \eqref{eq:TMatrixKernelEq} we obtain a set of coupled equations for the coefficients $t_{mm'}$.
These are given by
\begin{align}
\label{eq:TMatrixKernelAngExpCoupledEqs}
      &t_{mm'}(\dimlessMarker{k}, \dimlessMarker{k}', \dimlessMarker{K}, \dimlessMarker{E}) = \delta_{mm'} \sqrt{2 \pi} \hspace{0.3mm} v_{m}(\dimlessMarker{k}, \dimlessMarker{k}') \\
      &+ \paramIntStrengthVSMoatBarrier \int_{0}^{\infty} \frac{\dimlessMarker{p} \hspace{0.3mm} \mathrm{d} \dimlessMarker{p}}{2 \pi} \sum_{l} v_{m}(\dimlessMarker{k}, \dimlessMarker{p}) g^{(2)}_{m - l}(\dimlessMarker{p}, \dimlessMarker{K}, \dimlessMarker{E}) t_{lm'}(\dimlessMarker{p}, \dimlessMarker{k}', \dimlessMarker{K}, \dimlessMarker{E}) , \nonumber
\end{align}
where the angular components of the bare interaction and the two-boson propagator respectively read
\begin{align}
   v_{m}(\dimlessMarker{k}, \dimlessMarker{k}') &= \int_{0}^{2 \pi} \frac{\mathrm{d} \phi}{\sqrt{2 \pi}} \, v_{\dimlessMarker{q}(\phi)} \, \ec^{{-} \iu m \phi} , \\
   g^{(2)}_{m}(\dimlessMarker{p}, \dimlessMarker{K}, \dimlessMarker{E}) &= \int_{0}^{2 \pi} \frac{\mathrm{d} \phi_{\dimlessMarker{\vec{p}}}}{\sqrt{2 \pi}} \, \frac{\ec^{{-} \iu m \phi_{\dimlessMarker{\vec{p}}}}}{\dimlessMarker{E} - \dimlessMarker{\varepsilon}_{\dimlessMarker{K} \hat{\mathbf{x}}/2 + \dimlessMarker{\vec{p}}} - \dimlessMarker{\varepsilon}_{\dimlessMarker{K} \hat{\mathbf{x}}/2 - \dimlessMarker{\vec{p}}}} ,
\end{align}
with $\dimlessMarker{q}(\phi) \equiv \sqrt{\dimlessMarker{k}^{2} + \dimlessMarker{k}'^{2} - 2 \dimlessMarker{k} \dimlessMarker{k}' \cos \phi}$.
We then solve the coupled set of Eqs.\ \eqref{eq:TMatrixKernelAngExpCoupledEqs} by keeping partial waves with $|m| \leq 4$ and discretizing the momentum space with $N_{\dimlessMarker{k}} = 1000$ points while imposing a momentum cutoff $\dimlessMarker{k}_{\text{max}} = \max(10, 3 \dimlessMoatMomentum + 1)$ to correctly resolve the infrared features of both the dispersion and the bare interaction.
We have checked that these parameters ensure convergence in all our calculations.

%% file: app_CSL.tex

The interaction energy per particle of the CSL state in the fully filled $l$th LL can be expressed as \cite{girvin2002quantum,sodemann2013landau}
\begin{subequations}
	\begin{align}
		\label{eq:EintWithgl}
		e^{(l)}_{\text{int}, \text{CSL}} &= \frac{n}{2} \int \mathrm{d}^{2} r \, V(r) g_{l}(r) \\
		\label{eq:EintWithFn}
		&= \frac{n}{2} \int \mathrm{d}^{2} r \, V(r) - \frac{1}{2} \int \frac{\mathrm{d}^{2} q}{(2 \pi)^{2}} \, V_{\vec{q}} \hspace{0.3mm} |\mathcal{F}_{l}(q)|^{2} .
	\end{align}
\end{subequations}
In the first line we have written it in terms of the pair-correlation function $g_{l}(r)$ corresponding to the fully filled LLL, while the expression in the second line involves the form factor
\begin{equation}
	\mathcal{F}_{l}(q) = L_{l}\big(q^{2} \ell_{n}^{2} / 2\big) \hspace{0.3mm} \ec^{{-} q^{2} \ell_{n}^{2} / 4} ,
\end{equation}
with $L_{l}(x)$ the $l$th order Laguerre polynomial and $\ell_{n} \equiv (2 \pi n)^{{-} 1/2}$ the magnetic length.
To correctly compare with the BEC phases, it is important to keep the zero-momentum component of $V$ corresponding to the first term of Eq.\ \eqref{eq:EintWithFn}.
This is because the calculation of the CSL correlation energy is performed with the bare interaction, while that of the BEC phases requires the associated $T$ matrix, whose zero-point energy is in general slightly different.
We note that our expression in Eq.\ \eqref{eq:EintWithgl} involves $g_{l}(r)$ only, not the combination $g_{l}(r) - 1$ like in the usual theory for the electronic QHE.
This is because for excitons we do not need to subtract any neutralizing background, as the system is already charge-neutral and the interaction is integrable.
However, in order to use the LL form-factor formalism, we must recall that the latter is derived by including the $-1$ term, so that expression \eqref{eq:EintWithFn} results in our case.
For the $l = 0$ case one can explicitly show that Eqs.\ \eqref{eq:EintWithgl} and \eqref{eq:EintWithFn} both lead to the same interaction energy by using the analytically known correlation function
\begin{equation}
	g_{l = 0}(r) = 1 - \ec^{{-} r^{2} / 2 \ell_{n}^{2}} .
\end{equation}

To bring Eq.\ \eqref{eq:EintWithFn} to the form of Eq.\ \eqref{eq:CSLEintBilayerXXPot} in the main text we must change to our set of dimensionless variables introduced in Sec.\ \ref{sec:dimlessUnits}, thus we must divide the interaction energy by $b / \intRange^{4}$.
In particular we note that $\int \mathrm{d}^{2} r \, V(r) = 2 \pi \intStrength \intRange^{2} v_{0}$ is the zero-momentum Fourier transform of the interaction, thus the term $(n / 2) \int \mathrm{d}^{2} r \, V(r)$ becomes $(\pi n \intRange^{2}) (\intStrength \intRange^{4} / b) v_{0} = \paramDensityVSIntRange^{2} \paramIntStrengthVSMoatBarrier v_{0}$.
Assuming without loss of generality that the interaction kernel satisfies $v_{0} = 1$ we obtain the first term of Eq.\ \eqref{eq:CSLEintBilayerXXPot}.
The second term is obtained in a similar fashion after performing the change of variables $u = q^{2} \ell^{2}_{n} / 2$ inside the second integral of Eq.\ \eqref{eq:EintWithFn}, and we obtain
\begin{equation}
\label{eq:EintCSLArbitraryKernel}
	\dimlessMarker{e}^{(l)}_{\text{int}, \text{CSL}} = \paramDensityVSIntRange^{2} \paramIntStrengthVSMoatBarrier \hspace{0.3mm} \bigg(1 - \int_{0}^{\infty} \mathrm{d} u \, v_{2 \paramDensityVSIntRange \sqrt{u}} \hspace{0.3mm} [L_{l}(u)]^{2} \hspace{0.3mm} \ec^{{-} u} \bigg)
\end{equation}
for an arbitrary long-range potential.

Let us now consider the interaction energy of Eq.\ \eqref{eq:EintCSLArbitraryKernel} in the limit of small $\paramDensityVSIntRange$, in which case the order $l$ of the LL becomes very high according to $\paramDensityVSIntRange \sim \dimlessMoatMomentum / \sqrt{2(2l + 1)}$.
While it is tempting to expand $v_{2 \paramDensityVSIntRange \sqrt{u}} = 1 + \mathcal{O}(\paramDensityVSIntRange)$ and use the identity $\int_{0}^{\infty} \mathrm{d} u \, [L_{l}(u)]^{2} \hspace{0.3mm} \ec^{{-}u} = 1$ to conclude that the interaction energy scales as $\paramDensityVSIntRange^{3} \paramIntStrengthVSMoatBarrier$, this reasoning is incorrect.
Indeed, we must be careful because the remaining integral still depends on $l \sim \paramDensityVSIntRange^{-2}$ via the Laguerre polynomial.
To extract the correct scaling of the correlation energy with $l$ we use the asymptotic Plancherel--Rotach expansion for the Laguerre polynomials, which is valid for large $l$.
Putting $u = (4l + 2) \cos^{2} \varphi$ with $0 < \varphi < \pi / 2$ it holds that \cite{szego1975orthogonal}
\begin{equation}
	\ec^{{-}u/2} L_{l}(u) = \frac{(-1)^{l}}{(\pi \sin \varphi)^{1/2} (lu)^{1/4}} \big[{\sin \phi_{l}(\varphi)} + \mathcal{O}(1 / \sqrt{lu}) \big] ,
\end{equation}
where
\begin{equation}
	\phi_{l}(\varphi) = (l + 1/2)(\sin 2 \varphi - 2 \varphi) + 3 \pi / 4 .
\end{equation}
Using this asymptotic form in Eq.\ \eqref{eq:EintCSLArbitraryKernel} and changing variables from $u$ to $\varphi$ yields the interaction energy for large $l$ as
\begin{equation}
\label{eq:EintCSLLargelExp}
	\begin{split}
		\dimlessMarker{e}^{(l \gg 1)}_{\text{int}, \text{CSL}} \approx \paramDensityVSIntRange^{2} \paramIntStrengthVSMoatBarrier \hspace{0.3mm} \bigg(1 - \frac{4}{\pi} \int_{0}^{\pi/2} \mathrm{d} \varphi \, v_{2 \dimlessMoatMomentum \cos \varphi} \sin^{2} \phi_{l}(\varphi) \bigg) ,
	\end{split}
\end{equation}
where we have used that $\paramDensityVSIntRange \approx \dimlessMoatMomentum / \sqrt{4l}$ for $l \gg 1$.
This analysis shows that the integral of Eq.\ \eqref{eq:EintCSLArbitraryKernel} does not scale with $\paramDensityVSIntRange$, as we have $\mathrm{d} u / \sqrt{l u} \sim l / \sqrt{l^2} \sim \mathcal{O}(1)$ upon performing the asymptotic expansion.
For completeness we also give the dimensionful version of the correlation energy at large $l$, which reads
\begin{equation}
\label{eq:limitEintCSLlargelDimful}
	e^{(l \gg 1)}_{\text{int}, \text{CSL}} \approx \frac{n V_{0}}{2} \bigg(1 - \frac{4}{\pi} \int_{0}^{\pi/2} \mathrm{d} \varphi \, \frac{V_{2 \moatMomentum \cos \varphi}}{V_{0}} \sin^{2} \phi_{l}(\varphi)\bigg) ,
\end{equation}
with $V_{0}$ the zero-momentum Fourier transform of $V(r)$.
Equations\ \eqref{eq:EintCSLLargelExp} and \eqref{eq:limitEintCSLlargelDimful} approach a $\moatMomentum$-dependent limiting value for $l \rightarrow \infty$, which is obtained by observing that $\lim_{l \rightarrow \infty} \sin^{2} \phi_{l}(\varphi) = \frac{1}{2}$ in the distributional sense.

Equation \eqref{eq:EintCSLLargelExp} is in excellent agreement with our bilayer numerical calculations in Fig.\ \ref{FIG_CSLvsFFvsLO_phaseDiagBilayer} when $\paramDensityVSIntRange$ is small.
Furthermore, it shows a crucial point, namely that the interaction energy of the CSL scales as $\paramDensityVSIntRange^{2} \paramIntStrengthVSMoatBarrier$ at small $\paramDensityVSIntRange$.
In other words, the correlation energy for an arbitrary long-range interaction is linear in the density, which is precisely the same scaling as that of the BEC phases according to Eqs.\ \eqref{eq:intEnergyFFDimlessTMatrix} and \eqref{eq:intEnergyLOFullTMatrix}.
The origin of this scaling is the long-range nature of the interaction potential.
Indeed, for a contact interaction with $v_{\dimlessMarker{q}} = 1$ for all momenta the integral in Eq.\ \eqref{eq:EintCSLArbitraryKernel} is $1$ and the right-hand side vanishes as a result, consistent with the idea that the fermionized bosons are noninteracting due to the Pauli exclusion principle.
Thus, as stated in the main text, we conclude that the long-range interaction introduces a rich competition between the CSL and BEC phases even at low densities.

%% file: main.bbl
\begin{thebibliography}{132}%
\makeatletter
\providecommand \@ifxundefined [1]{%
 \@ifx{#1\undefined}
}%
\providecommand \@ifnum [1]{%
 \ifnum #1\expandafter \@firstoftwo
 \else \expandafter \@secondoftwo
 \fi
}%
\providecommand \@ifx [1]{%
 \ifx #1\expandafter \@firstoftwo
 \else \expandafter \@secondoftwo
 \fi
}%
\providecommand \natexlab [1]{#1}%
\providecommand \enquote  [1]{``#1''}%
\providecommand \bibnamefont  [1]{#1}%
\providecommand \bibfnamefont [1]{#1}%
\providecommand \citenamefont [1]{#1}%
\providecommand \href@noop [0]{\@secondoftwo}%
\providecommand \href [0]{\begingroup \@sanitize@url \@href}%
\providecommand \@href[1]{\@@startlink{#1}\@@href}%
\providecommand \@@href[1]{\endgroup#1\@@endlink}%
\providecommand \@sanitize@url [0]{\catcode `\\12\catcode `\$12\catcode
  `\&12\catcode `\#12\catcode `\^12\catcode `\_12\catcode `\%12\relax}%
\providecommand \@@startlink[1]{}%
\providecommand \@@endlink[0]{}%
\providecommand \url  [0]{\begingroup\@sanitize@url \@url }%
\providecommand \@url [1]{\endgroup\@href {#1}{\urlprefix }}%
\providecommand \urlprefix  [0]{URL }%
\providecommand \Eprint [0]{\href }%
\providecommand \doibase [0]{https://doi.org/}%
\providecommand \selectlanguage [0]{\@gobble}%
\providecommand \bibinfo  [0]{\@secondoftwo}%
\providecommand \bibfield  [0]{\@secondoftwo}%
\providecommand \translation [1]{[#1]}%
\providecommand \BibitemOpen [0]{}%
\providecommand \bibitemStop [0]{}%
\providecommand \bibitemNoStop [0]{.\EOS\space}%
\providecommand \EOS [0]{\spacefactor3000\relax}%
\providecommand \BibitemShut  [1]{\csname bibitem#1\endcsname}%
\let\auto@bib@innerbib\@empty
\bibitem [{\citenamefont {Blatt}\ \emph {et~al.}(1962)\citenamefont {Blatt},
  \citenamefont {B{\"o}er},\ and\ \citenamefont {Brandt}}]{blatt1962bose}%
  \BibitemOpen
  \bibfield  {author} {\bibinfo {author} {\bibfnamefont {J.~M.}\ \bibnamefont
  {Blatt}}, \bibinfo {author} {\bibfnamefont {K.}~\bibnamefont {B{\"o}er}},\
  and\ \bibinfo {author} {\bibfnamefont {W.}~\bibnamefont {Brandt}},\
  }\bibfield  {title} {\bibinfo {title} {Bose-{E}instein condensation of
  excitons},\ }\href {https://doi.org/10.1103/PhysRev.126.1691} {\bibfield
  {journal} {\bibinfo  {journal} {Physical Review}\ }\textbf {\bibinfo {volume}
  {126}},\ \bibinfo {pages} {1691} (\bibinfo {year} {1962})}\BibitemShut
  {NoStop}%
\bibitem [{\citenamefont {Keldysh}\ and\ \citenamefont
  {Kozlov}(1968)}]{keldysh1968collective}%
  \BibitemOpen
  \bibfield  {author} {\bibinfo {author} {\bibfnamefont {L.~V.}\ \bibnamefont
  {Keldysh}}\ and\ \bibinfo {author} {\bibfnamefont {A.~N.}\ \bibnamefont
  {Kozlov}},\ }\bibfield  {title} {\bibinfo {title} {Collective properties of
  excitons in semiconductors},\ }\href
  {https://doi.org/10.1142/9789811279461_0011} {\bibfield  {journal} {\bibinfo
  {journal} {Sov. Phys. JETP}\ }\textbf {\bibinfo {volume} {27}},\ \bibinfo
  {pages} {521} (\bibinfo {year} {1968})}\BibitemShut {NoStop}%
\bibitem [{\citenamefont {Keldysh}\ and\ \citenamefont
  {Kopaev}(1965)}]{keldysh1965possible}%
  \BibitemOpen
  \bibfield  {author} {\bibinfo {author} {\bibfnamefont {L.~V.}\ \bibnamefont
  {Keldysh}}\ and\ \bibinfo {author} {\bibfnamefont {Y.~V.}\ \bibnamefont
  {Kopaev}},\ }\bibfield  {title} {\bibinfo {title} {Possible instability of
  the semimetallic state toward {Coulomb} interaction},\ }\href
  {https://doi.org/10.1142/9789811279461_0006} {\bibfield  {journal} {\bibinfo
  {journal} {Sov. Phys. Solid State}\ }\textbf {\bibinfo {volume} {6}},\
  \bibinfo {pages} {2219} (\bibinfo {year} {1965})}\BibitemShut {NoStop}%
\bibitem [{\citenamefont {Jerome}\ \emph {et~al.}(1967)\citenamefont {Jerome},
  \citenamefont {Rice},\ and\ \citenamefont {Kohn}}]{jerome1967excitonic}%
  \BibitemOpen
  \bibfield  {author} {\bibinfo {author} {\bibfnamefont {D.}~\bibnamefont
  {Jerome}}, \bibinfo {author} {\bibfnamefont {T.}~\bibnamefont {Rice}},\ and\
  \bibinfo {author} {\bibfnamefont {W.}~\bibnamefont {Kohn}},\ }\bibfield
  {title} {\bibinfo {title} {Excitonic insulator},\ }\href
  {https://doi.org/10.1103/PhysRev.158.462} {\bibfield  {journal} {\bibinfo
  {journal} {Physical Review}\ }\textbf {\bibinfo {volume} {158}},\ \bibinfo
  {pages} {462} (\bibinfo {year} {1967})}\BibitemShut {NoStop}%
\bibitem [{\citenamefont {Halperin}\ and\ \citenamefont
  {Rice}(1968)}]{halperin1968possible}%
  \BibitemOpen
  \bibfield  {author} {\bibinfo {author} {\bibfnamefont {B.}~\bibnamefont
  {Halperin}}\ and\ \bibinfo {author} {\bibfnamefont {T.}~\bibnamefont
  {Rice}},\ }\bibfield  {title} {\bibinfo {title} {Possible anomalies at a
  semimetal-semiconductor transistion},\ }\href
  {https://doi.org/10.1103/RevModPhys.40.755} {\bibfield  {journal} {\bibinfo
  {journal} {Reviews of Modern Physics}\ }\textbf {\bibinfo {volume} {40}},\
  \bibinfo {pages} {755} (\bibinfo {year} {1968})}\BibitemShut {NoStop}%
\bibitem [{\citenamefont {Butov}\ \emph {et~al.}(1994)\citenamefont {Butov},
  \citenamefont {Zrenner}, \citenamefont {Abstreiter}, \citenamefont
  {B{\"o}hm},\ and\ \citenamefont {Weimann}}]{butov1994condensation}%
  \BibitemOpen
  \bibfield  {author} {\bibinfo {author} {\bibfnamefont {L.~V.}\ \bibnamefont
  {Butov}}, \bibinfo {author} {\bibfnamefont {A.}~\bibnamefont {Zrenner}},
  \bibinfo {author} {\bibfnamefont {G.}~\bibnamefont {Abstreiter}}, \bibinfo
  {author} {\bibfnamefont {G.}~\bibnamefont {B{\"o}hm}},\ and\ \bibinfo
  {author} {\bibfnamefont {G.}~\bibnamefont {Weimann}},\ }\bibfield  {title}
  {\bibinfo {title} {Condensation of indirect excitons in coupled
  \ce{AlAs}/\ce{GaAs} quantum wells},\ }\href
  {https://doi.org/10.1103/PhysRevLett.73.304} {\bibfield  {journal} {\bibinfo
  {journal} {Physical Review Letters}\ }\textbf {\bibinfo {volume} {73}},\
  \bibinfo {pages} {304} (\bibinfo {year} {1994})}\BibitemShut {NoStop}%
\bibitem [{\citenamefont {Butov}\ \emph {et~al.}(2001)\citenamefont {Butov},
  \citenamefont {Ivanov}, \citenamefont {Imamoglu}, \citenamefont {Littlewood},
  \citenamefont {Shashkin}, \citenamefont {Dolgopolov}, \citenamefont
  {Campman},\ and\ \citenamefont {Gossard}}]{butov2001stimulated}%
  \BibitemOpen
  \bibfield  {author} {\bibinfo {author} {\bibfnamefont {L.~V.}\ \bibnamefont
  {Butov}}, \bibinfo {author} {\bibfnamefont {A.~L.}\ \bibnamefont {Ivanov}},
  \bibinfo {author} {\bibfnamefont {A.}~\bibnamefont {Imamoglu}}, \bibinfo
  {author} {\bibfnamefont {P.~B.}\ \bibnamefont {Littlewood}}, \bibinfo
  {author} {\bibfnamefont {A.~A.}\ \bibnamefont {Shashkin}}, \bibinfo {author}
  {\bibfnamefont {V.~T.}\ \bibnamefont {Dolgopolov}}, \bibinfo {author}
  {\bibfnamefont {K.~L.}\ \bibnamefont {Campman}},\ and\ \bibinfo {author}
  {\bibfnamefont {A.~C.}\ \bibnamefont {Gossard}},\ }\bibfield  {title}
  {\bibinfo {title} {Stimulated scattering of indirect excitons in coupled
  quantum wells: {S}ignature of a degenerate {B}ose-gas of excitons},\ }\href
  {https://doi.org/10.1103/PhysRevLett.86.5608} {\bibfield  {journal} {\bibinfo
   {journal} {Physical Review Letters}\ }\textbf {\bibinfo {volume} {86}},\
  \bibinfo {pages} {5608} (\bibinfo {year} {2001})}\BibitemShut {NoStop}%
\bibitem [{\citenamefont {Butov}\ \emph
  {et~al.}(2002{\natexlab{a}})\citenamefont {Butov}, \citenamefont {Gossard},\
  and\ \citenamefont {Chemla}}]{butov2002macroscopically}%
  \BibitemOpen
  \bibfield  {author} {\bibinfo {author} {\bibfnamefont {L.~V.}\ \bibnamefont
  {Butov}}, \bibinfo {author} {\bibfnamefont {A.~C.}\ \bibnamefont {Gossard}},\
  and\ \bibinfo {author} {\bibfnamefont {D.~S.}\ \bibnamefont {Chemla}},\
  }\bibfield  {title} {\bibinfo {title} {Macroscopically ordered state in an
  exciton system},\ }\href {https://doi.org/10.1038/nature00943} {\bibfield
  {journal} {\bibinfo  {journal} {Nature}\ }\textbf {\bibinfo {volume} {418}},\
  \bibinfo {pages} {751} (\bibinfo {year} {2002}{\natexlab{a}})}\BibitemShut
  {NoStop}%
\bibitem [{\citenamefont {Butov}\ \emph
  {et~al.}(2002{\natexlab{b}})\citenamefont {Butov}, \citenamefont {Lai},
  \citenamefont {Ivanov}, \citenamefont {Gossard},\ and\ \citenamefont
  {Chemla}}]{butov2002towards}%
  \BibitemOpen
  \bibfield  {author} {\bibinfo {author} {\bibfnamefont {L.~V.}\ \bibnamefont
  {Butov}}, \bibinfo {author} {\bibfnamefont {C.~W.}\ \bibnamefont {Lai}},
  \bibinfo {author} {\bibfnamefont {A.~L.}\ \bibnamefont {Ivanov}}, \bibinfo
  {author} {\bibfnamefont {A.~C.}\ \bibnamefont {Gossard}},\ and\ \bibinfo
  {author} {\bibfnamefont {D.~S.}\ \bibnamefont {Chemla}},\ }\bibfield  {title}
  {\bibinfo {title} {Towards {B}ose--{E}instein condensation of excitons in
  potential traps},\ }\href {https://doi.org/10.1038/417047a} {\bibfield
  {journal} {\bibinfo  {journal} {Nature}\ }\textbf {\bibinfo {volume} {417}},\
  \bibinfo {pages} {47} (\bibinfo {year} {2002}{\natexlab{b}})}\BibitemShut
  {NoStop}%
\bibitem [{\citenamefont {Eisenstein}\ and\ \citenamefont
  {MacDonald}(2004)}]{eisenstein2004bose}%
  \BibitemOpen
  \bibfield  {author} {\bibinfo {author} {\bibfnamefont {J.}~\bibnamefont
  {Eisenstein}}\ and\ \bibinfo {author} {\bibfnamefont {A.~H.}\ \bibnamefont
  {MacDonald}},\ }\bibfield  {title} {\bibinfo {title} {Bose--{E}instein
  condensation of excitons in bilayer electron systems},\ }\href
  {https://doi.org/10.1038/nature03081} {\bibfield  {journal} {\bibinfo
  {journal} {Nature}\ }\textbf {\bibinfo {volume} {432}},\ \bibinfo {pages}
  {691} (\bibinfo {year} {2004})}\BibitemShut {NoStop}%
\bibitem [{\citenamefont {Eisenstein}(2014)}]{eisenstein2014exciton}%
  \BibitemOpen
  \bibfield  {author} {\bibinfo {author} {\bibfnamefont {J.}~\bibnamefont
  {Eisenstein}},\ }\bibfield  {title} {\bibinfo {title} {Exciton condensation
  in bilayer quantum {H}all systems},\ }\href
  {https://doi.org/10.1146/annurev-conmatphys-031113-133832} {\bibfield
  {journal} {\bibinfo  {journal} {Annu. Rev. Condens. Matter Phys.}\ }\textbf
  {\bibinfo {volume} {5}},\ \bibinfo {pages} {159} (\bibinfo {year}
  {2014})}\BibitemShut {NoStop}%
\bibitem [{\citenamefont {Brem}\ \emph {et~al.}(2020)\citenamefont {Brem},
  \citenamefont {Linder{\"a}lv}, \citenamefont {Erhart},\ and\ \citenamefont
  {Malic}}]{brem2020tunable}%
  \BibitemOpen
  \bibfield  {author} {\bibinfo {author} {\bibfnamefont {S.}~\bibnamefont
  {Brem}}, \bibinfo {author} {\bibfnamefont {C.}~\bibnamefont {Linder{\"a}lv}},
  \bibinfo {author} {\bibfnamefont {P.}~\bibnamefont {Erhart}},\ and\ \bibinfo
  {author} {\bibfnamefont {E.}~\bibnamefont {Malic}},\ }\bibfield  {title}
  {\bibinfo {title} {Tunable phases of moir{\'e} excitons in van der {W}aals
  heterostructures},\ }\href {https://doi.org/10.1021/acs.nanolett.0c03019}
  {\bibfield  {journal} {\bibinfo  {journal} {Nano Letters}\ }\textbf {\bibinfo
  {volume} {20}},\ \bibinfo {pages} {8534} (\bibinfo {year}
  {2020})}\BibitemShut {NoStop}%
\bibitem [{\citenamefont {Wilson}\ \emph {et~al.}(2021)\citenamefont {Wilson},
  \citenamefont {Yao}, \citenamefont {Shan},\ and\ \citenamefont
  {Xu}}]{wilson2021excitons}%
  \BibitemOpen
  \bibfield  {author} {\bibinfo {author} {\bibfnamefont {N.~P.}\ \bibnamefont
  {Wilson}}, \bibinfo {author} {\bibfnamefont {W.}~\bibnamefont {Yao}},
  \bibinfo {author} {\bibfnamefont {J.}~\bibnamefont {Shan}},\ and\ \bibinfo
  {author} {\bibfnamefont {X.}~\bibnamefont {Xu}},\ }\bibfield  {title}
  {\bibinfo {title} {Excitons and emergent quantum phenomena in stacked {2D}
  semiconductors},\ }\href {https://doi.org/10.1038/s41586-021-03979-1}
  {\bibfield  {journal} {\bibinfo  {journal} {Nature}\ }\textbf {\bibinfo
  {volume} {599}},\ \bibinfo {pages} {383} (\bibinfo {year}
  {2021})}\BibitemShut {NoStop}%
\bibitem [{\citenamefont {Regan}\ \emph {et~al.}(2022)\citenamefont {Regan},
  \citenamefont {Wang}, \citenamefont {Paik}, \citenamefont {Zeng},
  \citenamefont {Zhang}, \citenamefont {Zhu}, \citenamefont {MacDonald},
  \citenamefont {Deng},\ and\ \citenamefont {Wang}}]{regan2022emerging}%
  \BibitemOpen
  \bibfield  {author} {\bibinfo {author} {\bibfnamefont {E.~C.}\ \bibnamefont
  {Regan}}, \bibinfo {author} {\bibfnamefont {D.}~\bibnamefont {Wang}},
  \bibinfo {author} {\bibfnamefont {E.~Y.}\ \bibnamefont {Paik}}, \bibinfo
  {author} {\bibfnamefont {Y.}~\bibnamefont {Zeng}}, \bibinfo {author}
  {\bibfnamefont {L.}~\bibnamefont {Zhang}}, \bibinfo {author} {\bibfnamefont
  {J.}~\bibnamefont {Zhu}}, \bibinfo {author} {\bibfnamefont {A.~H.}\
  \bibnamefont {MacDonald}}, \bibinfo {author} {\bibfnamefont {H.}~\bibnamefont
  {Deng}},\ and\ \bibinfo {author} {\bibfnamefont {F.}~\bibnamefont {Wang}},\
  }\bibfield  {title} {\bibinfo {title} {Emerging exciton physics in transition
  metal dichalcogenide heterobilayers},\ }\href
  {https://doi.org/10.1038/s41578-022-00440-1} {\bibfield  {journal} {\bibinfo
  {journal} {Nature Reviews Materials}\ }\textbf {\bibinfo {volume} {7}},\
  \bibinfo {pages} {778} (\bibinfo {year} {2022})}\BibitemShut {NoStop}%
\bibitem [{\citenamefont {Moon}\ \emph {et~al.}(2025)\citenamefont {Moon},
  \citenamefont {Mondal}, \citenamefont {Efimkin},\ and\ \citenamefont
  {Lee}}]{moon2025exciton}%
  \BibitemOpen
  \bibfield  {author} {\bibinfo {author} {\bibfnamefont {B.~H.}\ \bibnamefont
  {Moon}}, \bibinfo {author} {\bibfnamefont {A.}~\bibnamefont {Mondal}},
  \bibinfo {author} {\bibfnamefont {D.~K.}\ \bibnamefont {Efimkin}},\ and\
  \bibinfo {author} {\bibfnamefont {Y.~H.}\ \bibnamefont {Lee}},\ }\bibfield
  {title} {\bibinfo {title} {Exciton condensate in van der {W}aals layered
  materials},\ }\href {https://doi.org/10.1038/s42254-025-00834-4} {\bibfield
  {journal} {\bibinfo  {journal} {Nature Reviews Physics}\ }\textbf {\bibinfo
  {volume} {7}},\ \bibinfo {pages} {388} (\bibinfo {year} {2025})}\BibitemShut
  {NoStop}%
\bibitem [{\citenamefont {Qi}\ \emph {et~al.}(2025)\citenamefont {Qi},
  \citenamefont {Li}, \citenamefont {Zhang}, \citenamefont {Nie}, \citenamefont
  {Zou}, \citenamefont {Cui}, \citenamefont {Kim}, \citenamefont {Sanborn},
  \citenamefont {Chen}, \citenamefont {Xie} \emph
  {et~al.}}]{qi2025competition}%
  \BibitemOpen
  \bibfield  {author} {\bibinfo {author} {\bibfnamefont {R.}~\bibnamefont
  {Qi}}, \bibinfo {author} {\bibfnamefont {Q.}~\bibnamefont {Li}}, \bibinfo
  {author} {\bibfnamefont {Z.}~\bibnamefont {Zhang}}, \bibinfo {author}
  {\bibfnamefont {J.}~\bibnamefont {Nie}}, \bibinfo {author} {\bibfnamefont
  {B.}~\bibnamefont {Zou}}, \bibinfo {author} {\bibfnamefont {Z.}~\bibnamefont
  {Cui}}, \bibinfo {author} {\bibfnamefont {H.}~\bibnamefont {Kim}}, \bibinfo
  {author} {\bibfnamefont {C.}~\bibnamefont {Sanborn}}, \bibinfo {author}
  {\bibfnamefont {S.}~\bibnamefont {Chen}}, \bibinfo {author} {\bibfnamefont
  {J.}~\bibnamefont {Xie}}, \emph {et~al.},\ }\bibfield  {title} {\bibinfo
  {title} {Competition between excitonic insulators and quantum {H}all states
  in correlated electron--hole bilayers},\ }\href
  {https://doi.org/10.1038/s41563-025-02316-5} {\bibfield  {journal} {\bibinfo
  {journal} {Nature Materials}\ ,\ \bibinfo {pages} {1}} (\bibinfo {year}
  {2025})}\BibitemShut {NoStop}%
\bibitem [{\citenamefont {Lozovik}\ and\ \citenamefont
  {Yudson}(1976)}]{lozovik1976new}%
  \BibitemOpen
  \bibfield  {author} {\bibinfo {author} {\bibfnamefont {Y.~E.}\ \bibnamefont
  {Lozovik}}\ and\ \bibinfo {author} {\bibfnamefont {V.}~\bibnamefont
  {Yudson}},\ }\bibfield  {title} {\bibinfo {title} {A new mechanism for
  superconductivity: pairing between spatially separated electrons and holes},\
  }\href@noop {} {\bibfield  {journal} {\bibinfo  {journal} {Zh. Eksp. Teor.
  Fiz}\ }\textbf {\bibinfo {volume} {71}},\ \bibinfo {pages} {738} (\bibinfo
  {year} {1976})}\BibitemShut {NoStop}%
\bibitem [{\citenamefont {Fogler}\ \emph {et~al.}(2014)\citenamefont {Fogler},
  \citenamefont {Butov},\ and\ \citenamefont {Novoselov}}]{fogler2014high}%
  \BibitemOpen
  \bibfield  {author} {\bibinfo {author} {\bibfnamefont {M.}~\bibnamefont
  {Fogler}}, \bibinfo {author} {\bibfnamefont {L.}~\bibnamefont {Butov}},\ and\
  \bibinfo {author} {\bibfnamefont {K.}~\bibnamefont {Novoselov}},\ }\bibfield
  {title} {\bibinfo {title} {High-temperature superfluidity with indirect
  excitons in van der {W}aals heterostructures},\ }\href
  {https://doi.org/10.1038/ncomms5555} {\bibfield  {journal} {\bibinfo
  {journal} {Nature Communications}\ }\textbf {\bibinfo {volume} {5}},\
  \bibinfo {pages} {4555} (\bibinfo {year} {2014})}\BibitemShut {NoStop}%
\bibitem [{\citenamefont {Li}\ \emph {et~al.}(2017)\citenamefont {Li},
  \citenamefont {Taniguchi}, \citenamefont {Watanabe}, \citenamefont {Hone},\
  and\ \citenamefont {Dean}}]{li2017excitonic}%
  \BibitemOpen
  \bibfield  {author} {\bibinfo {author} {\bibfnamefont {J.}~\bibnamefont
  {Li}}, \bibinfo {author} {\bibfnamefont {T.}~\bibnamefont {Taniguchi}},
  \bibinfo {author} {\bibfnamefont {K.}~\bibnamefont {Watanabe}}, \bibinfo
  {author} {\bibfnamefont {J.}~\bibnamefont {Hone}},\ and\ \bibinfo {author}
  {\bibfnamefont {C.}~\bibnamefont {Dean}},\ }\bibfield  {title} {\bibinfo
  {title} {Excitonic superfluid phase in double bilayer graphene},\ }\href
  {https://doi.org/10.1038/nphys4140} {\bibfield  {journal} {\bibinfo
  {journal} {Nature Physics}\ }\textbf {\bibinfo {volume} {13}},\ \bibinfo
  {pages} {751} (\bibinfo {year} {2017})}\BibitemShut {NoStop}%
\bibitem [{\citenamefont {Berman}\ and\ \citenamefont
  {Kezerashvili}(2016)}]{berman2016high}%
  \BibitemOpen
  \bibfield  {author} {\bibinfo {author} {\bibfnamefont {O.~L.}\ \bibnamefont
  {Berman}}\ and\ \bibinfo {author} {\bibfnamefont {R.~Y.}\ \bibnamefont
  {Kezerashvili}},\ }\bibfield  {title} {\bibinfo {title} {High-temperature
  superfluidity of the two-component {B}ose gas in a transition metal
  dichalcogenide bilayer},\ }\href {https://doi.org/10.1103/PhysRevB.93.245410}
  {\bibfield  {journal} {\bibinfo  {journal} {Physical Review B}\ }\textbf
  {\bibinfo {volume} {93}},\ \bibinfo {pages} {245410} (\bibinfo {year}
  {2016})}\BibitemShut {NoStop}%
\bibitem [{\citenamefont {Berman}\ \emph {et~al.}(2017)\citenamefont {Berman},
  \citenamefont {Gumbs},\ and\ \citenamefont {Kezerashvili}}]{berman2017bose}%
  \BibitemOpen
  \bibfield  {author} {\bibinfo {author} {\bibfnamefont {O.~L.}\ \bibnamefont
  {Berman}}, \bibinfo {author} {\bibfnamefont {G.}~\bibnamefont {Gumbs}},\ and\
  \bibinfo {author} {\bibfnamefont {R.~Y.}\ \bibnamefont {Kezerashvili}},\
  }\bibfield  {title} {\bibinfo {title} {{B}ose-{E}instein condensation and
  superfluidity of dipolar excitons in a phosphorene double layer},\ }\href
  {https://doi.org/10.1103/PhysRevB.96.014505} {\bibfield  {journal} {\bibinfo
  {journal} {Physical Review B}\ }\textbf {\bibinfo {volume} {96}},\ \bibinfo
  {pages} {014505} (\bibinfo {year} {2017})}\BibitemShut {NoStop}%
\bibitem [{\citenamefont {Berman}\ and\ \citenamefont
  {Kezerashvili}(2017)}]{berman2017superfluidity}%
  \BibitemOpen
  \bibfield  {author} {\bibinfo {author} {\bibfnamefont {O.~L.}\ \bibnamefont
  {Berman}}\ and\ \bibinfo {author} {\bibfnamefont {R.~Y.}\ \bibnamefont
  {Kezerashvili}},\ }\bibfield  {title} {\bibinfo {title} {Superfluidity of
  dipolar excitons in a transition metal dichalcogenide double layer},\ }\href
  {https://doi.org/10.1103/PhysRevB.96.094502} {\bibfield  {journal} {\bibinfo
  {journal} {Physical Review B}\ }\textbf {\bibinfo {volume} {96}},\ \bibinfo
  {pages} {094502} (\bibinfo {year} {2017})}\BibitemShut {NoStop}%
\bibitem [{\citenamefont {Julku}(2022)}]{julku2022nonlocal}%
  \BibitemOpen
  \bibfield  {author} {\bibinfo {author} {\bibfnamefont {A.}~\bibnamefont
  {Julku}},\ }\bibfield  {title} {\bibinfo {title} {Nonlocal interactions and
  supersolidity of moir{\'e} excitons},\ }\href
  {https://doi.org/10.1103/PhysRevB.106.035406} {\bibfield  {journal} {\bibinfo
   {journal} {Physical Review B}\ }\textbf {\bibinfo {volume} {106}},\ \bibinfo
  {pages} {035406} (\bibinfo {year} {2022})}\BibitemShut {NoStop}%
\bibitem [{\citenamefont {Huang}\ \emph {et~al.}(2023)\citenamefont {Huang},
  \citenamefont {Chou}, \citenamefont {Baldwin}, \citenamefont {Wu},\ and\
  \citenamefont {Hafezi}}]{huang2023mott}%
  \BibitemOpen
  \bibfield  {author} {\bibinfo {author} {\bibfnamefont {T.-S.}\ \bibnamefont
  {Huang}}, \bibinfo {author} {\bibfnamefont {Y.-Z.}\ \bibnamefont {Chou}},
  \bibinfo {author} {\bibfnamefont {C.~L.}\ \bibnamefont {Baldwin}}, \bibinfo
  {author} {\bibfnamefont {F.}~\bibnamefont {Wu}},\ and\ \bibinfo {author}
  {\bibfnamefont {M.}~\bibnamefont {Hafezi}},\ }\bibfield  {title} {\bibinfo
  {title} {Mott-moir{\'e} excitons},\ }\href
  {https://doi.org/10.1103/PhysRevB.107.195151} {\bibfield  {journal} {\bibinfo
   {journal} {Physical Review B}\ }\textbf {\bibinfo {volume} {107}},\ \bibinfo
  {pages} {195151} (\bibinfo {year} {2023})}\BibitemShut {NoStop}%
\bibitem [{\citenamefont {Dai}\ and\ \citenamefont {Fu}(2024)}]{dai2024strong}%
  \BibitemOpen
  \bibfield  {author} {\bibinfo {author} {\bibfnamefont {D.~D.}\ \bibnamefont
  {Dai}}\ and\ \bibinfo {author} {\bibfnamefont {L.}~\bibnamefont {Fu}},\
  }\bibfield  {title} {\bibinfo {title} {Strong-coupling phases of trions and
  excitons in electron-hole bilayers at commensurate densities},\ }\href
  {https://doi.org/10.1103/PhysRevLett.132.196202} {\bibfield  {journal}
  {\bibinfo  {journal} {Physical Review Letters}\ }\textbf {\bibinfo {volume}
  {132}},\ \bibinfo {pages} {196202} (\bibinfo {year} {2024})}\BibitemShut
  {NoStop}%
\bibitem [{\citenamefont {Froese}\ \emph {et~al.}(2025)\citenamefont {Froese},
  \citenamefont {Neupert},\ and\ \citenamefont
  {Wagner}}]{froese2025topological}%
  \BibitemOpen
  \bibfield  {author} {\bibinfo {author} {\bibfnamefont {P.}~\bibnamefont
  {Froese}}, \bibinfo {author} {\bibfnamefont {T.}~\bibnamefont {Neupert}},\
  and\ \bibinfo {author} {\bibfnamefont {G.}~\bibnamefont {Wagner}},\
  }\bibfield  {title} {\bibinfo {title} {Topological excitons in moir{\'e}
  \ce{MoTe2}/\ce{WSe2} heterobilayers},\ }\href
  {https://doi.org/10.1103/PhysRevResearch.7.023047} {\bibfield  {journal}
  {\bibinfo  {journal} {Physical Review Research}\ }\textbf {\bibinfo {volume}
  {7}},\ \bibinfo {pages} {023047} (\bibinfo {year} {2025})}\BibitemShut
  {NoStop}%
\bibitem [{\citenamefont {You}\ \emph {et~al.}(2025)\citenamefont {You},
  \citenamefont {Hsu}, \citenamefont {Zhu}, \citenamefont {Zhang},
  \citenamefont {Ye}, \citenamefont {Naik}, \citenamefont {Cao}, \citenamefont
  {Hsueh}, \citenamefont {Louie}, \citenamefont {Del~Ben} \emph
  {et~al.}}]{you2025moire}%
  \BibitemOpen
  \bibfield  {author} {\bibinfo {author} {\bibfnamefont {J.-Y.}\ \bibnamefont
  {You}}, \bibinfo {author} {\bibfnamefont {C.-E.}\ \bibnamefont {Hsu}},
  \bibinfo {author} {\bibfnamefont {Z.}~\bibnamefont {Zhu}}, \bibinfo {author}
  {\bibfnamefont {B.}~\bibnamefont {Zhang}}, \bibinfo {author} {\bibfnamefont
  {Z.}~\bibnamefont {Ye}}, \bibinfo {author} {\bibfnamefont {M.~H.}\
  \bibnamefont {Naik}}, \bibinfo {author} {\bibfnamefont {T.}~\bibnamefont
  {Cao}}, \bibinfo {author} {\bibfnamefont {H.-C.}\ \bibnamefont {Hsueh}},
  \bibinfo {author} {\bibfnamefont {S.~G.}\ \bibnamefont {Louie}}, \bibinfo
  {author} {\bibfnamefont {M.}~\bibnamefont {Del~Ben}}, \emph {et~al.},\
  }\bibfield  {title} {\bibinfo {title} {Moir{\'e} excitons in generalized
  {W}igner crystals},\ }\href@noop {} {\bibfield  {journal} {\bibinfo
  {journal} {arXiv preprint arXiv:2509.08211}\ } (\bibinfo {year}
  {2025})}\BibitemShut {NoStop}%
\bibitem [{\citenamefont {Xiong}\ \emph {et~al.}(2023)\citenamefont {Xiong},
  \citenamefont {Nie}, \citenamefont {Brantly}, \citenamefont {Hays},
  \citenamefont {Sailus}, \citenamefont {Watanabe}, \citenamefont {Taniguchi},
  \citenamefont {Tongay},\ and\ \citenamefont {Jin}}]{xiong2023correlated}%
  \BibitemOpen
  \bibfield  {author} {\bibinfo {author} {\bibfnamefont {R.}~\bibnamefont
  {Xiong}}, \bibinfo {author} {\bibfnamefont {J.~H.}\ \bibnamefont {Nie}},
  \bibinfo {author} {\bibfnamefont {S.~L.}\ \bibnamefont {Brantly}}, \bibinfo
  {author} {\bibfnamefont {P.}~\bibnamefont {Hays}}, \bibinfo {author}
  {\bibfnamefont {R.}~\bibnamefont {Sailus}}, \bibinfo {author} {\bibfnamefont
  {K.}~\bibnamefont {Watanabe}}, \bibinfo {author} {\bibfnamefont
  {T.}~\bibnamefont {Taniguchi}}, \bibinfo {author} {\bibfnamefont
  {S.}~\bibnamefont {Tongay}},\ and\ \bibinfo {author} {\bibfnamefont
  {C.}~\bibnamefont {Jin}},\ }\bibfield  {title} {\bibinfo {title} {Correlated
  insulator of excitons in \ce{WSe2}/\ce{WS2} moir{\'e} superlattices},\ }\href
  {https://doi.org/10.1126/science.add557} {\bibfield  {journal} {\bibinfo
  {journal} {Science}\ }\textbf {\bibinfo {volume} {380}},\ \bibinfo {pages}
  {860} (\bibinfo {year} {2023})}\BibitemShut {NoStop}%
\bibitem [{\citenamefont {Gao}\ \emph {et~al.}(2024)\citenamefont {Gao},
  \citenamefont {Su{\'a}rez-Forero}, \citenamefont {Sarkar}, \citenamefont
  {Huang}, \citenamefont {Session}, \citenamefont {Mehrabad}, \citenamefont
  {Ni}, \citenamefont {Xie}, \citenamefont {Upadhyay}, \citenamefont {Vannucci}
  \emph {et~al.}}]{gao2024excitonic}%
  \BibitemOpen
  \bibfield  {author} {\bibinfo {author} {\bibfnamefont {B.}~\bibnamefont
  {Gao}}, \bibinfo {author} {\bibfnamefont {D.~G.}\ \bibnamefont
  {Su{\'a}rez-Forero}}, \bibinfo {author} {\bibfnamefont {S.}~\bibnamefont
  {Sarkar}}, \bibinfo {author} {\bibfnamefont {T.-S.}\ \bibnamefont {Huang}},
  \bibinfo {author} {\bibfnamefont {D.}~\bibnamefont {Session}}, \bibinfo
  {author} {\bibfnamefont {M.~J.}\ \bibnamefont {Mehrabad}}, \bibinfo {author}
  {\bibfnamefont {R.}~\bibnamefont {Ni}}, \bibinfo {author} {\bibfnamefont
  {M.}~\bibnamefont {Xie}}, \bibinfo {author} {\bibfnamefont {P.}~\bibnamefont
  {Upadhyay}}, \bibinfo {author} {\bibfnamefont {J.}~\bibnamefont {Vannucci}},
  \emph {et~al.},\ }\bibfield  {title} {\bibinfo {title} {Excitonic {M}ott
  insulator in a {B}ose-{F}ermi-{H}ubbard system of moir{\'e}
  \ce{WS2}/\ce{WSe2} heterobilayer},\ }\href
  {https://doi.org/10.1038/s41467-024-46616-x} {\bibfield  {journal} {\bibinfo
  {journal} {Nature Communications}\ }\textbf {\bibinfo {volume} {15}},\
  \bibinfo {pages} {2305} (\bibinfo {year} {2024})}\BibitemShut {NoStop}%
\bibitem [{\citenamefont {Lian}\ \emph {et~al.}(2024)\citenamefont {Lian},
  \citenamefont {Meng}, \citenamefont {Ma}, \citenamefont {Maity},
  \citenamefont {Yan}, \citenamefont {Wu}, \citenamefont {Huang}, \citenamefont
  {Chen}, \citenamefont {Chen}, \citenamefont {Chen} \emph
  {et~al.}}]{lian2024valley}%
  \BibitemOpen
  \bibfield  {author} {\bibinfo {author} {\bibfnamefont {Z.}~\bibnamefont
  {Lian}}, \bibinfo {author} {\bibfnamefont {Y.}~\bibnamefont {Meng}}, \bibinfo
  {author} {\bibfnamefont {L.}~\bibnamefont {Ma}}, \bibinfo {author}
  {\bibfnamefont {I.}~\bibnamefont {Maity}}, \bibinfo {author} {\bibfnamefont
  {L.}~\bibnamefont {Yan}}, \bibinfo {author} {\bibfnamefont {Q.}~\bibnamefont
  {Wu}}, \bibinfo {author} {\bibfnamefont {X.}~\bibnamefont {Huang}}, \bibinfo
  {author} {\bibfnamefont {D.}~\bibnamefont {Chen}}, \bibinfo {author}
  {\bibfnamefont {X.}~\bibnamefont {Chen}}, \bibinfo {author} {\bibfnamefont
  {X.}~\bibnamefont {Chen}}, \emph {et~al.},\ }\bibfield  {title} {\bibinfo
  {title} {Valley-polarized excitonic {M}ott insulator in \ce{WS2}/\ce{WSe2}
  moir{\'e} superlattice},\ }\href {https://doi.org/10.1038/s41567-023-02266-2}
  {\bibfield  {journal} {\bibinfo  {journal} {Nature Physics}\ }\textbf
  {\bibinfo {volume} {20}},\ \bibinfo {pages} {34} (\bibinfo {year}
  {2024})}\BibitemShut {NoStop}%
\bibitem [{\citenamefont {Cutshall}\ \emph {et~al.}(2025)\citenamefont
  {Cutshall}, \citenamefont {Mahdikhany}, \citenamefont {Roche}, \citenamefont
  {Shanks}, \citenamefont {Koehler}, \citenamefont {Mandrus}, \citenamefont
  {Taniguchi}, \citenamefont {Watanabe}, \citenamefont {Zhu}, \citenamefont
  {LeRoy} \emph {et~al.}}]{cutshall2025imaging}%
  \BibitemOpen
  \bibfield  {author} {\bibinfo {author} {\bibfnamefont {J.}~\bibnamefont
  {Cutshall}}, \bibinfo {author} {\bibfnamefont {F.}~\bibnamefont
  {Mahdikhany}}, \bibinfo {author} {\bibfnamefont {A.}~\bibnamefont {Roche}},
  \bibinfo {author} {\bibfnamefont {D.~N.}\ \bibnamefont {Shanks}}, \bibinfo
  {author} {\bibfnamefont {M.~R.}\ \bibnamefont {Koehler}}, \bibinfo {author}
  {\bibfnamefont {D.~G.}\ \bibnamefont {Mandrus}}, \bibinfo {author}
  {\bibfnamefont {T.}~\bibnamefont {Taniguchi}}, \bibinfo {author}
  {\bibfnamefont {K.}~\bibnamefont {Watanabe}}, \bibinfo {author}
  {\bibfnamefont {Q.}~\bibnamefont {Zhu}}, \bibinfo {author} {\bibfnamefont
  {B.~J.}\ \bibnamefont {LeRoy}}, \emph {et~al.},\ }\bibfield  {title}
  {\bibinfo {title} {Imaging interlayer exciton superfluidity in a {2D}
  semiconductor heterostructure},\ }\href
  {https://doi.org/10.1126/sciadv.adr1772} {\bibfield  {journal} {\bibinfo
  {journal} {Science Advances}\ }\textbf {\bibinfo {volume} {11}},\ \bibinfo
  {pages} {eadr1772} (\bibinfo {year} {2025})}\BibitemShut {NoStop}%
\bibitem [{\citenamefont {Miao}\ \emph {et~al.}(2021)\citenamefont {Miao},
  \citenamefont {Wang}, \citenamefont {Huang}, \citenamefont {Chen},
  \citenamefont {Lian}, \citenamefont {Wang}, \citenamefont {Blei},
  \citenamefont {Taniguchi}, \citenamefont {Watanabe}, \citenamefont {Tongay}
  \emph {et~al.}}]{miao2021strong}%
  \BibitemOpen
  \bibfield  {author} {\bibinfo {author} {\bibfnamefont {S.}~\bibnamefont
  {Miao}}, \bibinfo {author} {\bibfnamefont {T.}~\bibnamefont {Wang}}, \bibinfo
  {author} {\bibfnamefont {X.}~\bibnamefont {Huang}}, \bibinfo {author}
  {\bibfnamefont {D.}~\bibnamefont {Chen}}, \bibinfo {author} {\bibfnamefont
  {Z.}~\bibnamefont {Lian}}, \bibinfo {author} {\bibfnamefont {C.}~\bibnamefont
  {Wang}}, \bibinfo {author} {\bibfnamefont {M.}~\bibnamefont {Blei}}, \bibinfo
  {author} {\bibfnamefont {T.}~\bibnamefont {Taniguchi}}, \bibinfo {author}
  {\bibfnamefont {K.}~\bibnamefont {Watanabe}}, \bibinfo {author}
  {\bibfnamefont {S.}~\bibnamefont {Tongay}}, \emph {et~al.},\ }\bibfield
  {title} {\bibinfo {title} {Strong interaction between interlayer excitons and
  correlated electrons in \ce{WSe2}/\ce{WS2} moir{\'e} superlattice},\ }\href
  {https://doi.org/10.1038/s41467-021-23732-6} {\bibfield  {journal} {\bibinfo
  {journal} {Nature communications}\ }\textbf {\bibinfo {volume} {12}},\
  \bibinfo {pages} {3608} (\bibinfo {year} {2021})}\BibitemShut {NoStop}%
\bibitem [{\citenamefont {Yan}\ \emph {et~al.}(2025)\citenamefont {Yan},
  \citenamefont {Ma}, \citenamefont {Meng}, \citenamefont {Xiao}, \citenamefont
  {Chen}, \citenamefont {Wu}, \citenamefont {Cui}, \citenamefont {Cao},
  \citenamefont {Banerjee}, \citenamefont {Taniguchi} \emph
  {et~al.}}]{yan2025anomalously}%
  \BibitemOpen
  \bibfield  {author} {\bibinfo {author} {\bibfnamefont {L.}~\bibnamefont
  {Yan}}, \bibinfo {author} {\bibfnamefont {L.}~\bibnamefont {Ma}}, \bibinfo
  {author} {\bibfnamefont {Y.}~\bibnamefont {Meng}}, \bibinfo {author}
  {\bibfnamefont {C.}~\bibnamefont {Xiao}}, \bibinfo {author} {\bibfnamefont
  {B.}~\bibnamefont {Chen}}, \bibinfo {author} {\bibfnamefont {Q.}~\bibnamefont
  {Wu}}, \bibinfo {author} {\bibfnamefont {J.}~\bibnamefont {Cui}}, \bibinfo
  {author} {\bibfnamefont {Q.}~\bibnamefont {Cao}}, \bibinfo {author}
  {\bibfnamefont {R.}~\bibnamefont {Banerjee}}, \bibinfo {author}
  {\bibfnamefont {T.}~\bibnamefont {Taniguchi}}, \emph {et~al.},\ }\bibfield
  {title} {\bibinfo {title} {Anomalously enhanced diffusivity of moir{\'e}
  excitons via manipulating the interplay with correlated electrons},\ }\href
  {https://doi.org/10.1038/s41467-025-65602-5} {\bibfield  {journal} {\bibinfo
  {journal} {Nature Communications}\ }\textbf {\bibinfo {volume} {16}},\
  \bibinfo {pages} {10569} (\bibinfo {year} {2025})}\BibitemShut {NoStop}%
\bibitem [{\citenamefont {Zhang}\ \emph {et~al.}(2025)\citenamefont {Zhang},
  \citenamefont {Nguyen}, \citenamefont {Batra}, \citenamefont {Liu},
  \citenamefont {Watanabe}, \citenamefont {Taniguchi}, \citenamefont
  {Feldman},\ and\ \citenamefont {Li}}]{zhang2025excitons}%
  \BibitemOpen
  \bibfield  {author} {\bibinfo {author} {\bibfnamefont {N.~J.}\ \bibnamefont
  {Zhang}}, \bibinfo {author} {\bibfnamefont {R.~Q.}\ \bibnamefont {Nguyen}},
  \bibinfo {author} {\bibfnamefont {N.}~\bibnamefont {Batra}}, \bibinfo
  {author} {\bibfnamefont {X.}~\bibnamefont {Liu}}, \bibinfo {author}
  {\bibfnamefont {K.}~\bibnamefont {Watanabe}}, \bibinfo {author}
  {\bibfnamefont {T.}~\bibnamefont {Taniguchi}}, \bibinfo {author}
  {\bibfnamefont {D.}~\bibnamefont {Feldman}},\ and\ \bibinfo {author}
  {\bibfnamefont {J.}~\bibnamefont {Li}},\ }\bibfield  {title} {\bibinfo
  {title} {Excitons in the fractional quantum {H}all effect},\ }\href
  {https://doi.org/10.1038/s41586-024-08274-3} {\bibfield  {journal} {\bibinfo
  {journal} {Nature}\ }\textbf {\bibinfo {volume} {637}},\ \bibinfo {pages}
  {327} (\bibinfo {year} {2025})}\BibitemShut {NoStop}%
\bibitem [{\citenamefont {Nguyen}\ \emph {et~al.}(2026)\citenamefont {Nguyen},
  \citenamefont {Zhang}, \citenamefont {Khurana-Batra}, \citenamefont
  {Alkidim}, \citenamefont {Liu}, \citenamefont {Watanabe}, \citenamefont
  {Taniguchi}, \citenamefont {Feldman},\ and\ \citenamefont
  {Li}}]{nguyen2026bilayer}%
  \BibitemOpen
  \bibfield  {author} {\bibinfo {author} {\bibfnamefont {R.~Q.}\ \bibnamefont
  {Nguyen}}, \bibinfo {author} {\bibfnamefont {N.~J.}\ \bibnamefont {Zhang}},
  \bibinfo {author} {\bibfnamefont {N.}~\bibnamefont {Khurana-Batra}}, \bibinfo
  {author} {\bibfnamefont {S.}~\bibnamefont {Alkidim}}, \bibinfo {author}
  {\bibfnamefont {X.}~\bibnamefont {Liu}}, \bibinfo {author} {\bibfnamefont
  {K.}~\bibnamefont {Watanabe}}, \bibinfo {author} {\bibfnamefont
  {T.}~\bibnamefont {Taniguchi}}, \bibinfo {author} {\bibfnamefont
  {D.}~\bibnamefont {Feldman}},\ and\ \bibinfo {author} {\bibfnamefont
  {J.}~\bibnamefont {Li}},\ }\bibfield  {title} {\bibinfo {title} {Bilayer
  excitons in the {L}aughlin fractional quantum {H}all state},\ }\href
  {https://doi.org/10.1038/s41567-026-03325-0} {\bibfield  {journal} {\bibinfo
  {journal} {Nature Physics}\ }\textbf {\bibinfo {volume} {22}},\ \bibinfo
  {pages} {1} (\bibinfo {year} {2026})}\BibitemShut {NoStop}%
\bibitem [{\citenamefont {Varney}\ \emph {et~al.}(2011)\citenamefont {Varney},
  \citenamefont {Sun}, \citenamefont {Galitski},\ and\ \citenamefont
  {Rigol}}]{varney2011kaleidoscope}%
  \BibitemOpen
  \bibfield  {author} {\bibinfo {author} {\bibfnamefont {C.~N.}\ \bibnamefont
  {Varney}}, \bibinfo {author} {\bibfnamefont {K.}~\bibnamefont {Sun}},
  \bibinfo {author} {\bibfnamefont {V.}~\bibnamefont {Galitski}},\ and\
  \bibinfo {author} {\bibfnamefont {M.}~\bibnamefont {Rigol}},\ }\bibfield
  {title} {\bibinfo {title} {Kaleidoscope of {E}xotic {Q}uantum {P}hases in a
  {F}rustrated ${XY}$ {M}odel},\ }\href
  {https://doi.org/10.1103/PhysRevLett.107.077201} {\bibfield  {journal}
  {\bibinfo  {journal} {Physical Review Letters}\ }\textbf {\bibinfo {volume}
  {107}},\ \bibinfo {pages} {077201} (\bibinfo {year} {2011})}\BibitemShut
  {NoStop}%
\bibitem [{\citenamefont {Sedrakyan}\ \emph {et~al.}(2014)\citenamefont
  {Sedrakyan}, \citenamefont {Glazman},\ and\ \citenamefont
  {Kamenev}}]{sedrakyan2014absence}%
  \BibitemOpen
  \bibfield  {author} {\bibinfo {author} {\bibfnamefont {T.~A.}\ \bibnamefont
  {Sedrakyan}}, \bibinfo {author} {\bibfnamefont {L.~I.}\ \bibnamefont
  {Glazman}},\ and\ \bibinfo {author} {\bibfnamefont {A.}~\bibnamefont
  {Kamenev}},\ }\bibfield  {title} {\bibinfo {title} {Absence of {B}ose
  condensation on lattices with moat bands},\ }\href
  {https://doi.org/10.1103/PhysRevB.89.201112} {\bibfield  {journal} {\bibinfo
  {journal} {Physical Review B}\ }\textbf {\bibinfo {volume} {89}},\ \bibinfo
  {pages} {201112} (\bibinfo {year} {2014})}\BibitemShut {NoStop}%
\bibitem [{\citenamefont {Sedrakyan}\ \emph
  {et~al.}(2015{\natexlab{a}})\citenamefont {Sedrakyan}, \citenamefont
  {Glazman},\ and\ \citenamefont {Kamenev}}]{sedrakyan2015spontaneous}%
  \BibitemOpen
  \bibfield  {author} {\bibinfo {author} {\bibfnamefont {T.~A.}\ \bibnamefont
  {Sedrakyan}}, \bibinfo {author} {\bibfnamefont {L.~I.}\ \bibnamefont
  {Glazman}},\ and\ \bibinfo {author} {\bibfnamefont {A.}~\bibnamefont
  {Kamenev}},\ }\bibfield  {title} {\bibinfo {title} {Spontaneous formation of
  a nonuniform chiral spin liquid in a moat-band lattice},\ }\href
  {https://doi.org/10.1103/PhysRevLett.114.037203} {\bibfield  {journal}
  {\bibinfo  {journal} {Physical Review Letters}\ }\textbf {\bibinfo {volume}
  {114}},\ \bibinfo {pages} {037203} (\bibinfo {year}
  {2015}{\natexlab{a}})}\BibitemShut {NoStop}%
\bibitem [{\citenamefont {Campbell}\ \emph {et~al.}(2011)\citenamefont
  {Campbell}, \citenamefont {Juzeli{\=u}nas},\ and\ \citenamefont
  {Spielman}}]{campbell2011realistic}%
  \BibitemOpen
  \bibfield  {author} {\bibinfo {author} {\bibfnamefont {D.~L.}\ \bibnamefont
  {Campbell}}, \bibinfo {author} {\bibfnamefont {G.}~\bibnamefont
  {Juzeli{\=u}nas}},\ and\ \bibinfo {author} {\bibfnamefont {I.~B.}\
  \bibnamefont {Spielman}},\ }\bibfield  {title} {\bibinfo {title} {Realistic
  {R}ashba and {D}resselhaus spin-orbit coupling for neutral atoms},\ }\href
  {https://doi.org/10.1103/PhysRevA.84.025602} {\bibfield  {journal} {\bibinfo
  {journal} {Physical Review A}\ }\textbf {\bibinfo {volume} {84}},\ \bibinfo
  {pages} {025602} (\bibinfo {year} {2011})}\BibitemShut {NoStop}%
\bibitem [{\citenamefont {Gopalakrishnan}\ \emph {et~al.}(2011)\citenamefont
  {Gopalakrishnan}, \citenamefont {Lamacraft},\ and\ \citenamefont
  {Goldbart}}]{gopalakrishnan2011universal}%
  \BibitemOpen
  \bibfield  {author} {\bibinfo {author} {\bibfnamefont {S.}~\bibnamefont
  {Gopalakrishnan}}, \bibinfo {author} {\bibfnamefont {A.}~\bibnamefont
  {Lamacraft}},\ and\ \bibinfo {author} {\bibfnamefont {P.~M.}\ \bibnamefont
  {Goldbart}},\ }\bibfield  {title} {\bibinfo {title} {Universal phase
  structure of dilute {B}ose gases with {R}ashba spin-orbit coupling},\ }\href
  {https://doi.org/10.1103/PhysRevA.84.061604} {\bibfield  {journal} {\bibinfo
  {journal} {Physical Review A}\ }\textbf {\bibinfo {volume} {84}},\ \bibinfo
  {pages} {061604} (\bibinfo {year} {2011})}\BibitemShut {NoStop}%
\bibitem [{\citenamefont {Berg}\ \emph {et~al.}(2012)\citenamefont {Berg},
  \citenamefont {Rudner},\ and\ \citenamefont {Kivelson}}]{berg2012electronic}%
  \BibitemOpen
  \bibfield  {author} {\bibinfo {author} {\bibfnamefont {E.}~\bibnamefont
  {Berg}}, \bibinfo {author} {\bibfnamefont {M.~S.}\ \bibnamefont {Rudner}},\
  and\ \bibinfo {author} {\bibfnamefont {S.~A.}\ \bibnamefont {Kivelson}},\
  }\bibfield  {title} {\bibinfo {title} {Electronic liquid crystalline phases
  in a spin-orbit coupled two-dimensional electron gas},\ }\href
  {https://doi.org/10.1103/PhysRevB.85.035116} {\bibfield  {journal} {\bibinfo
  {journal} {Physical Review B}\ }\textbf {\bibinfo {volume} {85}},\ \bibinfo
  {pages} {035116} (\bibinfo {year} {2012})}\BibitemShut {NoStop}%
\bibitem [{\citenamefont {Ozawa}\ and\ \citenamefont
  {Baym}(2012)}]{ozawa2012stability}%
  \BibitemOpen
  \bibfield  {author} {\bibinfo {author} {\bibfnamefont {T.}~\bibnamefont
  {Ozawa}}\ and\ \bibinfo {author} {\bibfnamefont {G.}~\bibnamefont {Baym}},\
  }\bibfield  {title} {\bibinfo {title} {Stability of ultracold atomic bose
  condensates with rashba spin-orbit coupling against quantum and thermal
  fluctuations},\ }\href {https://doi.org/10.1103/PhysRevLett.109.025301}
  {\bibfield  {journal} {\bibinfo  {journal} {Physical Review Letters}\
  }\textbf {\bibinfo {volume} {109}},\ \bibinfo {pages} {025301} (\bibinfo
  {year} {2012})}\BibitemShut {NoStop}%
\bibitem [{\citenamefont {Galitski}\ and\ \citenamefont
  {Spielman}(2013)}]{galitski2013spin}%
  \BibitemOpen
  \bibfield  {author} {\bibinfo {author} {\bibfnamefont {V.}~\bibnamefont
  {Galitski}}\ and\ \bibinfo {author} {\bibfnamefont {I.~B.}\ \bibnamefont
  {Spielman}},\ }\bibfield  {title} {\bibinfo {title} {Spin--orbit coupling in
  quantum gases},\ }\href {https://doi.org/10.1038/nature11841} {\bibfield
  {journal} {\bibinfo  {journal} {Nature}\ }\textbf {\bibinfo {volume} {494}},\
  \bibinfo {pages} {49} (\bibinfo {year} {2013})}\BibitemShut {NoStop}%
\bibitem [{\citenamefont {Zhai}(2015)}]{zhai2015degenerate}%
  \BibitemOpen
  \bibfield  {author} {\bibinfo {author} {\bibfnamefont {H.}~\bibnamefont
  {Zhai}},\ }\bibfield  {title} {\bibinfo {title} {Degenerate quantum gases
  with spin--orbit coupling: a review},\ }\href
  {https://doi.org/10.1088/0034-4885/78/2/026001} {\bibfield  {journal}
  {\bibinfo  {journal} {Reports on Progress in Physics}\ }\textbf {\bibinfo
  {volume} {78}},\ \bibinfo {pages} {026001} (\bibinfo {year}
  {2015})}\BibitemShut {NoStop}%
\bibitem [{\citenamefont {Bracamontes}\ \emph {et~al.}(2022)\citenamefont
  {Bracamontes}, \citenamefont {Maslek},\ and\ \citenamefont
  {Porto}}]{bracamontes2022realization}%
  \BibitemOpen
  \bibfield  {author} {\bibinfo {author} {\bibfnamefont {C.}~\bibnamefont
  {Bracamontes}}, \bibinfo {author} {\bibfnamefont {J.}~\bibnamefont
  {Maslek}},\ and\ \bibinfo {author} {\bibfnamefont {J.}~\bibnamefont
  {Porto}},\ }\bibfield  {title} {\bibinfo {title} {Realization of a
  floquet-engineered moat band for ultracold atoms},\ }\href
  {https://doi.org/10.1103/PhysRevLett.128.213401} {\bibfield  {journal}
  {\bibinfo  {journal} {Physical Review Letters}\ }\textbf {\bibinfo {volume}
  {128}},\ \bibinfo {pages} {213401} (\bibinfo {year} {2022})}\BibitemShut
  {NoStop}%
\bibitem [{\citenamefont {Maisel~Licer{\'a}n}\ \emph
  {et~al.}(2023)\citenamefont {Maisel~Licer{\'a}n}, \citenamefont
  {Garc{\'\i}a~Fl{\'o}rez}, \citenamefont {Siebbeles},\ and\ \citenamefont
  {Stoof}}]{maisel2023single}%
  \BibitemOpen
  \bibfield  {author} {\bibinfo {author} {\bibfnamefont {L.}~\bibnamefont
  {Maisel~Licer{\'a}n}}, \bibinfo {author} {\bibfnamefont {F.}~\bibnamefont
  {Garc{\'\i}a~Fl{\'o}rez}}, \bibinfo {author} {\bibfnamefont {L.~D.}\
  \bibnamefont {Siebbeles}},\ and\ \bibinfo {author} {\bibfnamefont {H.~T.}\
  \bibnamefont {Stoof}},\ }\bibfield  {title} {\bibinfo {title}
  {Single-particle properties of topological {W}annier excitons in bismuth
  chalcogenide nanosheets},\ }\href
  {https://doi.org/10.1038/s41598-023-32740-z} {\bibfield  {journal} {\bibinfo
  {journal} {Scientific Reports}\ }\textbf {\bibinfo {volume} {13}},\ \bibinfo
  {pages} {6337} (\bibinfo {year} {2023})}\BibitemShut {NoStop}%
\bibitem [{\citenamefont {Sedrakyan}\ \emph {et~al.}(2013)\citenamefont
  {Sedrakyan}, \citenamefont {Kamenev},\ and\ \citenamefont
  {Glazman}}]{sedrakyan2013composite}%
  \BibitemOpen
  \bibfield  {author} {\bibinfo {author} {\bibfnamefont {T.~A.}\ \bibnamefont
  {Sedrakyan}}, \bibinfo {author} {\bibfnamefont {A.}~\bibnamefont {Kamenev}},\
  and\ \bibinfo {author} {\bibfnamefont {L.~I.}\ \bibnamefont {Glazman}},\
  }\bibfield  {title} {\bibinfo {title} {Composite fermion state of
  spin-orbit-coupled bosons},\ }\href
  {https://doi.org/10.1103/PhysRevA.86.063639} {\bibfield  {journal} {\bibinfo
  {journal} {Physical Review A}\ }\textbf {\bibinfo {volume} {86}},\ \bibinfo
  {pages} {063639} (\bibinfo {year} {2013})}\BibitemShut {NoStop}%
\bibitem [{\citenamefont {Sedrakyan}\ \emph
  {et~al.}(2015{\natexlab{b}})\citenamefont {Sedrakyan}, \citenamefont
  {Galitski},\ and\ \citenamefont {Kamenev}}]{sedrakyan2015statistical}%
  \BibitemOpen
  \bibfield  {author} {\bibinfo {author} {\bibfnamefont {T.~A.}\ \bibnamefont
  {Sedrakyan}}, \bibinfo {author} {\bibfnamefont {V.~M.}\ \bibnamefont
  {Galitski}},\ and\ \bibinfo {author} {\bibfnamefont {A.}~\bibnamefont
  {Kamenev}},\ }\bibfield  {title} {\bibinfo {title} {Statistical transmutation
  in {F}loquet driven optical lattices},\ }\href
  {https://doi.org/10.1103/PhysRevLett.115.195301} {\bibfield  {journal}
  {\bibinfo  {journal} {Physical Review Letters}\ }\textbf {\bibinfo {volume}
  {115}},\ \bibinfo {pages} {195301} (\bibinfo {year}
  {2015}{\natexlab{b}})}\BibitemShut {NoStop}%
\bibitem [{\citenamefont {Wei}\ and\ \citenamefont
  {Sedrakyan}(2023)}]{wei2023chiral}%
  \BibitemOpen
  \bibfield  {author} {\bibinfo {author} {\bibfnamefont {C.}~\bibnamefont
  {Wei}}\ and\ \bibinfo {author} {\bibfnamefont {T.~A.}\ \bibnamefont
  {Sedrakyan}},\ }\bibfield  {title} {\bibinfo {title} {Chiral spin liquid
  state of strongly interacting bosons with a moat dispersion: {A} {M}onte
  {C}arlo simulation},\ }\href {https://doi.org/10.1016/j.aop.2023.169354}
  {\bibfield  {journal} {\bibinfo  {journal} {Annals of Physics}\ }\textbf
  {\bibinfo {volume} {456}},\ \bibinfo {pages} {169354} (\bibinfo {year}
  {2023})}\BibitemShut {NoStop}%
\bibitem [{\citenamefont {Wang}\ \emph {et~al.}(2022)\citenamefont {Wang},
  \citenamefont {Xie}, \citenamefont {Wang},\ and\ \citenamefont
  {Sedrakyan}}]{wang2022emergent}%
  \BibitemOpen
  \bibfield  {author} {\bibinfo {author} {\bibfnamefont {R.}~\bibnamefont
  {Wang}}, \bibinfo {author} {\bibfnamefont {Z.}~\bibnamefont {Xie}}, \bibinfo
  {author} {\bibfnamefont {B.}~\bibnamefont {Wang}},\ and\ \bibinfo {author}
  {\bibfnamefont {T.}~\bibnamefont {Sedrakyan}},\ }\bibfield  {title} {\bibinfo
  {title} {Emergent topological orders and phase transitions in lattice
  {C}hern-{S}imons theory of quantum magnets},\ }\href
  {https://doi.org/10.1103/PhysRevB.106.L121117} {\bibfield  {journal}
  {\bibinfo  {journal} {Physical Review B}\ }\textbf {\bibinfo {volume}
  {106}},\ \bibinfo {pages} {L121117} (\bibinfo {year} {2022})}\BibitemShut
  {NoStop}%
\bibitem [{\citenamefont {Wang}\ \emph {et~al.}(2024)\citenamefont {Wang},
  \citenamefont {Yang}, \citenamefont {Xie}, \citenamefont {Wang},\ and\
  \citenamefont {Xie}}]{wang2024susceptibility}%
  \BibitemOpen
  \bibfield  {author} {\bibinfo {author} {\bibfnamefont {R.}~\bibnamefont
  {Wang}}, \bibinfo {author} {\bibfnamefont {T.}~\bibnamefont {Yang}}, \bibinfo
  {author} {\bibfnamefont {Z.}~\bibnamefont {Xie}}, \bibinfo {author}
  {\bibfnamefont {B.}~\bibnamefont {Wang}},\ and\ \bibinfo {author}
  {\bibfnamefont {X.}~\bibnamefont {Xie}},\ }\bibfield  {title} {\bibinfo
  {title} {Susceptibility indicator for chiral topological orders emergent from
  correlated fermions},\ }\href {https://doi.org/10.1103/PhysRevB.109.L241113}
  {\bibfield  {journal} {\bibinfo  {journal} {Physical Review B}\ }\textbf
  {\bibinfo {volume} {109}},\ \bibinfo {pages} {L241113} (\bibinfo {year}
  {2024})}\BibitemShut {NoStop}%
\bibitem [{\citenamefont {Wang}\ \emph {et~al.}(2023)\citenamefont {Wang},
  \citenamefont {Sedrakyan}, \citenamefont {Wang}, \citenamefont {Du},\ and\
  \citenamefont {Du}}]{wang2023excitonic}%
  \BibitemOpen
  \bibfield  {author} {\bibinfo {author} {\bibfnamefont {R.}~\bibnamefont
  {Wang}}, \bibinfo {author} {\bibfnamefont {T.~A.}\ \bibnamefont {Sedrakyan}},
  \bibinfo {author} {\bibfnamefont {B.}~\bibnamefont {Wang}}, \bibinfo {author}
  {\bibfnamefont {L.}~\bibnamefont {Du}},\ and\ \bibinfo {author}
  {\bibfnamefont {R.-R.}\ \bibnamefont {Du}},\ }\bibfield  {title} {\bibinfo
  {title} {Excitonic topological order in imbalanced electron--hole bilayers},\
  }\href {https://doi.org/10.1038/s41586-023-06065-w} {\bibfield  {journal}
  {\bibinfo  {journal} {Nature}\ }\textbf {\bibinfo {volume} {619}},\ \bibinfo
  {pages} {57} (\bibinfo {year} {2023})}\BibitemShut {NoStop}%
\bibitem [{\citenamefont {Chester}(1970)}]{PhysRevA.2.256}%
  \BibitemOpen
  \bibfield  {author} {\bibinfo {author} {\bibfnamefont {G.~V.}\ \bibnamefont
  {Chester}},\ }\bibfield  {title} {\bibinfo {title} {Speculations on
  {B}ose--{E}instein condensation and quantum crystals},\ }\href
  {https://doi.org/10.1103/PhysRevA.2.256} {\bibfield  {journal} {\bibinfo
  {journal} {Physical Review A}\ }\textbf {\bibinfo {volume} {2}},\ \bibinfo
  {pages} {256} (\bibinfo {year} {1970})}\BibitemShut {NoStop}%
\bibitem [{\citenamefont {{Andreev}}\ and\ \citenamefont
  {{Lifshitz}}(1969)}]{1969JETP...29.1107A}%
  \BibitemOpen
  \bibfield  {author} {\bibinfo {author} {\bibfnamefont {A.~F.}\ \bibnamefont
  {{Andreev}}}\ and\ \bibinfo {author} {\bibfnamefont {I.~M.}\ \bibnamefont
  {{Lifshitz}}},\ }\bibfield  {title} {\bibinfo {title} {{Quantum Theory of
  Defects in Crystals}},\ }\href@noop {} {\bibfield  {journal} {\bibinfo
  {journal} {Soviet Journal of Experimental and Theoretical Physics}\ }\textbf
  {\bibinfo {volume} {29}},\ \bibinfo {pages} {1107} (\bibinfo {year}
  {1969})}\BibitemShut {NoStop}%
\bibitem [{\citenamefont {Leggett}(1970)}]{PhysRevLett.25.1543}%
  \BibitemOpen
  \bibfield  {author} {\bibinfo {author} {\bibfnamefont {A.~J.}\ \bibnamefont
  {Leggett}},\ }\bibfield  {title} {\bibinfo {title} {Can a solid be
  ``superfluid''?},\ }\href {https://doi.org/10.1103/PhysRevLett.25.1543}
  {\bibfield  {journal} {\bibinfo  {journal} {Physical Review Letters}\
  }\textbf {\bibinfo {volume} {25}},\ \bibinfo {pages} {1543} (\bibinfo {year}
  {1970})}\BibitemShut {NoStop}%
\bibitem [{\citenamefont {Balibar}(2010)}]{nature08913}%
  \BibitemOpen
  \bibfield  {author} {\bibinfo {author} {\bibfnamefont {S.}~\bibnamefont
  {Balibar}},\ }\bibfield  {title} {\bibinfo {title} {The enigma of
  supersolidity},\ }\href {https://doi.org/10.1038/nature08913} {\bibfield
  {journal} {\bibinfo  {journal} {Nature}\ }\textbf {\bibinfo {volume} {464}},\
  \bibinfo {pages} {176} (\bibinfo {year} {2010})}\BibitemShut {NoStop}%
\bibitem [{\citenamefont {Joglekar}\ \emph {et~al.}(2006)\citenamefont
  {Joglekar}, \citenamefont {Balatsky},\ and\ \citenamefont
  {Das~Sarma}}]{joglekar2006wigner}%
  \BibitemOpen
  \bibfield  {author} {\bibinfo {author} {\bibfnamefont {Y.~N.}\ \bibnamefont
  {Joglekar}}, \bibinfo {author} {\bibfnamefont {A.~V.}\ \bibnamefont
  {Balatsky}},\ and\ \bibinfo {author} {\bibfnamefont {S.}~\bibnamefont
  {Das~Sarma}},\ }\bibfield  {title} {\bibinfo {title} {Wigner supersolid of
  excitons in electron-hole bilayers},\ }\href
  {https://doi.org/10.1103/PhysRevB.74.233302} {\bibfield  {journal} {\bibinfo
  {journal} {Physical Review B}\ }\textbf {\bibinfo {volume} {74}},\ \bibinfo
  {pages} {233302} (\bibinfo {year} {2006})}\BibitemShut {NoStop}%
\bibitem [{\citenamefont {Conti}\ \emph {et~al.}(2023)\citenamefont {Conti},
  \citenamefont {Perali}, \citenamefont {Hamilton}, \citenamefont
  {Milo{\v{s}}evi{\'c}}, \citenamefont {Peeters},\ and\ \citenamefont
  {Neilson}}]{conti2023chester}%
  \BibitemOpen
  \bibfield  {author} {\bibinfo {author} {\bibfnamefont {S.}~\bibnamefont
  {Conti}}, \bibinfo {author} {\bibfnamefont {A.}~\bibnamefont {Perali}},
  \bibinfo {author} {\bibfnamefont {A.~R.}\ \bibnamefont {Hamilton}}, \bibinfo
  {author} {\bibfnamefont {M.~V.}\ \bibnamefont {Milo{\v{s}}evi{\'c}}},
  \bibinfo {author} {\bibfnamefont {F.~M.}\ \bibnamefont {Peeters}},\ and\
  \bibinfo {author} {\bibfnamefont {D.}~\bibnamefont {Neilson}},\ }\bibfield
  {title} {\bibinfo {title} {Chester supersolid of spatially indirect excitons
  in double-layer semiconductor heterostructures},\ }\href
  {https://doi.org/10.1103/PhysRevLett.130.057001} {\bibfield  {journal}
  {\bibinfo  {journal} {Physical Review Letters}\ }\textbf {\bibinfo {volume}
  {130}},\ \bibinfo {pages} {057001} (\bibinfo {year} {2023})}\BibitemShut
  {NoStop}%
\bibitem [{\citenamefont {Conti}\ \emph {et~al.}(2025)\citenamefont {Conti},
  \citenamefont {Chaves}, \citenamefont {Hamilton}, \citenamefont {Tempere},
  \citenamefont {Milosevic},\ and\ \citenamefont {Neilson}}]{conti2025gross}%
  \BibitemOpen
  \bibfield  {author} {\bibinfo {author} {\bibfnamefont {S.}~\bibnamefont
  {Conti}}, \bibinfo {author} {\bibfnamefont {A.}~\bibnamefont {Chaves}},
  \bibinfo {author} {\bibfnamefont {A.~R.}\ \bibnamefont {Hamilton}}, \bibinfo
  {author} {\bibfnamefont {J.}~\bibnamefont {Tempere}}, \bibinfo {author}
  {\bibfnamefont {M.~V.}\ \bibnamefont {Milosevic}},\ and\ \bibinfo {author}
  {\bibfnamefont {D.}~\bibnamefont {Neilson}},\ }\bibfield  {title} {\bibinfo
  {title} {A {G}ross-{P}itaevskii theory for an excitonic incompressible {B}ose
  solid},\ }\href@noop {} {\bibfield  {journal} {\bibinfo  {journal} {arXiv
  preprint arXiv:2507.20236}\ } (\bibinfo {year} {2025})}\BibitemShut {NoStop}%
\bibitem [{\citenamefont {de~Oliveira~Neto}\ \emph {et~al.}(2025)\citenamefont
  {de~Oliveira~Neto}, \citenamefont {Guimar{\~a}es}, \citenamefont {Dantas},
  \citenamefont {Peeters}, \citenamefont {Milo{\v{s}}evi{\'c}},\ and\
  \citenamefont {Chaves}}]{deoliveira2025striped}%
  \BibitemOpen
  \bibfield  {author} {\bibinfo {author} {\bibfnamefont {J.}~\bibnamefont
  {de~Oliveira~Neto}}, \bibinfo {author} {\bibfnamefont {F.}~\bibnamefont
  {Guimar{\~a}es}}, \bibinfo {author} {\bibfnamefont {D.~S.}\ \bibnamefont
  {Dantas}}, \bibinfo {author} {\bibfnamefont {F.}~\bibnamefont {Peeters}},
  \bibinfo {author} {\bibfnamefont {M.}~\bibnamefont {Milo{\v{s}}evi{\'c}}},\
  and\ \bibinfo {author} {\bibfnamefont {A.}~\bibnamefont {Chaves}},\
  }\bibfield  {title} {\bibinfo {title} {Striped excitonic (super) solid in
  anisotropic semiconductors with screened exciton interactions},\ }\href
  {https://doi.org/10.1103/PhysRevB.111.L180506} {\bibfield  {journal}
  {\bibinfo  {journal} {Physical Review B}\ }\textbf {\bibinfo {volume}
  {111}},\ \bibinfo {pages} {L180506} (\bibinfo {year} {2025})}\BibitemShut
  {NoStop}%
\bibitem [{\citenamefont {Noordman}\ \emph {et~al.}(2026)\citenamefont
  {Noordman}, \citenamefont {Maisel~Licer{\'a}n},\ and\ \citenamefont
  {Stoof}}]{noordman2026variational}%
  \BibitemOpen
  \bibfield  {author} {\bibinfo {author} {\bibfnamefont {P.~A.}\ \bibnamefont
  {Noordman}}, \bibinfo {author} {\bibfnamefont {L.}~\bibnamefont
  {Maisel~Licer{\'a}n}},\ and\ \bibinfo {author} {\bibfnamefont {H.~T.~C.}\
  \bibnamefont {Stoof}},\ }\bibfield  {title} {\bibinfo {title} {Variational
  and field-theoretical approach to exciton-exciton interactions and biexcitons
  in semiconductors},\ }\href {https://doi.org/10.1103/hmkm-jxzl} {\bibfield
  {journal} {\bibinfo  {journal} {Physical Review B}\ }\textbf {\bibinfo
  {volume} {113}},\ \bibinfo {pages} {155440} (\bibinfo {year}
  {2026})}\BibitemShut {NoStop}%
\bibitem [{\citenamefont {Bijlsma}\ and\ \citenamefont
  {Stoof}(1997)}]{bijlsma1997variational}%
  \BibitemOpen
  \bibfield  {author} {\bibinfo {author} {\bibfnamefont {M.}~\bibnamefont
  {Bijlsma}}\ and\ \bibinfo {author} {\bibfnamefont {H.}~\bibnamefont
  {Stoof}},\ }\bibfield  {title} {\bibinfo {title} {Variational approach to the
  dilute {B}ose gas},\ }\href {https://doi.org/10.1103/PhysRevA.55.498}
  {\bibfield  {journal} {\bibinfo  {journal} {Physical Review A}\ }\textbf
  {\bibinfo {volume} {55}},\ \bibinfo {pages} {498} (\bibinfo {year}
  {1997})}\BibitemShut {NoStop}%
\bibitem [{\citenamefont {Proukakis}\ \emph {et~al.}(1998)\citenamefont
  {Proukakis}, \citenamefont {Morgan}, \citenamefont {Choi},\ and\
  \citenamefont {Burnett}}]{proukakis1998comparison}%
  \BibitemOpen
  \bibfield  {author} {\bibinfo {author} {\bibfnamefont {N.}~\bibnamefont
  {Proukakis}}, \bibinfo {author} {\bibfnamefont {S.}~\bibnamefont {Morgan}},
  \bibinfo {author} {\bibfnamefont {S.}~\bibnamefont {Choi}},\ and\ \bibinfo
  {author} {\bibfnamefont {K.}~\bibnamefont {Burnett}},\ }\bibfield  {title}
  {\bibinfo {title} {Comparison of gapless mean-field theories for trapped
  {B}ose-{E}instein condensates},\ }\href
  {https://doi.org/10.1103/PhysRevA.58.2435} {\bibfield  {journal} {\bibinfo
  {journal} {Physical Review A}\ }\textbf {\bibinfo {volume} {58}},\ \bibinfo
  {pages} {2435} (\bibinfo {year} {1998})}\BibitemShut {NoStop}%
\bibitem [{\citenamefont {Lee}\ \emph {et~al.}(2002)\citenamefont {Lee},
  \citenamefont {Morgan}, \citenamefont {Davis},\ and\ \citenamefont
  {Burnett}}]{lee2002energy}%
  \BibitemOpen
  \bibfield  {author} {\bibinfo {author} {\bibfnamefont {M.~D.}\ \bibnamefont
  {Lee}}, \bibinfo {author} {\bibfnamefont {S.~A.}\ \bibnamefont {Morgan}},
  \bibinfo {author} {\bibfnamefont {M.~J.}\ \bibnamefont {Davis}},\ and\
  \bibinfo {author} {\bibfnamefont {K.}~\bibnamefont {Burnett}},\ }\bibfield
  {title} {\bibinfo {title} {Energy-dependent scattering and the
  {G}ross--{P}itaevskii equation in two-dimensional {B}ose--{E}instein
  condensates},\ }\href {https://doi.org/10.1103/PhysRevA.65.043617} {\bibfield
   {journal} {\bibinfo  {journal} {Physical Review A}\ }\textbf {\bibinfo
  {volume} {65}},\ \bibinfo {pages} {043617} (\bibinfo {year}
  {2002})}\BibitemShut {NoStop}%
\bibitem [{\citenamefont {Gies}\ \emph {et~al.}(2005)\citenamefont {Gies},
  \citenamefont {Lee},\ and\ \citenamefont {Hutchinson}}]{gies2005many}%
  \BibitemOpen
  \bibfield  {author} {\bibinfo {author} {\bibfnamefont {C.}~\bibnamefont
  {Gies}}, \bibinfo {author} {\bibfnamefont {M.}~\bibnamefont {Lee}},\ and\
  \bibinfo {author} {\bibfnamefont {D.}~\bibnamefont {Hutchinson}},\ }\bibfield
   {title} {\bibinfo {title} {Many-body {$T$}-matrix of a two-dimensional
  {B}ose--{E}instein condensate within the {H}artree--{F}ock--{B}ogoliubov
  formalism},\ }\href {https://doi.org/10.1088/0953-4075/38/11/019} {\bibfield
  {journal} {\bibinfo  {journal} {Journal of Physics B}\ }\textbf {\bibinfo
  {volume} {38}},\ \bibinfo {pages} {1797} (\bibinfo {year}
  {2005})}\BibitemShut {NoStop}%
\bibitem [{\citenamefont {Stoof}\ and\ \citenamefont
  {Bijlsma}(1993)}]{stoof1993kosterlitz}%
  \BibitemOpen
  \bibfield  {author} {\bibinfo {author} {\bibfnamefont {H.~T.~C.}\
  \bibnamefont {Stoof}}\ and\ \bibinfo {author} {\bibfnamefont
  {M.}~\bibnamefont {Bijlsma}},\ }\bibfield  {title} {\bibinfo {title}
  {Kosterlitz--{T}houless transition in a dilute {B}ose gas},\ }\href
  {https://doi.org/10.1103/PhysRevE.47.939} {\bibfield  {journal} {\bibinfo
  {journal} {Physical Review E}\ }\textbf {\bibinfo {volume} {47}},\ \bibinfo
  {pages} {939} (\bibinfo {year} {1993})}\BibitemShut {NoStop}%
\bibitem [{\citenamefont {Andersen}\ \emph {et~al.}(2002)\citenamefont
  {Andersen}, \citenamefont {Al~Khawaja},\ and\ \citenamefont
  {Stoof}}]{andersen2002phase}%
  \BibitemOpen
  \bibfield  {author} {\bibinfo {author} {\bibfnamefont {J.}~\bibnamefont
  {Andersen}}, \bibinfo {author} {\bibfnamefont {U.}~\bibnamefont
  {Al~Khawaja}},\ and\ \bibinfo {author} {\bibfnamefont {H.~T.~C.}\
  \bibnamefont {Stoof}},\ }\bibfield  {title} {\bibinfo {title} {Phase
  fluctuations in atomic {B}ose gases},\ }\href
  {https://doi.org/10.1103/PhysRevLett.88.070407} {\bibfield  {journal}
  {\bibinfo  {journal} {Physical Review Letters}\ }\textbf {\bibinfo {volume}
  {88}},\ \bibinfo {pages} {070407} (\bibinfo {year} {2002})}\BibitemShut
  {NoStop}%
\bibitem [{\citenamefont {Al~Khawaja}\ \emph {et~al.}(2002)\citenamefont
  {Al~Khawaja}, \citenamefont {Andersen}, \citenamefont {Proukakis},\ and\
  \citenamefont {Stoof}}]{alkhawaja2002low}%
  \BibitemOpen
  \bibfield  {author} {\bibinfo {author} {\bibfnamefont {U.}~\bibnamefont
  {Al~Khawaja}}, \bibinfo {author} {\bibfnamefont {J.}~\bibnamefont
  {Andersen}}, \bibinfo {author} {\bibfnamefont {N.}~\bibnamefont
  {Proukakis}},\ and\ \bibinfo {author} {\bibfnamefont {H.~C.}\ \bibnamefont
  {Stoof}},\ }\bibfield  {title} {\bibinfo {title} {Low dimensional {B}ose
  gases},\ }\href {https://doi.org/10.1103/PhysRevA.66.013615} {\bibfield
  {journal} {\bibinfo  {journal} {Physical Review A}\ }\textbf {\bibinfo
  {volume} {66}},\ \bibinfo {pages} {013615} (\bibinfo {year}
  {2002})}\BibitemShut {NoStop}%
\bibitem [{\citenamefont {Rajagopal}\ \emph {et~al.}(2004)\citenamefont
  {Rajagopal}, \citenamefont {Vignolo},\ and\ \citenamefont
  {Tosi}}]{rajagopal2004density}%
  \BibitemOpen
  \bibfield  {author} {\bibinfo {author} {\bibfnamefont {K.}~\bibnamefont
  {Rajagopal}}, \bibinfo {author} {\bibfnamefont {P.}~\bibnamefont {Vignolo}},\
  and\ \bibinfo {author} {\bibfnamefont {M.}~\bibnamefont {Tosi}},\ }\bibfield
  {title} {\bibinfo {title} {Density profile of a strictly two-dimensional
  {B}ose gas at finite temperature},\ }\href
  {https://doi.org/10.1016/j.physb.2003.09.251} {\bibfield  {journal} {\bibinfo
   {journal} {Physica B}\ }\textbf {\bibinfo {volume} {344}},\ \bibinfo {pages}
  {157} (\bibinfo {year} {2004})}\BibitemShut {NoStop}%
\bibitem [{Note1()}]{Note1}%
  \BibitemOpen
  \bibinfo {note} {In Ref.\ \cite {lee2002energy}, a somewhat more detailed
  (but not fully rigorous) argument is used to justify that the 2D many-body
  $T$ matrix should be replaced by the two-body one at an energy ${-}\mu $, as
  opposed to ${-}2 \mu $. However, the calculation depends on the precise form
  of the single-particle dispersion and is therefore not universal. In view of
  this uncertainty we choose here to consider the ${-}2 \mu $ energy arising
  from the intuitive physical picture. In any case, we do not expect the
  results to be affected significantly by this particular aspect of the
  theory.}\BibitemShut {Stop}%
\bibitem [{\citenamefont {G{\"o}tting}\ \emph {et~al.}(2022)\citenamefont
  {G{\"o}tting}, \citenamefont {Lohof},\ and\ \citenamefont
  {Gies}}]{gotting2022moire}%
  \BibitemOpen
  \bibfield  {author} {\bibinfo {author} {\bibfnamefont {N.}~\bibnamefont
  {G{\"o}tting}}, \bibinfo {author} {\bibfnamefont {F.}~\bibnamefont {Lohof}},\
  and\ \bibinfo {author} {\bibfnamefont {C.}~\bibnamefont {Gies}},\ }\bibfield
  {title} {\bibinfo {title} {Moir{\'e}-{B}ose-{H}ubbard model for interlayer
  excitons in twisted transition metal dichalcogenide heterostructures},\
  }\href {https://doi.org/10.1103/PhysRevB.105.165419} {\bibfield  {journal}
  {\bibinfo  {journal} {Physical Review B}\ }\textbf {\bibinfo {volume}
  {105}},\ \bibinfo {pages} {165419} (\bibinfo {year} {2022})}\BibitemShut
  {NoStop}%
\bibitem [{\citenamefont {Kezerashvili}\ and\ \citenamefont
  {Kezerashvili}(2022)}]{kezerashvili2022charge}%
  \BibitemOpen
  \bibfield  {author} {\bibinfo {author} {\bibfnamefont {R.~Y.}\ \bibnamefont
  {Kezerashvili}}\ and\ \bibinfo {author} {\bibfnamefont {V.~Y.}\ \bibnamefont
  {Kezerashvili}},\ }\bibfield  {title} {\bibinfo {title} {Charge-dipole and
  dipole-dipole interactions in two-dimensional materials},\ }\href
  {https://doi.org/10.1103/PhysRevB.105.205416} {\bibfield  {journal} {\bibinfo
   {journal} {Physical Review B}\ }\textbf {\bibinfo {volume} {105}},\ \bibinfo
  {pages} {205416} (\bibinfo {year} {2022})}\BibitemShut {NoStop}%
\bibitem [{\citenamefont {Fulde}\ and\ \citenamefont
  {Ferrell}(1964)}]{fulde1964superconductivity}%
  \BibitemOpen
  \bibfield  {author} {\bibinfo {author} {\bibfnamefont {P.}~\bibnamefont
  {Fulde}}\ and\ \bibinfo {author} {\bibfnamefont {R.~A.}\ \bibnamefont
  {Ferrell}},\ }\bibfield  {title} {\bibinfo {title} {Superconductivity in a
  strong spin-exchange field},\ }\href
  {https://doi.org/10.1103/PhysRev.135.A550} {\bibfield  {journal} {\bibinfo
  {journal} {Physical Review}\ }\textbf {\bibinfo {volume} {135}},\ \bibinfo
  {pages} {A550} (\bibinfo {year} {1964})}\BibitemShut {NoStop}%
\bibitem [{\citenamefont {Larkin}\ and\ \citenamefont
  {Ovchinnikov}(1965)}]{larkin1965inhomogeneous}%
  \BibitemOpen
  \bibfield  {author} {\bibinfo {author} {\bibfnamefont {A.~I.}\ \bibnamefont
  {Larkin}}\ and\ \bibinfo {author} {\bibfnamefont {Y.~N.}\ \bibnamefont
  {Ovchinnikov}},\ }\bibfield  {title} {\bibinfo {title} {Inhomogeneous state
  of superconductors},\ }\href@noop {} {\bibfield  {journal} {\bibinfo
  {journal} {Sov. Phys. JETP}\ }\textbf {\bibinfo {volume} {20}},\ \bibinfo
  {pages} {762} (\bibinfo {year} {1965})}\BibitemShut {NoStop}%
\bibitem [{Note2()}]{Note2}%
  \BibitemOpen
  \bibinfo {note} {Note that here we follow the nomenclature of Refs.\ \cite
  {matsuda2007fulde,quan2010interplay} and make the distinction between the FF
  and LO phases, which are often collectively denoted as the ``FFLO'' phase in
  the literature. This distinction is important because only the LO state
  corresponds to a spatially inhomogeneous state, and furthermore their
  superfluid properties are completely different.}\BibitemShut {Stop}%
\bibitem [{\citenamefont {Lee}\ and\ \citenamefont {Yang}(1957)}]{lee1957many}%
  \BibitemOpen
  \bibfield  {author} {\bibinfo {author} {\bibfnamefont {T.}~\bibnamefont
  {Lee}}\ and\ \bibinfo {author} {\bibfnamefont {C.}~\bibnamefont {Yang}},\
  }\bibfield  {title} {\bibinfo {title} {Many-body problem in quantum mechanics
  and quantum statistical mechanics},\ }\href
  {https://doi.org/10.1103/PhysRev.105.1119} {\bibfield  {journal} {\bibinfo
  {journal} {Physical Review}\ }\textbf {\bibinfo {volume} {105}},\ \bibinfo
  {pages} {1119} (\bibinfo {year} {1957})}\BibitemShut {NoStop}%
\bibitem [{\citenamefont {Sep{\'u}lveda}\ \emph {et~al.}(2008)\citenamefont
  {Sep{\'u}lveda}, \citenamefont {Josserand},\ and\ \citenamefont
  {Rica}}]{sepulveda2008nonclassical}%
  \BibitemOpen
  \bibfield  {author} {\bibinfo {author} {\bibfnamefont {N.}~\bibnamefont
  {Sep{\'u}lveda}}, \bibinfo {author} {\bibfnamefont {C.}~\bibnamefont
  {Josserand}},\ and\ \bibinfo {author} {\bibfnamefont {S.}~\bibnamefont
  {Rica}},\ }\bibfield  {title} {\bibinfo {title} {Nonclassical rotational
  inertia fraction in a one-dimensional model of a supersolid},\ }\href
  {https://doi.org/10.1103/PhysRevB.77.054513} {\bibfield  {journal} {\bibinfo
  {journal} {Physical Review B}\ }\textbf {\bibinfo {volume} {77}},\ \bibinfo
  {pages} {054513} (\bibinfo {year} {2008})}\BibitemShut {NoStop}%
\bibitem [{\citenamefont {Sepulveda}\ \emph {et~al.}(2010)\citenamefont
  {Sepulveda}, \citenamefont {Josserand},\ and\ \citenamefont
  {Rica}}]{sepulveda2010superfluid}%
  \BibitemOpen
  \bibfield  {author} {\bibinfo {author} {\bibfnamefont {N.}~\bibnamefont
  {Sepulveda}}, \bibinfo {author} {\bibfnamefont {C.}~\bibnamefont
  {Josserand}},\ and\ \bibinfo {author} {\bibfnamefont {S.}~\bibnamefont
  {Rica}},\ }\bibfield  {title} {\bibinfo {title} {Superfluid density in a
  two-dimensional model of supersolid},\ }\href
  {https://doi.org/10.1140/epjb/e2010-10176-y} {\bibfield  {journal} {\bibinfo
  {journal} {The European Physical Journal B}\ }\textbf {\bibinfo {volume}
  {78}},\ \bibinfo {pages} {439} (\bibinfo {year} {2010})}\BibitemShut
  {NoStop}%
\bibitem [{\citenamefont {Yang}\ and\ \citenamefont
  {Sachdev}(2006)}]{yang2006quantum}%
  \BibitemOpen
  \bibfield  {author} {\bibinfo {author} {\bibfnamefont {K.}~\bibnamefont
  {Yang}}\ and\ \bibinfo {author} {\bibfnamefont {S.}~\bibnamefont {Sachdev}},\
  }\bibfield  {title} {\bibinfo {title} {Quantum criticality of a {F}ermi gas
  with a spherical dispersion minimum},\ }\href
  {https://doi.org/10.1103/PhysRevLett.96.187001} {\bibfield  {journal}
  {\bibinfo  {journal} {Physical Review Letters}\ }\textbf {\bibinfo {volume}
  {96}},\ \bibinfo {pages} {187001} (\bibinfo {year} {2006})}\BibitemShut
  {NoStop}%
\bibitem [{\citenamefont {Stoof}\ \emph {et~al.}(2009)\citenamefont {Stoof},
  \citenamefont {Gubbels},\ and\ \citenamefont
  {Dickerscheid}}]{stoof2009ultracold}%
  \BibitemOpen
  \bibfield  {author} {\bibinfo {author} {\bibfnamefont {H.~T.}\ \bibnamefont
  {Stoof}}, \bibinfo {author} {\bibfnamefont {K.~B.}\ \bibnamefont {Gubbels}},\
  and\ \bibinfo {author} {\bibfnamefont {D.~B.}\ \bibnamefont {Dickerscheid}},\
  }\href {https://doi.org/10.1007/978-1-4020-8763-9} {\emph {\bibinfo {title}
  {Ultracold quantum fields}}}\ (\bibinfo  {publisher} {Springer},\ \bibinfo
  {year} {2009})\BibitemShut {NoStop}%
\bibitem [{\citenamefont {Leggett}(2006)}]{leggett2006quantum}%
  \BibitemOpen
  \bibfield  {author} {\bibinfo {author} {\bibfnamefont {A.~J.}\ \bibnamefont
  {Leggett}},\ }\href
  {https://doi.org/10.1093/acprof:oso/9780198526438.001.0001} {\emph {\bibinfo
  {title} {Quantum liquids: {B}ose condensation and {C}ooper pairing in
  condensed-matter systems}}}\ (\bibinfo  {publisher} {Oxford University
  Press},\ \bibinfo {year} {2006})\BibitemShut {NoStop}%
\bibitem [{\citenamefont {Maiti}\ and\ \citenamefont
  {Sedrakyan}(2019)}]{maiti2019fermionization}%
  \BibitemOpen
  \bibfield  {author} {\bibinfo {author} {\bibfnamefont {S.}~\bibnamefont
  {Maiti}}\ and\ \bibinfo {author} {\bibfnamefont {T.}~\bibnamefont
  {Sedrakyan}},\ }\bibfield  {title} {\bibinfo {title} {Fermionization of
  bosons in a flat band},\ }\href {https://doi.org/10.1103/PhysRevB.99.174418}
  {\bibfield  {journal} {\bibinfo  {journal} {Physical Review B}\ }\textbf
  {\bibinfo {volume} {99}},\ \bibinfo {pages} {174418} (\bibinfo {year}
  {2019})}\BibitemShut {NoStop}%
\bibitem [{\citenamefont {Zhang}\ \emph {et~al.}(1989)\citenamefont {Zhang},
  \citenamefont {Hansson},\ and\ \citenamefont
  {Kivelson}}]{zhang1989effective}%
  \BibitemOpen
  \bibfield  {author} {\bibinfo {author} {\bibfnamefont {S.~C.}\ \bibnamefont
  {Zhang}}, \bibinfo {author} {\bibfnamefont {T.~H.}\ \bibnamefont {Hansson}},\
  and\ \bibinfo {author} {\bibfnamefont {S.}~\bibnamefont {Kivelson}},\
  }\bibfield  {title} {\bibinfo {title} {Effective-field-theory model for the
  fractional quantum {H}all effect},\ }\href
  {https://doi.org/10.1103/PhysRevLett.62.82} {\bibfield  {journal} {\bibinfo
  {journal} {Physical Review Letters}\ }\textbf {\bibinfo {volume} {62}},\
  \bibinfo {pages} {82} (\bibinfo {year} {1989})}\BibitemShut {NoStop}%
\bibitem [{\citenamefont {Lopez}\ and\ \citenamefont
  {Fradkin}(1991)}]{lopez1991fractional}%
  \BibitemOpen
  \bibfield  {author} {\bibinfo {author} {\bibfnamefont {A.}~\bibnamefont
  {Lopez}}\ and\ \bibinfo {author} {\bibfnamefont {E.}~\bibnamefont
  {Fradkin}},\ }\bibfield  {title} {\bibinfo {title} {Fractional quantum {H}all
  effect and {C}hern-{S}imons gauge theories},\ }\href
  {https://doi.org/10.1103/PhysRevB.44.5246} {\bibfield  {journal} {\bibinfo
  {journal} {Physical Review B}\ }\textbf {\bibinfo {volume} {44}},\ \bibinfo
  {pages} {5246} (\bibinfo {year} {1991})}\BibitemShut {NoStop}%
\bibitem [{\citenamefont {Zhang}(1992)}]{zhang1992chern}%
  \BibitemOpen
  \bibfield  {author} {\bibinfo {author} {\bibfnamefont {S.~C.}\ \bibnamefont
  {Zhang}},\ }\bibfield  {title} {\bibinfo {title} {The
  {C}hern--{S}imons--{L}andau--{G}inzburg theory of the fractional quantum
  {H}all effect},\ }\href {https://doi.org/10.1142/S0217979292000037}
  {\bibfield  {journal} {\bibinfo  {journal} {International Journal of Modern
  Physics B}\ }\textbf {\bibinfo {volume} {6}},\ \bibinfo {pages} {25}
  (\bibinfo {year} {1992})}\BibitemShut {NoStop}%
\bibitem [{\citenamefont {Wang}\ \emph {et~al.}(2021)\citenamefont {Wang},
  \citenamefont {Shi}, \citenamefont {Shih}, \citenamefont {Zhou},
  \citenamefont {Wu}, \citenamefont {Bai}, \citenamefont {Rhodes},
  \citenamefont {Barmak}, \citenamefont {Hone}, \citenamefont {Dean} \emph
  {et~al.}}]{wang2021diffusivity}%
  \BibitemOpen
  \bibfield  {author} {\bibinfo {author} {\bibfnamefont {J.}~\bibnamefont
  {Wang}}, \bibinfo {author} {\bibfnamefont {Q.}~\bibnamefont {Shi}}, \bibinfo
  {author} {\bibfnamefont {E.-M.}\ \bibnamefont {Shih}}, \bibinfo {author}
  {\bibfnamefont {L.}~\bibnamefont {Zhou}}, \bibinfo {author} {\bibfnamefont
  {W.}~\bibnamefont {Wu}}, \bibinfo {author} {\bibfnamefont {Y.}~\bibnamefont
  {Bai}}, \bibinfo {author} {\bibfnamefont {D.}~\bibnamefont {Rhodes}},
  \bibinfo {author} {\bibfnamefont {K.}~\bibnamefont {Barmak}}, \bibinfo
  {author} {\bibfnamefont {J.}~\bibnamefont {Hone}}, \bibinfo {author}
  {\bibfnamefont {C.~R.}\ \bibnamefont {Dean}}, \emph {et~al.},\ }\bibfield
  {title} {\bibinfo {title} {Diffusivity reveals three distinct phases of
  interlayer excitons in \ce{MoSe2}/\ce{WSe2} heterobilayers},\ }\href
  {https://doi.org/10.1103/PhysRevLett.126.106804} {\bibfield  {journal}
  {\bibinfo  {journal} {Physical Review Letters}\ }\textbf {\bibinfo {volume}
  {126}},\ \bibinfo {pages} {106804} (\bibinfo {year} {2021})}\BibitemShut
  {NoStop}%
\bibitem [{\citenamefont {Qi}\ \emph {et~al.}(2022)\citenamefont {Qi},
  \citenamefont {Dai}, \citenamefont {Luo}, \citenamefont {Tao}, \citenamefont
  {Qian}, \citenamefont {Zhang}, \citenamefont {Zhang}, \citenamefont {Zhang},
  \citenamefont {Lin}, \citenamefont {Liu} \emph {et~al.}}]{qi2022molding}%
  \BibitemOpen
  \bibfield  {author} {\bibinfo {author} {\bibfnamefont {P.}~\bibnamefont
  {Qi}}, \bibinfo {author} {\bibfnamefont {Y.}~\bibnamefont {Dai}}, \bibinfo
  {author} {\bibfnamefont {Y.}~\bibnamefont {Luo}}, \bibinfo {author}
  {\bibfnamefont {G.}~\bibnamefont {Tao}}, \bibinfo {author} {\bibfnamefont
  {W.}~\bibnamefont {Qian}}, \bibinfo {author} {\bibfnamefont {Z.}~\bibnamefont
  {Zhang}}, \bibinfo {author} {\bibfnamefont {Z.}~\bibnamefont {Zhang}},
  \bibinfo {author} {\bibfnamefont {T.~H.}\ \bibnamefont {Zhang}}, \bibinfo
  {author} {\bibfnamefont {L.}~\bibnamefont {Lin}}, \bibinfo {author}
  {\bibfnamefont {W.}~\bibnamefont {Liu}}, \emph {et~al.},\ }\bibfield  {title}
  {\bibinfo {title} {Molding 2d exciton flux toward room temperature excitonic
  devices},\ }\href {https://doi.org/10.1002/admt.202200032} {\bibfield
  {journal} {\bibinfo  {journal} {Advanced Materials Technologies}\ }\textbf
  {\bibinfo {volume} {7}},\ \bibinfo {pages} {2200032} (\bibinfo {year}
  {2022})}\BibitemShut {NoStop}%
\bibitem [{\citenamefont {Bai}\ \emph {et~al.}(2023)\citenamefont {Bai},
  \citenamefont {Li}, \citenamefont {Liu}, \citenamefont {Guo}, \citenamefont
  {Pack}, \citenamefont {Wang}, \citenamefont {Dean}, \citenamefont {Hone},\
  and\ \citenamefont {Zhu}}]{bai2023evidence}%
  \BibitemOpen
  \bibfield  {author} {\bibinfo {author} {\bibfnamefont {Y.}~\bibnamefont
  {Bai}}, \bibinfo {author} {\bibfnamefont {Y.}~\bibnamefont {Li}}, \bibinfo
  {author} {\bibfnamefont {S.}~\bibnamefont {Liu}}, \bibinfo {author}
  {\bibfnamefont {Y.}~\bibnamefont {Guo}}, \bibinfo {author} {\bibfnamefont
  {J.}~\bibnamefont {Pack}}, \bibinfo {author} {\bibfnamefont {J.}~\bibnamefont
  {Wang}}, \bibinfo {author} {\bibfnamefont {C.~R.}\ \bibnamefont {Dean}},
  \bibinfo {author} {\bibfnamefont {J.}~\bibnamefont {Hone}},\ and\ \bibinfo
  {author} {\bibfnamefont {X.}~\bibnamefont {Zhu}},\ }\bibfield  {title}
  {\bibinfo {title} {Evidence for exciton crystals in a {2D} semiconductor
  heterotrilayer},\ }\href {https://doi.org/10.1021/acs.nanolett.3c03453}
  {\bibfield  {journal} {\bibinfo  {journal} {Nano Letters}\ }\textbf {\bibinfo
  {volume} {23}},\ \bibinfo {pages} {11621} (\bibinfo {year}
  {2023})}\BibitemShut {NoStop}%
\bibitem [{\citenamefont {Laikhtman}\ and\ \citenamefont
  {Rapaport}(2009)}]{laikhtman2009exciton}%
  \BibitemOpen
  \bibfield  {author} {\bibinfo {author} {\bibfnamefont {B.}~\bibnamefont
  {Laikhtman}}\ and\ \bibinfo {author} {\bibfnamefont {R.}~\bibnamefont
  {Rapaport}},\ }\bibfield  {title} {\bibinfo {title} {Exciton correlations in
  coupled quantum wells and their luminescence blue shift},\ }\href
  {https://doi.org/10.1103/PhysRevB.80.195313} {\bibfield  {journal} {\bibinfo
  {journal} {Physical Review B}\ }\textbf {\bibinfo {volume} {80}},\ \bibinfo
  {pages} {195313} (\bibinfo {year} {2009})}\BibitemShut {NoStop}%
\bibitem [{\citenamefont {Jiang}\ \emph {et~al.}(2021)\citenamefont {Jiang},
  \citenamefont {Chen}, \citenamefont {Zheng}, \citenamefont {Zheng},\ and\
  \citenamefont {Pan}}]{jiang2021interlayer}%
  \BibitemOpen
  \bibfield  {author} {\bibinfo {author} {\bibfnamefont {Y.}~\bibnamefont
  {Jiang}}, \bibinfo {author} {\bibfnamefont {S.}~\bibnamefont {Chen}},
  \bibinfo {author} {\bibfnamefont {W.}~\bibnamefont {Zheng}}, \bibinfo
  {author} {\bibfnamefont {B.}~\bibnamefont {Zheng}},\ and\ \bibinfo {author}
  {\bibfnamefont {A.}~\bibnamefont {Pan}},\ }\bibfield  {title} {\bibinfo
  {title} {Interlayer exciton formation, relaxation, and transport in {TMD} van
  der waals heterostructures},\ }\href
  {https://doi.org/10.1038/s41377-021-00500-1} {\bibfield  {journal} {\bibinfo
  {journal} {Light: Science \& Applications}\ }\textbf {\bibinfo {volume}
  {10}},\ \bibinfo {pages} {72} (\bibinfo {year} {2021})}\BibitemShut {NoStop}%
\bibitem [{\citenamefont {Chen}\ \emph {et~al.}(2025)\citenamefont {Chen},
  \citenamefont {Lagarde}, \citenamefont {Hemmen}, \citenamefont {Lombez},
  \citenamefont {Renucci}, \citenamefont {Mauguet}, \citenamefont {Ren},
  \citenamefont {Robert}, \citenamefont {Grandjean},\ and\ \citenamefont
  {Marie}}]{chen2025interlayer}%
  \BibitemOpen
  \bibfield  {author} {\bibinfo {author} {\bibfnamefont {D.}~\bibnamefont
  {Chen}}, \bibinfo {author} {\bibfnamefont {D.}~\bibnamefont {Lagarde}},
  \bibinfo {author} {\bibfnamefont {L.}~\bibnamefont {Hemmen}}, \bibinfo
  {author} {\bibfnamefont {L.}~\bibnamefont {Lombez}}, \bibinfo {author}
  {\bibfnamefont {P.}~\bibnamefont {Renucci}}, \bibinfo {author} {\bibfnamefont
  {M.}~\bibnamefont {Mauguet}}, \bibinfo {author} {\bibfnamefont
  {L.}~\bibnamefont {Ren}}, \bibinfo {author} {\bibfnamefont {C.}~\bibnamefont
  {Robert}}, \bibinfo {author} {\bibfnamefont {N.}~\bibnamefont {Grandjean}},\
  and\ \bibinfo {author} {\bibfnamefont {X.}~\bibnamefont {Marie}},\ }\bibfield
   {title} {\bibinfo {title} {Interlayer coupling and exciton dynamics in
  two-dimensional hybrid structures based on an (\ce{In}, \ce{Ga})\ce{N}
  quantum well coupled to a \ce{MoSe2} monolayer},\ }\href
  {https://doi.org/10.1103/xmk4-zy5q} {\bibfield  {journal} {\bibinfo
  {journal} {Physical Review Applied}\ }\textbf {\bibinfo {volume} {24}},\
  \bibinfo {pages} {034021} (\bibinfo {year} {2025})}\BibitemShut {NoStop}%
\bibitem [{\citenamefont {Liu}\ \emph {et~al.}(2025)\citenamefont {Liu},
  \citenamefont {Tao}, \citenamefont {Luo},\ and\ \citenamefont
  {Chakraborty}}]{liu2025interlayer}%
  \BibitemOpen
  \bibfield  {author} {\bibinfo {author} {\bibfnamefont {X.}~\bibnamefont
  {Liu}}, \bibinfo {author} {\bibfnamefont {Z.}~\bibnamefont {Tao}}, \bibinfo
  {author} {\bibfnamefont {W.}~\bibnamefont {Luo}},\ and\ \bibinfo {author}
  {\bibfnamefont {T.}~\bibnamefont {Chakraborty}},\ }\bibfield  {title}
  {\bibinfo {title} {Interlayer excitons in double-layer transition metal
  dichalcogenide quantum dots},\ }\href
  {https://doi.org/10.1103/PhysRevB.111.085424} {\bibfield  {journal} {\bibinfo
   {journal} {Physical Review B}\ }\textbf {\bibinfo {volume} {111}},\ \bibinfo
  {pages} {085424} (\bibinfo {year} {2025})}\BibitemShut {NoStop}%
\bibitem [{\citenamefont {Zhou}\ \emph {et~al.}(2026)\citenamefont {Zhou},
  \citenamefont {Szwed}, \citenamefont {Brunner}, \citenamefont {Henstridge},
  \citenamefont {Fowler-Gerace},\ and\ \citenamefont {Butov}}]{zhou2026long}%
  \BibitemOpen
  \bibfield  {author} {\bibinfo {author} {\bibfnamefont {Z.}~\bibnamefont
  {Zhou}}, \bibinfo {author} {\bibfnamefont {E.}~\bibnamefont {Szwed}},
  \bibinfo {author} {\bibfnamefont {W.}~\bibnamefont {Brunner}}, \bibinfo
  {author} {\bibfnamefont {H.}~\bibnamefont {Henstridge}}, \bibinfo {author}
  {\bibfnamefont {L.}~\bibnamefont {Fowler-Gerace}},\ and\ \bibinfo {author}
  {\bibfnamefont {L.}~\bibnamefont {Butov}},\ }\bibfield  {title} {\bibinfo
  {title} {Long-range spatial extension of exciton states in van der {W}aals
  heterostructure},\ }\href {https://doi.org/10.1038/s41467-026-70218-4}
  {\bibfield  {journal} {\bibinfo  {journal} {Nature Communications}\ }\textbf
  {\bibinfo {volume} {17}},\ \bibinfo {pages} {3503} (\bibinfo {year}
  {2026})}\BibitemShut {NoStop}%
\bibitem [{\citenamefont {Humphreys}\ \emph {et~al.}(1978)\citenamefont
  {Humphreys}, \citenamefont {R{\"o}ssler},\ and\ \citenamefont
  {Cardona}}]{humphreys1978indirect}%
  \BibitemOpen
  \bibfield  {author} {\bibinfo {author} {\bibfnamefont {R.}~\bibnamefont
  {Humphreys}}, \bibinfo {author} {\bibfnamefont {U.}~\bibnamefont
  {R{\"o}ssler}},\ and\ \bibinfo {author} {\bibfnamefont {M.}~\bibnamefont
  {Cardona}},\ }\bibfield  {title} {\bibinfo {title} {Indirect exciton fine
  structure in \ce{GaP} and the effect of uniaxial stress},\ }\href
  {https://doi.org/10.1103/PhysRevB.18.5590} {\bibfield  {journal} {\bibinfo
  {journal} {Physical Review B}\ }\textbf {\bibinfo {volume} {18}},\ \bibinfo
  {pages} {5590} (\bibinfo {year} {1978})}\BibitemShut {NoStop}%
\bibitem [{\citenamefont {Efros}\ and\ \citenamefont
  {Gelmont}(1984)}]{efros1984exciton}%
  \BibitemOpen
  \bibfield  {author} {\bibinfo {author} {\bibfnamefont {A.~L.}\ \bibnamefont
  {Efros}}\ and\ \bibinfo {author} {\bibfnamefont {B.}~\bibnamefont
  {Gelmont}},\ }\bibfield  {title} {\bibinfo {title} {Exciton dispersion law in
  diamond-like semiconductors},\ }\href
  {https://doi.org/10.1016/0038-1098(84)90445-9} {\bibfield  {journal}
  {\bibinfo  {journal} {Solid State Communications}\ }\textbf {\bibinfo
  {volume} {49}},\ \bibinfo {pages} {883} (\bibinfo {year} {1984})}\BibitemShut
  {NoStop}%
\bibitem [{\citenamefont {Siarkos}\ \emph {et~al.}(2000)\citenamefont
  {Siarkos}, \citenamefont {Runge},\ and\ \citenamefont
  {Zimmermann}}]{siarkos2000center}%
  \BibitemOpen
  \bibfield  {author} {\bibinfo {author} {\bibfnamefont {A.}~\bibnamefont
  {Siarkos}}, \bibinfo {author} {\bibfnamefont {E.}~\bibnamefont {Runge}},\
  and\ \bibinfo {author} {\bibfnamefont {R.}~\bibnamefont {Zimmermann}},\
  }\bibfield  {title} {\bibinfo {title} {Center-of-mass properties of the
  exciton in quantum wells},\ }\href
  {https://doi.org/10.1103/PhysRevB.61.10854} {\bibfield  {journal} {\bibinfo
  {journal} {Physical Review B}\ }\textbf {\bibinfo {volume} {61}},\ \bibinfo
  {pages} {10854} (\bibinfo {year} {2000})}\BibitemShut {NoStop}%
\bibitem [{\citenamefont {Takeda}\ \emph {et~al.}(2005)\citenamefont {Takeda},
  \citenamefont {Higashi},\ and\ \citenamefont
  {Daimon}}]{takeda2005visualization}%
  \BibitemOpen
  \bibfield  {author} {\bibinfo {author} {\bibfnamefont {S.~N.}\ \bibnamefont
  {Takeda}}, \bibinfo {author} {\bibfnamefont {N.}~\bibnamefont {Higashi}},\
  and\ \bibinfo {author} {\bibfnamefont {H.}~\bibnamefont {Daimon}},\
  }\bibfield  {title} {\bibinfo {title} {Visualization of in-plane dispersion
  of hole subbands by photoelectron spectroscopy},\ }\href
  {https://doi.org/10.1103/PhysRevLett.94.037401} {\bibfield  {journal}
  {\bibinfo  {journal} {Physical Review Letters}\ }\textbf {\bibinfo {volume}
  {94}},\ \bibinfo {pages} {037401} (\bibinfo {year} {2005})}\BibitemShut
  {NoStop}%
\bibitem [{\citenamefont {Minkov}\ \emph {et~al.}(2013)\citenamefont {Minkov},
  \citenamefont {Germanenko}, \citenamefont {Rut}, \citenamefont
  {Sherstobitov}, \citenamefont {Dvoretski},\ and\ \citenamefont
  {Mikhailov}}]{minkov2013two}%
  \BibitemOpen
  \bibfield  {author} {\bibinfo {author} {\bibfnamefont {G.}~\bibnamefont
  {Minkov}}, \bibinfo {author} {\bibfnamefont {A.}~\bibnamefont {Germanenko}},
  \bibinfo {author} {\bibfnamefont {O.}~\bibnamefont {Rut}}, \bibinfo {author}
  {\bibfnamefont {A.}~\bibnamefont {Sherstobitov}}, \bibinfo {author}
  {\bibfnamefont {S.}~\bibnamefont {Dvoretski}},\ and\ \bibinfo {author}
  {\bibfnamefont {N.}~\bibnamefont {Mikhailov}},\ }\bibfield  {title} {\bibinfo
  {title} {Two-dimensional semimetal in a wide \ce{HgTe} quantum well:
  {M}agnetotransport and energy spectrum},\ }\href
  {https://doi.org/10.1103/PhysRevB.88.155306} {\bibfield  {journal} {\bibinfo
  {journal} {Physical Review B}\ }\textbf {\bibinfo {volume} {88}},\ \bibinfo
  {pages} {155306} (\bibinfo {year} {2013})}\BibitemShut {NoStop}%
\bibitem [{\citenamefont {Minkov}\ \emph {et~al.}(2014)\citenamefont {Minkov},
  \citenamefont {Germanenko}, \citenamefont {Rut}, \citenamefont
  {Sherstobitov}, \citenamefont {Dvoretski},\ and\ \citenamefont
  {Mikhailov}}]{minkov2014hole}%
  \BibitemOpen
  \bibfield  {author} {\bibinfo {author} {\bibfnamefont {G.}~\bibnamefont
  {Minkov}}, \bibinfo {author} {\bibfnamefont {A.}~\bibnamefont {Germanenko}},
  \bibinfo {author} {\bibfnamefont {O.}~\bibnamefont {Rut}}, \bibinfo {author}
  {\bibfnamefont {A.}~\bibnamefont {Sherstobitov}}, \bibinfo {author}
  {\bibfnamefont {S.}~\bibnamefont {Dvoretski}},\ and\ \bibinfo {author}
  {\bibfnamefont {N.}~\bibnamefont {Mikhailov}},\ }\bibfield  {title} {\bibinfo
  {title} {Hole transport and valence-band dispersion law in a \ce{HgTe}
  quantum well with a normal energy spectrum},\ }\href
  {https://doi.org/10.1103/PhysRevB.89.165311} {\bibfield  {journal} {\bibinfo
  {journal} {Physical Review B}\ }\textbf {\bibinfo {volume} {89}},\ \bibinfo
  {pages} {165311} (\bibinfo {year} {2014})}\BibitemShut {NoStop}%
\bibitem [{\citenamefont {Ceferino}\ \emph {et~al.}(2020)\citenamefont
  {Ceferino}, \citenamefont {Song}, \citenamefont {Magorrian}, \citenamefont
  {Z{\'o}lyomi},\ and\ \citenamefont {Fal'ko}}]{ceferino2020crossover}%
  \BibitemOpen
  \bibfield  {author} {\bibinfo {author} {\bibfnamefont {A.}~\bibnamefont
  {Ceferino}}, \bibinfo {author} {\bibfnamefont {K.~W.}\ \bibnamefont {Song}},
  \bibinfo {author} {\bibfnamefont {S.~J.}\ \bibnamefont {Magorrian}}, \bibinfo
  {author} {\bibfnamefont {V.}~\bibnamefont {Z{\'o}lyomi}},\ and\ \bibinfo
  {author} {\bibfnamefont {V.~I.}\ \bibnamefont {Fal'ko}},\ }\bibfield  {title}
  {\bibinfo {title} {Crossover from weakly indirect to direct excitons in
  atomically thin films of \ce{InSe}},\ }\href
  {https://doi.org/10.1103/PhysRevB.101.245432} {\bibfield  {journal} {\bibinfo
   {journal} {Physical Review B}\ }\textbf {\bibinfo {volume} {101}},\ \bibinfo
  {pages} {245432} (\bibinfo {year} {2020})}\BibitemShut {NoStop}%
\bibitem [{\citenamefont {Golub}\ \emph {et~al.}(1990)\citenamefont {Golub},
  \citenamefont {Kash}, \citenamefont {Harbison},\ and\ \citenamefont
  {Florez}}]{golub1990long}%
  \BibitemOpen
  \bibfield  {author} {\bibinfo {author} {\bibfnamefont {J.~E.}\ \bibnamefont
  {Golub}}, \bibinfo {author} {\bibfnamefont {K.}~\bibnamefont {Kash}},
  \bibinfo {author} {\bibfnamefont {J.~P.}\ \bibnamefont {Harbison}},\ and\
  \bibinfo {author} {\bibfnamefont {L.~T.}\ \bibnamefont {Florez}},\ }\bibfield
   {title} {\bibinfo {title} {Long-lived spatially indirect excitons in coupled
  \ce{GaAs}/\ce{Al_{x}Ga_{1-x}As} quantum wells},\ }\href
  {https://doi.org/10.1103/PhysRevB.41.8564} {\bibfield  {journal} {\bibinfo
  {journal} {Physical Review B}\ }\textbf {\bibinfo {volume} {41}},\ \bibinfo
  {pages} {8564} (\bibinfo {year} {1990})}\BibitemShut {NoStop}%
\bibitem [{\citenamefont {High}\ \emph {et~al.}(2008)\citenamefont {High},
  \citenamefont {Novitskaya}, \citenamefont {Butov}, \citenamefont {Hanson},\
  and\ \citenamefont {Gossard}}]{high2008control}%
  \BibitemOpen
  \bibfield  {author} {\bibinfo {author} {\bibfnamefont {A.~A.}\ \bibnamefont
  {High}}, \bibinfo {author} {\bibfnamefont {E.~E.}\ \bibnamefont
  {Novitskaya}}, \bibinfo {author} {\bibfnamefont {L.~V.}\ \bibnamefont
  {Butov}}, \bibinfo {author} {\bibfnamefont {M.}~\bibnamefont {Hanson}},\ and\
  \bibinfo {author} {\bibfnamefont {A.~C.}\ \bibnamefont {Gossard}},\
  }\bibfield  {title} {\bibinfo {title} {Control of exciton fluxes in an
  excitonic integrated circuit},\ }\href
  {https://doi.org/10.1126/science.1157845} {\bibfield  {journal} {\bibinfo
  {journal} {Science}\ }\textbf {\bibinfo {volume} {321}},\ \bibinfo {pages}
  {229} (\bibinfo {year} {2008})}\BibitemShut {NoStop}%
\bibitem [{\citenamefont {High}\ \emph {et~al.}(2012)\citenamefont {High},
  \citenamefont {Leonard}, \citenamefont {Hammack}, \citenamefont {Fogler},
  \citenamefont {Butov}, \citenamefont {Kavokin}, \citenamefont {Campman},\
  and\ \citenamefont {Gossard}}]{high2012spontaneous}%
  \BibitemOpen
  \bibfield  {author} {\bibinfo {author} {\bibfnamefont {A.~A.}\ \bibnamefont
  {High}}, \bibinfo {author} {\bibfnamefont {J.~R.}\ \bibnamefont {Leonard}},
  \bibinfo {author} {\bibfnamefont {A.~T.}\ \bibnamefont {Hammack}}, \bibinfo
  {author} {\bibfnamefont {M.~M.}\ \bibnamefont {Fogler}}, \bibinfo {author}
  {\bibfnamefont {L.~V.}\ \bibnamefont {Butov}}, \bibinfo {author}
  {\bibfnamefont {A.~V.}\ \bibnamefont {Kavokin}}, \bibinfo {author}
  {\bibfnamefont {K.~L.}\ \bibnamefont {Campman}},\ and\ \bibinfo {author}
  {\bibfnamefont {A.~C.}\ \bibnamefont {Gossard}},\ }\bibfield  {title}
  {\bibinfo {title} {Spontaneous coherence in a cold exciton gas},\ }\href
  {https://doi.org/10.1038/nature10903} {\bibfield  {journal} {\bibinfo
  {journal} {Nature}\ }\textbf {\bibinfo {volume} {483}},\ \bibinfo {pages}
  {584} (\bibinfo {year} {2012})}\BibitemShut {NoStop}%
\bibitem [{\citenamefont {Robert}\ \emph {et~al.}(2017)\citenamefont {Robert},
  \citenamefont {Amand}, \citenamefont {Cadiz}, \citenamefont {Lagarde},
  \citenamefont {Courtade}, \citenamefont {Manca}, \citenamefont {Taniguchi},
  \citenamefont {Watanabe}, \citenamefont {Urbaszek},\ and\ \citenamefont
  {Marie}}]{robert2017fine}%
  \BibitemOpen
  \bibfield  {author} {\bibinfo {author} {\bibfnamefont {C.}~\bibnamefont
  {Robert}}, \bibinfo {author} {\bibfnamefont {T.}~\bibnamefont {Amand}},
  \bibinfo {author} {\bibfnamefont {F.}~\bibnamefont {Cadiz}}, \bibinfo
  {author} {\bibfnamefont {D.}~\bibnamefont {Lagarde}}, \bibinfo {author}
  {\bibfnamefont {E.}~\bibnamefont {Courtade}}, \bibinfo {author}
  {\bibfnamefont {M.}~\bibnamefont {Manca}}, \bibinfo {author} {\bibfnamefont
  {T.}~\bibnamefont {Taniguchi}}, \bibinfo {author} {\bibfnamefont
  {K.}~\bibnamefont {Watanabe}}, \bibinfo {author} {\bibfnamefont
  {B.}~\bibnamefont {Urbaszek}},\ and\ \bibinfo {author} {\bibfnamefont
  {X.}~\bibnamefont {Marie}},\ }\bibfield  {title} {\bibinfo {title} {Fine
  structure and lifetime of dark excitons in transition metal dichalcogenide
  monolayers},\ }\href {https://doi.org/10.1103/PhysRevB.96.155423} {\bibfield
  {journal} {\bibinfo  {journal} {Physical review B}\ }\textbf {\bibinfo
  {volume} {96}},\ \bibinfo {pages} {155423} (\bibinfo {year}
  {2017})}\BibitemShut {NoStop}%
\bibitem [{\citenamefont {Schmidt}\ \emph {et~al.}(2019)\citenamefont
  {Schmidt}, \citenamefont {Berger}, \citenamefont {Kahlert}, \citenamefont
  {Bayer}, \citenamefont {Schneider}, \citenamefont {H{\"o}fling},
  \citenamefont {Sedov}, \citenamefont {Kavokin},\ and\ \citenamefont
  {A{\ss}mann}}]{schmidt2019tracking}%
  \BibitemOpen
  \bibfield  {author} {\bibinfo {author} {\bibfnamefont {D.}~\bibnamefont
  {Schmidt}}, \bibinfo {author} {\bibfnamefont {B.}~\bibnamefont {Berger}},
  \bibinfo {author} {\bibfnamefont {M.}~\bibnamefont {Kahlert}}, \bibinfo
  {author} {\bibfnamefont {M.}~\bibnamefont {Bayer}}, \bibinfo {author}
  {\bibfnamefont {C.}~\bibnamefont {Schneider}}, \bibinfo {author}
  {\bibfnamefont {S.}~\bibnamefont {H{\"o}fling}}, \bibinfo {author}
  {\bibfnamefont {E.}~\bibnamefont {Sedov}}, \bibinfo {author} {\bibfnamefont
  {A.}~\bibnamefont {Kavokin}},\ and\ \bibinfo {author} {\bibfnamefont
  {M.}~\bibnamefont {A{\ss}mann}},\ }\bibfield  {title} {\bibinfo {title}
  {Tracking dark excitons with exciton polaritons in semiconductor
  microcavities},\ }\href {https://doi.org/10.1103/PhysRevLett.122.047403}
  {\bibfield  {journal} {\bibinfo  {journal} {Physical Review Letters}\
  }\textbf {\bibinfo {volume} {122}},\ \bibinfo {pages} {047403} (\bibinfo
  {year} {2019})}\BibitemShut {NoStop}%
\bibitem [{\citenamefont {Chand}\ \emph {et~al.}(2023)\citenamefont {Chand},
  \citenamefont {Woods}, \citenamefont {Quan}, \citenamefont {Mejia},
  \citenamefont {Taniguchi}, \citenamefont {Watanabe}, \citenamefont
  {Al{\`u}},\ and\ \citenamefont {Grosso}}]{chand2023interaction}%
  \BibitemOpen
  \bibfield  {author} {\bibinfo {author} {\bibfnamefont {S.~B.}\ \bibnamefont
  {Chand}}, \bibinfo {author} {\bibfnamefont {J.~M.}\ \bibnamefont {Woods}},
  \bibinfo {author} {\bibfnamefont {J.}~\bibnamefont {Quan}}, \bibinfo {author}
  {\bibfnamefont {E.}~\bibnamefont {Mejia}}, \bibinfo {author} {\bibfnamefont
  {T.}~\bibnamefont {Taniguchi}}, \bibinfo {author} {\bibfnamefont
  {K.}~\bibnamefont {Watanabe}}, \bibinfo {author} {\bibfnamefont
  {A.}~\bibnamefont {Al{\`u}}},\ and\ \bibinfo {author} {\bibfnamefont
  {G.}~\bibnamefont {Grosso}},\ }\bibfield  {title} {\bibinfo {title}
  {Interaction-driven transport of dark excitons in {2D} semiconductors with
  phonon-mediated optical readout},\ }\href
  {https://doi.org/10.1038/s41467-023-39339-y} {\bibfield  {journal} {\bibinfo
  {journal} {Nature Communications}\ }\textbf {\bibinfo {volume} {14}},\
  \bibinfo {pages} {3712} (\bibinfo {year} {2023})}\BibitemShut {NoStop}%
\bibitem [{\citenamefont {Fisher}\ \emph {et~al.}(1989)\citenamefont {Fisher},
  \citenamefont {Weichman}, \citenamefont {Grinstein},\ and\ \citenamefont
  {Fisher}}]{fisher1989boson}%
  \BibitemOpen
  \bibfield  {author} {\bibinfo {author} {\bibfnamefont {M.~P.~A.}\
  \bibnamefont {Fisher}}, \bibinfo {author} {\bibfnamefont {P.~B.}\
  \bibnamefont {Weichman}}, \bibinfo {author} {\bibfnamefont {G.}~\bibnamefont
  {Grinstein}},\ and\ \bibinfo {author} {\bibfnamefont {D.~S.}\ \bibnamefont
  {Fisher}},\ }\bibfield  {title} {\bibinfo {title} {Boson localization and the
  superfluid-insulator transition},\ }\href
  {https://doi.org/10.1103/PhysRevB.40.546} {\bibfield  {journal} {\bibinfo
  {journal} {Physical Review B}\ }\textbf {\bibinfo {volume} {40}},\ \bibinfo
  {pages} {546} (\bibinfo {year} {1989})}\BibitemShut {NoStop}%
\bibitem [{\citenamefont {Sanchez-Palencia}\ and\ \citenamefont
  {Lewenstein}(2010)}]{sanchez2010disordered}%
  \BibitemOpen
  \bibfield  {author} {\bibinfo {author} {\bibfnamefont {L.}~\bibnamefont
  {Sanchez-Palencia}}\ and\ \bibinfo {author} {\bibfnamefont {M.}~\bibnamefont
  {Lewenstein}},\ }\bibfield  {title} {\bibinfo {title} {Disordered quantum
  gases under control},\ }\href {https://doi.org/10.1038/nphys1507} {\bibfield
  {journal} {\bibinfo  {journal} {Nature Physics}\ }\textbf {\bibinfo {volume}
  {6}},\ \bibinfo {pages} {87} (\bibinfo {year} {2010})}\BibitemShut {NoStop}%
\bibitem [{\citenamefont {Fu}(2009)}]{fu2009hexagonal}%
  \BibitemOpen
  \bibfield  {author} {\bibinfo {author} {\bibfnamefont {L.}~\bibnamefont
  {Fu}},\ }\bibfield  {title} {\bibinfo {title} {Hexagonal warping effects in
  the surface states of the topological insulator \ce{Bi2Te3}},\ }\href
  {https://doi.org/10.1103/PhysRevLett.103.266801} {\bibfield  {journal}
  {\bibinfo  {journal} {Physical Review Letters}\ }\textbf {\bibinfo {volume}
  {103}},\ \bibinfo {pages} {266801} (\bibinfo {year} {2009})}\BibitemShut
  {NoStop}%
\bibitem [{\citenamefont {Liu}\ \emph {et~al.}(2010)\citenamefont {Liu},
  \citenamefont {Qi}, \citenamefont {Zhang}, \citenamefont {Dai}, \citenamefont
  {Fang},\ and\ \citenamefont {Zhang}}]{liu2010model}%
  \BibitemOpen
  \bibfield  {author} {\bibinfo {author} {\bibfnamefont {C.-X.}\ \bibnamefont
  {Liu}}, \bibinfo {author} {\bibfnamefont {X.-L.}\ \bibnamefont {Qi}},
  \bibinfo {author} {\bibfnamefont {H.}~\bibnamefont {Zhang}}, \bibinfo
  {author} {\bibfnamefont {X.}~\bibnamefont {Dai}}, \bibinfo {author}
  {\bibfnamefont {Z.}~\bibnamefont {Fang}},\ and\ \bibinfo {author}
  {\bibfnamefont {S.-C.}\ \bibnamefont {Zhang}},\ }\bibfield  {title} {\bibinfo
  {title} {Model {H}amiltonian for topological insulators},\ }\href
  {https://doi.org/10.1103/PhysRevB.82.045122} {\bibfield  {journal} {\bibinfo
  {journal} {Physical Review B}\ }\textbf {\bibinfo {volume} {82}},\ \bibinfo
  {pages} {045122} (\bibinfo {year} {2010})}\BibitemShut {NoStop}%
\bibitem [{\citenamefont {Brazovski{\u\i}}(1996)}]{brazovskiui1996phase}%
  \BibitemOpen
  \bibfield  {author} {\bibinfo {author} {\bibfnamefont {S.}~\bibnamefont
  {Brazovski{\u\i}}},\ }\bibfield  {title} {\bibinfo {title} {Phase transition
  of an isotropic system to a nonuniform state},\ }in\ \href
  {https://doi.org/10.1142/9789814317344_0016} {\emph {\bibinfo {booktitle} {30
  Years Of The Landau Institute—Selected Papers}}}\ (\bibinfo  {publisher}
  {World Scientific},\ \bibinfo {year} {1996})\ p.\ \bibinfo {pages}
  {109}\BibitemShut {NoStop}%
\bibitem [{\citenamefont {Rabec}\ \emph {et~al.}(2026)\citenamefont {Rabec},
  \citenamefont {Brochier}, \citenamefont {Wattellier}, \citenamefont
  {Chauveau}, \citenamefont {Li}, \citenamefont {Nascimbene}, \citenamefont
  {Dalibard},\ and\ \citenamefont {Beugnon}}]{rabec2026superfluid}%
  \BibitemOpen
  \bibfield  {author} {\bibinfo {author} {\bibfnamefont {F.}~\bibnamefont
  {Rabec}}, \bibinfo {author} {\bibfnamefont {G.}~\bibnamefont {Brochier}},
  \bibinfo {author} {\bibfnamefont {S.}~\bibnamefont {Wattellier}}, \bibinfo
  {author} {\bibfnamefont {G.}~\bibnamefont {Chauveau}}, \bibinfo {author}
  {\bibfnamefont {Y.}~\bibnamefont {Li}}, \bibinfo {author} {\bibfnamefont
  {S.}~\bibnamefont {Nascimbene}}, \bibinfo {author} {\bibfnamefont
  {J.}~\bibnamefont {Dalibard}},\ and\ \bibinfo {author} {\bibfnamefont
  {J.}~\bibnamefont {Beugnon}},\ }\bibfield  {title} {\bibinfo {title}
  {Superfluid {F}raction of a {2D} {B}ose-{E}instein {C}ondensate in a
  {T}riangular {L}attice},\ }\href {https://doi.org/10.1103/g494-rj5k}
  {\bibfield  {journal} {\bibinfo  {journal} {Physical Review Letters}\
  }\textbf {\bibinfo {volume} {136}},\ \bibinfo {pages} {133401} (\bibinfo
  {year} {2026})}\BibitemShut {NoStop}%
\bibitem [{Note3()}]{Note3}%
  \BibitemOpen
  \bibinfo {note} {Note that Leggett's bound in the form given here does not
  universally apply to the moat-band case due to the mixing of the ordering
  vectors and the components of the phase gradient in Eq.\ \protect \eqref
  {eq:approxLWLGradientPlaneWave}. However, for the LO phase in particular, the
  derived expression coincides with the one by Leggett as $n(\protect
  \boldsymbol {r}) \propto \cos ^{2} q_{\protect \mathrm {m}}x$.}\BibitemShut
  {Stop}%
\bibitem [{\citenamefont {Berezinskii}(1971)}]{berezinskii1971destruction}%
  \BibitemOpen
  \bibfield  {author} {\bibinfo {author} {\bibfnamefont {V.}~\bibnamefont
  {Berezinskii}},\ }\bibfield  {title} {\bibinfo {title} {Destruction of
  long-range order in one-dimensional and two-dimensional systems having a
  continuous symmetry group {I}. {C}lassical systems},\ }\href@noop {}
  {\bibfield  {journal} {\bibinfo  {journal} {Sov. Phys. JETP}\ }\textbf
  {\bibinfo {volume} {32}},\ \bibinfo {pages} {493} (\bibinfo {year}
  {1971})}\BibitemShut {NoStop}%
\bibitem [{\citenamefont {Berezinskii}(1972)}]{berezinskii1972destruction}%
  \BibitemOpen
  \bibfield  {author} {\bibinfo {author} {\bibfnamefont {V.}~\bibnamefont
  {Berezinskii}},\ }\bibfield  {title} {\bibinfo {title} {Destruction of
  long-range order in one-dimensional and two-dimensional systems possessing a
  continuous symmetry group. {II}. {Q}uantum systems},\ }\href@noop {}
  {\bibfield  {journal} {\bibinfo  {journal} {Sov. Phys. JETP}\ }\textbf
  {\bibinfo {volume} {34}},\ \bibinfo {pages} {610} (\bibinfo {year}
  {1972})}\BibitemShut {NoStop}%
\bibitem [{\citenamefont {Kosterlitz}\ and\ \citenamefont
  {Thouless}(1973)}]{kosterlitz1973ordering}%
  \BibitemOpen
  \bibfield  {author} {\bibinfo {author} {\bibfnamefont {J.~M.}\ \bibnamefont
  {Kosterlitz}}\ and\ \bibinfo {author} {\bibfnamefont {D.~J.}\ \bibnamefont
  {Thouless}},\ }\bibfield  {title} {\bibinfo {title} {Ordering, metastability
  and phase transitions in two-dimensional systems},\ }\href
  {https://doi.org/10.1088/0022-3719/6/7/010} {\bibfield  {journal} {\bibinfo
  {journal} {Journal of Physics C}\ }\textbf {\bibinfo {volume} {6}},\ \bibinfo
  {pages} {1181} (\bibinfo {year} {1973})}\BibitemShut {NoStop}%
\bibitem [{\citenamefont {Kosterlitz}(1974)}]{kosterlitz1974critical}%
  \BibitemOpen
  \bibfield  {author} {\bibinfo {author} {\bibfnamefont {J.~M.}\ \bibnamefont
  {Kosterlitz}},\ }\bibfield  {title} {\bibinfo {title} {The critical
  properties of the two-dimensional xy model},\ }\href
  {https://doi.org/10.1088/0022-3719/7/6/005} {\bibfield  {journal} {\bibinfo
  {journal} {Journal of Physics C: Solid State Physics}\ }\textbf {\bibinfo
  {volume} {7}},\ \bibinfo {pages} {1046} (\bibinfo {year} {1974})}\BibitemShut
  {NoStop}%
\bibitem [{\citenamefont {Bighin}\ and\ \citenamefont
  {Salasnich}(2018)}]{bighin2018renormalization}%
  \BibitemOpen
  \bibfield  {author} {\bibinfo {author} {\bibfnamefont {G.}~\bibnamefont
  {Bighin}}\ and\ \bibinfo {author} {\bibfnamefont {L.}~\bibnamefont
  {Salasnich}},\ }\bibfield  {title} {\bibinfo {title} {Renormalization of the
  superfluid density in the two-dimensional {BCS-BEC} crossover},\ }\href
  {https://doi.org/10.1142/S0217979218400222} {\bibfield  {journal} {\bibinfo
  {journal} {International Journal of Modern Physics B}\ }\textbf {\bibinfo
  {volume} {32}},\ \bibinfo {pages} {1840022} (\bibinfo {year}
  {2018})}\BibitemShut {NoStop}%
\bibitem [{\citenamefont {Kosterlitz}\ and\ \citenamefont
  {Thouless}(1972)}]{kosterlitz1972long}%
  \BibitemOpen
  \bibfield  {author} {\bibinfo {author} {\bibfnamefont {J.~M.}\ \bibnamefont
  {Kosterlitz}}\ and\ \bibinfo {author} {\bibfnamefont {D.}~\bibnamefont
  {Thouless}},\ }\bibfield  {title} {\bibinfo {title} {Long range order and
  metastability in two dimensional solids and superfluids. ({A}pplication of
  dislocation theory)},\ }\href {https://doi.org/10.1088/0022-3719/5/11/002}
  {\bibfield  {journal} {\bibinfo  {journal} {Journal of Physics C}\ }\textbf
  {\bibinfo {volume} {5}},\ \bibinfo {pages} {L124} (\bibinfo {year}
  {1972})}\BibitemShut {NoStop}%
\bibitem [{\citenamefont {Halperin}\ and\ \citenamefont
  {Nelson}(1978)}]{halperin1978theory}%
  \BibitemOpen
  \bibfield  {author} {\bibinfo {author} {\bibfnamefont {B.}~\bibnamefont
  {Halperin}}\ and\ \bibinfo {author} {\bibfnamefont {D.~R.}\ \bibnamefont
  {Nelson}},\ }\bibfield  {title} {\bibinfo {title} {Theory of two-dimensional
  melting},\ }\href {https://doi.org/10.1103/PhysRevLett.41.121} {\bibfield
  {journal} {\bibinfo  {journal} {Physical Review Letters}\ }\textbf {\bibinfo
  {volume} {41}},\ \bibinfo {pages} {121} (\bibinfo {year} {1978})}\BibitemShut
  {NoStop}%
\bibitem [{\citenamefont {Nelson}\ and\ \citenamefont
  {Halperin}(1979)}]{nelson1979dislocation}%
  \BibitemOpen
  \bibfield  {author} {\bibinfo {author} {\bibfnamefont {D.~R.}\ \bibnamefont
  {Nelson}}\ and\ \bibinfo {author} {\bibfnamefont {B.~I.}\ \bibnamefont
  {Halperin}},\ }\bibfield  {title} {\bibinfo {title} {Dislocation-mediated
  melting in two dimensions},\ }\href
  {https://doi.org/10.1103/PhysRevB.19.2457} {\bibfield  {journal} {\bibinfo
  {journal} {Physical Review B}\ }\textbf {\bibinfo {volume} {19}},\ \bibinfo
  {pages} {2457} (\bibinfo {year} {1979})}\BibitemShut {NoStop}%
\bibitem [{\citenamefont {De~Goey}\ \emph {et~al.}(1986)\citenamefont
  {De~Goey}, \citenamefont {van~den Berg}, \citenamefont {Mulders},
  \citenamefont {Stoof}, \citenamefont {Verhaar},\ and\ \citenamefont
  {Gl{\"o}ckle}}]{degoey1986three}%
  \BibitemOpen
  \bibfield  {author} {\bibinfo {author} {\bibfnamefont {L.}~\bibnamefont
  {De~Goey}}, \bibinfo {author} {\bibfnamefont {T.}~\bibnamefont {van~den
  Berg}}, \bibinfo {author} {\bibfnamefont {N.}~\bibnamefont {Mulders}},
  \bibinfo {author} {\bibfnamefont {H.}~\bibnamefont {Stoof}}, \bibinfo
  {author} {\bibfnamefont {B.}~\bibnamefont {Verhaar}},\ and\ \bibinfo {author}
  {\bibfnamefont {W.}~\bibnamefont {Gl{\"o}ckle}},\ }\bibfield  {title}
  {\bibinfo {title} {Three-body recombination in spin-polarized atomic
  hydrogen},\ }\href {https://doi.org/10.1103/PhysRevB.34.6183} {\bibfield
  {journal} {\bibinfo  {journal} {Physical Review B}\ }\textbf {\bibinfo
  {volume} {34}},\ \bibinfo {pages} {6183} (\bibinfo {year}
  {1986})}\BibitemShut {NoStop}%
\bibitem [{Note4()}]{Note4}%
  \BibitemOpen
  \bibinfo {note} {The reason for writing $\protect \tilde {g}_{1}$ and ${-}
  \protect \tilde {g}_{2}$ as $\pm \protect \tilde {G} / 2$ is that in our
  numerical procedure we obtain the order parameter as $\phi (x) = \protect
  \mathrm {e}^{\protect \mathrm {i}g x} (u_{0} + u_{1} \protect \mathrm
  {e}^{\protect \mathrm {i}G x})$, where $g = {-} G/2$ lies in the first
  Brillouin zone. In this way, the Bloch factor is identified with the term
  inside parentheses, which has the same periodicity as the emerging density
  lattice. Because in the Landau theory we group $g$ and $G$ together,
  ultimately $g_{1}$ and $g_{2}$ must be identified with $\pm G/2$ in order to
  compare with the numerical result.}\BibitemShut {Stop}%
\bibitem [{Note5()}]{Note5}%
  \BibitemOpen
  \bibinfo {note} {Writing the coefficients as $u_{i} = |u_{i}| z_{i}$, where
  $|z_{i}| = 1$, an optimal choice for the phases is $z_{i} = 1$ for all $i$
  when $t_{\protect \tilde {q}_{\protect \mathrm {m}}} + t_{\protect \sqrt {3}
  \protect \tilde {q}_{\protect \mathrm {m}}} < 0$, and $z_{1} z_{4} = 1$,
  $z_{2} z_{5} = \protect \mathrm {e}^{2 \protect \mathrm {i}\pi / 3}$, $z_{3}
  z_{6} = \protect \mathrm {e}^{4 \protect \mathrm {i}\pi / 3}$ when
  $t_{\protect \tilde {q}_{\protect \mathrm {m}}} + t_{\protect \sqrt {3}
  \protect \tilde {q}_{\protect \mathrm {m}}} > 0$. This choice leads to the
  expression of Eq.\ \protect \eqref {eq:fint6}}\BibitemShut {NoStop}%
\bibitem [{\citenamefont {Mkhonta}\ \emph {et~al.}(2013)\citenamefont
  {Mkhonta}, \citenamefont {Elder},\ and\ \citenamefont
  {Huang}}]{mkhonta2013exploring}%
  \BibitemOpen
  \bibfield  {author} {\bibinfo {author} {\bibfnamefont {S.}~\bibnamefont
  {Mkhonta}}, \bibinfo {author} {\bibfnamefont {K.}~\bibnamefont {Elder}},\
  and\ \bibinfo {author} {\bibfnamefont {Z.-F.}\ \bibnamefont {Huang}},\
  }\bibfield  {title} {\bibinfo {title} {Exploring the complex world of
  two-dimensional ordering with three modes},\ }\href
  {https://doi.org/10.1103/PhysRevLett.111.035501} {\bibfield  {journal}
  {\bibinfo  {journal} {Physical Review Letters}\ }\textbf {\bibinfo {volume}
  {111}},\ \bibinfo {pages} {035501} (\bibinfo {year} {2013})}\BibitemShut
  {NoStop}%
\bibitem [{\citenamefont {Heinonen}\ \emph {et~al.}(2019)\citenamefont
  {Heinonen}, \citenamefont {Burns},\ and\ \citenamefont
  {Dunkel}}]{heinonen2019quantum}%
  \BibitemOpen
  \bibfield  {author} {\bibinfo {author} {\bibfnamefont {V.}~\bibnamefont
  {Heinonen}}, \bibinfo {author} {\bibfnamefont {K.~J.}\ \bibnamefont
  {Burns}},\ and\ \bibinfo {author} {\bibfnamefont {J.}~\bibnamefont
  {Dunkel}},\ }\bibfield  {title} {\bibinfo {title} {Quantum hydrodynamics for
  supersolid crystals and quasicrystals},\ }\href
  {https://doi.org/10.1103/PhysRevA.99.063621} {\bibfield  {journal} {\bibinfo
  {journal} {Physical Review A}\ }\textbf {\bibinfo {volume} {99}},\ \bibinfo
  {pages} {063621} (\bibinfo {year} {2019})}\BibitemShut {NoStop}%
\bibitem [{\citenamefont {Mendoza-Coto}\ \emph {et~al.}(2022)\citenamefont
  {Mendoza-Coto}, \citenamefont {Turcati}, \citenamefont {Zampronio},
  \citenamefont {D{\'\i}az-M{\'e}ndez}, \citenamefont {Macr{\`\i}},\ and\
  \citenamefont {Cinti}}]{mendoza2022exploring}%
  \BibitemOpen
  \bibfield  {author} {\bibinfo {author} {\bibfnamefont {A.}~\bibnamefont
  {Mendoza-Coto}}, \bibinfo {author} {\bibfnamefont {R.}~\bibnamefont
  {Turcati}}, \bibinfo {author} {\bibfnamefont {V.}~\bibnamefont {Zampronio}},
  \bibinfo {author} {\bibfnamefont {R.}~\bibnamefont {D{\'\i}az-M{\'e}ndez}},
  \bibinfo {author} {\bibfnamefont {T.}~\bibnamefont {Macr{\`\i}}},\ and\
  \bibinfo {author} {\bibfnamefont {F.}~\bibnamefont {Cinti}},\ }\bibfield
  {title} {\bibinfo {title} {Exploring quantum quasicrystal patterns: {A}
  variational study},\ }\href {https://doi.org/10.1103/PhysRevB.105.134521}
  {\bibfield  {journal} {\bibinfo  {journal} {Physical Review B}\ }\textbf
  {\bibinfo {volume} {105}},\ \bibinfo {pages} {134521} (\bibinfo {year}
  {2022})}\BibitemShut {NoStop}%
\bibitem [{\citenamefont {Grossklags}\ \emph {et~al.}(2024)\citenamefont
  {Grossklags}, \citenamefont {Ciardi}, \citenamefont {Zampronio},
  \citenamefont {Cinti},\ and\ \citenamefont
  {Mendoza-Coto}}]{grossklags2024self}%
  \BibitemOpen
  \bibfield  {author} {\bibinfo {author} {\bibfnamefont {M.}~\bibnamefont
  {Grossklags}}, \bibinfo {author} {\bibfnamefont {M.}~\bibnamefont {Ciardi}},
  \bibinfo {author} {\bibfnamefont {V.}~\bibnamefont {Zampronio}}, \bibinfo
  {author} {\bibfnamefont {F.}~\bibnamefont {Cinti}},\ and\ \bibinfo {author}
  {\bibfnamefont {A.}~\bibnamefont {Mendoza-Coto}},\ }\bibfield  {title}
  {\bibinfo {title} {Self-induced {B}ose glass phase in quantum
  quasicrystals},\ }\href {https://doi.org/10.1016/j.rinp.2024.107991}
  {\bibfield  {journal} {\bibinfo  {journal} {Results in Physics}\ }\textbf
  {\bibinfo {volume} {65}},\ \bibinfo {pages} {107991} (\bibinfo {year}
  {2024})}\BibitemShut {NoStop}%
\bibitem [{\citenamefont {Jastrow}(1955)}]{jastrow1955many}%
  \BibitemOpen
  \bibfield  {author} {\bibinfo {author} {\bibfnamefont {R.}~\bibnamefont
  {Jastrow}},\ }\bibfield  {title} {\bibinfo {title} {Many-body problem with
  strong forces},\ }\href {https://doi.org/10.1103/PhysRev.98.1479} {\bibfield
  {journal} {\bibinfo  {journal} {Physical Review}\ }\textbf {\bibinfo {volume}
  {98}},\ \bibinfo {pages} {1479} (\bibinfo {year} {1955})}\BibitemShut
  {NoStop}%
\bibitem [{\citenamefont {Girvin}(2002)}]{girvin2002quantum}%
  \BibitemOpen
  \bibfield  {author} {\bibinfo {author} {\bibfnamefont {S.~M.}\ \bibnamefont
  {Girvin}},\ }\bibfield  {title} {\bibinfo {title} {The quantum {H}all effect:
  {N}ovel excitations and broken symmetries},\ }in\ \href
  {https://doi.org/10.1007/3-540-46637-1_2} {\emph {\bibinfo {booktitle}
  {Aspects topologiques de la physique en basse dimension. Topological aspects
  of low dimensional systems: Session LXIX. 7--31 July 1998}}}\ (\bibinfo
  {publisher} {Springer},\ \bibinfo {year} {2002})\ p.~\bibinfo {pages}
  {53}\BibitemShut {NoStop}%
\bibitem [{\citenamefont {Sodemann}\ and\ \citenamefont
  {MacDonald}(2013)}]{sodemann2013landau}%
  \BibitemOpen
  \bibfield  {author} {\bibinfo {author} {\bibfnamefont {I.}~\bibnamefont
  {Sodemann}}\ and\ \bibinfo {author} {\bibfnamefont {A.}~\bibnamefont
  {MacDonald}},\ }\bibfield  {title} {\bibinfo {title} {Landau level mixing and
  the fractional quantum {H}all effect},\ }\href
  {https://doi.org/10.1103/PhysRevB.87.245425} {\bibfield  {journal} {\bibinfo
  {journal} {Physical Review B}\ }\textbf {\bibinfo {volume} {87}},\ \bibinfo
  {pages} {245425} (\bibinfo {year} {2013})}\BibitemShut {NoStop}%
\bibitem [{\citenamefont {Szeg\H{o}}(1975)}]{szego1975orthogonal}%
  \BibitemOpen
  \bibfield  {author} {\bibinfo {author} {\bibfnamefont {G.}~\bibnamefont
  {Szeg\H{o}}},\ }\href@noop {} {\emph {\bibinfo {title} {Orthogonal
  polynomials}}}\ (\bibinfo  {publisher} {American Mathematical Society},\
  \bibinfo {year} {1975})\BibitemShut {NoStop}%
\end{thebibliography}%
